\documentclass[a4paper,12pt]{article}
\pdfoutput=1
\usepackage{a4wide}
\usepackage[latin1]{inputenc}
\usepackage{dsfont}
\usepackage{amsfonts}
\usepackage{amsmath}
\usepackage{amssymb}
\usepackage{amsthm}
\usepackage{graphicx}
\usepackage{booktabs}
\usepackage[numbers, sort]{natbib}
\allowdisplaybreaks[1]
\usepackage{color}
\usepackage{ulem}
\usepackage{tikz}
\usetikzlibrary{arrows}
\usepackage{floatrow}
\newfloatcommand{capbtabbox}{table}[][\FBwidth]
\usepackage{array}
\usepackage{cancel}

\usepackage[font=footnotesize,bf]{caption}
\usepackage{subcaption}

\def\tr{\mathop{\rm tr}}
\newcommand{\ci}{\text{i}}

\def\suT{\mathcal{T}}
\def\suF{\mathcal{F}}

\newcommand{\abs}[1]{\left| #1\right|}

\newcommand{\vev}[1]{\ensuremath{\langle #1 \rangle}}
\newcommand{\VEV}[1]{\ensuremath{\left\langle #1 \right\rangle}}
\newcommand{\GeV}{{\rm GeV}}
\newcommand{\TeV}{{\rm TeV}}
\newcommand{\gut}{{\text{GUT}}}
\newcommand{\susy}{{\text{SUSY}}}

\newcommand{\Mf}{{M^\text{dim=5}_{T}}}
\newcommand{\Mfoloop}{{M^\text{dim=5, 1-loop}_{T}}}
\newcommand{\Mftloop}{{M^\text{dim=5, 2-loop}_{T}}}
\newcommand{\Ms}{{M^\text{dim=6}_{T}}}
\newcommand{\Msoloop}{{M^\text{dim=6, 1-loop}_{T}}}
\newcommand{\Mstloop}{{M^\text{dim=6, 2-loop}_{T}}}
\newcommand{\MsT}{{M^\text{dim=6}_{T}}}
\newcommand{\MsTb}{{M^\text{dim=6}_{\bar T}}}

\numberwithin{equation}{section}

\begin{document}

\begin{titlepage}

\vspace*{-15mm}
\begin{flushright}
MPP-2014-240\\
TTP14-014
\end{flushright}
\vspace*{0.7cm}

\begin{center}
{
\bf\LARGE
Towards predictive flavour models in SUSY SU(5) GUTs
with doublet-triplet splitting
}
\\[8mm]
Stefan~Antusch$^{\star, \dagger}$
\footnote{E-mail: \texttt{stefan.antusch@unibas.ch}},
Ivo~de~Medeiros~Varzielas$^{\star}$
\footnote{E-mail: \texttt{ivo.de@unibas.ch}},
Vinzenz~Maurer$^{\star}$
\footnote{E-mail: \texttt{vinzenz.maurer@unibas.ch}},
Constantin~Sluka$^{\star}$~\footnote{E-mail: \texttt{constantin.sluka@unibas.ch}},
Martin~Spinrath$^{\ddag}$
\footnote{E-mail: \texttt{martin.spinrath@kit.edu}},
\\[1mm]
\end{center}
\vspace*{0.50cm}
\centerline{$^{\star}$ \it
 Department of Physics, University of Basel,}
\centerline{\it
Klingelbergstr.~82, CH-4056 Basel, Switzerland}
\vspace*{0.2cm}
\centerline{$^{\dagger}$ \it
Max-Planck-Institut f\"ur Physik (Werner-Heisenberg-Institut),}
\centerline{\it
F\"ohringer Ring 6, D-80805 M\"unchen, Germany}
\vspace*{0.2cm}
\centerline{$^{\ddag}$ \it
Institut f\"ur Theoretische Teilchenphysik, Karlsruhe Institute of Technology}
\centerline{\it
Engesserstra\ss{}e 7, D-76131 Karlsruhe, Germany }
\vspace*{1.20cm}

\begin{abstract}
\noindent

We discuss how the double missing partner mechanism solution to the
doublet-triplet splitting problem in four-dimensional supersymmetric
SU(5) Grand Unified Theories (GUTs) can be combined with predictive
models for the quark-lepton Yukawa coupling ratios at the GUT scale. 
It is argued that towards this goal a second SU(5) breaking Higgs
field in the adjoint representation is very useful and we discuss all
possible renormalizable superpotentials with two adjoint Higgs fields
and calculate the constraints on the GUT scale and effective triplet
mass from a two-loop gauge coupling unification analysis. Two explicit
flavour models with different predictions for the GUT scale Yukawa sector 
are presented, including shaping
symmetries and a renormalizable messenger sector. Towards calculating
the rates for proton decay induced by the exchange of colour triplets, 
the required Clebsch-Gordan coefficients for their couplings are calculated 
for the possible dimension five and six operators. They are provided in 
detailed tables in the appendix, together with additional helpful material
for GUT flavour model building. 

\end{abstract}

\end{titlepage}

\setcounter{footnote}{0}

\section{Introduction}

Grand Unified Theories (GUTs) offer one of the most attractive
extensions of the Standard Model of particle physics (SM), unifying
three of the four fundamental forces of nature. Towards a more fundamental
theory of nature it would, however, be desirable to also explain the pattern of fermion
masses and mixing with its plenty of parameters. 
Hence, successful models of flavour in GUTs
need to address two main issues, firstly, they have to achieve a working
mechanism for GUT symmetry breaking, including sufficient suppression of proton
decay without unacceptably large fine-tuning, and secondly, they should provide
correct predictions for the flavour observables, such as the
Yukawa coupling ratios and mixing angles. 

Existing models which manage to naturally suppress proton decay either predict the
unrealistic quark-lepton Yukawa relation $Y_e=Y_d^T$ (e.g.\ \cite{Berezhiani:1996nu}),
which may only be viable in the presence of extensive uncontrolled higher order
corrections (e.g.\ \cite{Altarelli:2000fu}) rendering the model non-predictive, 
or the experimentally disfavoured combination of the Georgi-Jarlskog relations
$y_\mu = - 3 y_s$ and $y_e = \frac{1}{3} y_d$ \cite{GJ}  (as e.g.\ in \cite{Murayama:1991ew}),
or rely on linear combinations of GUT Yukawa operators (e.g.\ \cite{Zhang:2012rc}),
which again implies the loss of predictivity.  Furthermore, there exists a large
number of GUT models which focus on the flavour sector, but do not include the
Higgs potential. 
So while there are several existing models focusing on one of these two concerns, we are not aware of any work to date capable of resolving the two challenges in full detail (and without invoking extra space-time dimensions) in a predictive setup, also given the present rather precise experimental data.

In this paper we employ the framework of supersymmetric (SUSY) GUTs,
as in the minimal supersymmetric extension of the SM (MSSM) the gauge couplings
unify to a surprising precision.
The scale where they unify (about $10^{16}$~GeV) is also large enough to sufficiently suppress proton decay from dimension six GUT operators \cite{Murayama:2001ur}.
However, in addition to proton decay mediated by additional heavy gauge bosons, in a GUT model the minimal embedding of the Higgs fields contains additional colour triplets which also lead to baryon number violating operators. Therefore the colour triplets have to be very heavy to suppress proton decay sufficiently or their couplings to the MSSM fields should be very strongly suppressed. Keeping the doublets of the SU(5) Higgs fields light, while generating a high enough mass for the colour triplets constitutes the so-called ``Doublet-Triplet Splitting (DTS) Problem''.

In SU(5) in four dimensions, a proposed solution to the DTS problem is the ``missing partner mechanism''
(MPM) \cite{Masiero:1982fe,Grinstein:1982um} or its improved version, the ``Double
Missing Partner Mechanism'' (DMPM) \cite{Hisano:1994fn} which we will
review briefly in the next section. In the existing models referred to above, either the MPM or the DMPM is applied.

Since GUTs not only unify the forces of the SM into a single GUT force, but also the fermions into joint GUT representations, they are indeed a promising starting point for addressing the flavour puzzle. More specifically, GUTs are capable of predicting the ratios between the Yukawa couplings of quarks and leptons at the GUT scale. After their renormalization group evolution to low energies (including supersymmetric 1-loop threshold corrections), these predictions can be compared to the experimental data for quark and lepton masses. As well known, the prediction of minimal SU(5) for the charged lepton
and down-type quark Yukawa matrix, the relation $Y_e=Y_d^T$ already mentioned above, is strongly disfavoured by the experimental results on the fermion masses. 
But also the ubiquitous proposal for more realistic ratios, 
the above mentioned Georgi-Jarlskog relations, 
obtained from the introduction of a 45-dimensional
Higgs representation of SU(5) and certain assumptions about the Yukawa
textures \cite{GJ}, are disfavoured by the recently improved data \cite{pdg}. 

To arrive at experimentally favoured predictions for GUT scale Yukawa coupling ratios, we employ an alternative approach \cite{Antusch:2009gu} involving higher-dimensional operators which contain a GUT breaking
Higgs field. Due to this, new Clebsch-Gordan (CG) factors appear as Yukawa coupling ratios, with interesting associated implications for the masses and mixing of the fermions, cf. \cite{Antusch:2011qg,Marzocca:2011dh} and  \cite{Antusch:2012fb,Meroni:2012ty,Antusch:2013kna,Antusch:2013tta}. The main goal of this paper is to show how GUT flavour models featuring these promising quark-lepton mass relations can be combined with a version of the DMPM for solving the DTS problem.  

As explicit examples, we will present two models with these properties that are ``UV complete'' in terms of messenger fields and 
employ sets of discrete Abelian symmetries (referred to as shaping symmetries)
such that only the desired effective GUT operators are generated when the heavy degrees of freedom are integrated out. The two models predict the GUT scale quark lepton Yukawa ratios but are not yet predictive for the fermion mixing parameters (although the experimentally observed values can be fitted by both of the models), so we only view them as existence proofs which show that DTS and predictive Yukawa coupling ratios can indeed be combined. The strategies discussed here, however, provide the tools for the construction of more ambitious GUT models of flavour, which should finally also predict the quark and lepton mixings and CP phases (and include the observed neutrino masses).  

The paper is organised as follows: We will start with a brief review of the MPM and DMPM and the recently proposed alternative Yukawa coupling ratios. We will also discuss the implications of replacing the 75-dimensional
representation used in the MPM and DMPM with an adjoint 24-dimensional representation
of SU(5). This choice is particularly well suited towards combining the DMPM with the novel 
CG factors. 
In section \ref{sec:GUT} we discuss the impact of the additional fields 
on gauge coupling unification and on their implications for the
colour triplet masses. We will especially focus here on the case of superpotentials with two
adjoint Higgs representations.
We then address the Yukawa sector in section \ref{sec:yukawas}, where we describe
the above-mentioned two predictive example models.
Before we summarise and conclude in section \ref{sec:conc},  in section \ref{sec:proton}
we briefly comment on proton decay, showing it is under
control in the proposed class of models.  We present additional helpful material for model building in the appendices.

\section{Strategy \label{sec:strat}}
In this section we present the general strategy that will be implemented in two example models.
We begin our presentation with the review of the MPM and DMPM and the origin of the CG coefficients
(coming from higher dimensional operators)
used in our models. Afterwards
we discuss a modification of the DMPM with a GUT Higgs field in the adjoint representation
and follow that with the actual DMPM realization implemented in our models, where a second Higgs field in the adjoint representation is added.

Throughout this section, for illustrative purposes, we consider that the bounds on proton decay rate require the effective mass of the colour triplets to be of at least \linebreak\mbox{$\Mf \approx 10^{17}~\GeV$} \cite{Goto:1998qg}, while the effective mass suppressing dimension six proton decay mediated by the colour triplets is required to be $\Ms \gtrsim 10^{12}~\GeV$ \cite{Nath:2006ut}.

\subsection{The missing partner mechanism}\label{subsec:mpm}
The basic idea of the missing partner mechanism (MPM) is the introduction of two
new superfields $Z_{50}$ and $\bar Z_{50}$ in $\mathbf{50}$ and $\mathbf{\overline{50}}$
representations of SU(5).
The decomposition of a $\mathbf{50}$ of SU(5) under the SM gauge group does
not contain an SU(2) doublet, but it includes an SU(3) triplet. Thus, using the
$50$-plets to generate an effective mass term keeps the electroweak doublets massless,
while the colour triplets acquire masses of the order of the GUT scale. The superpotential
for the MPM is given by\footnote{For simplicity we omit most order one coefficients in the superpotentials, except where they are relevant to the discussion.}
\begin{equation}
\label{eq:MPM75}
W_{\text{MPM}} = \bar H_5 H_{75} Z_{50} + \bar Z_{50}H_{75}H_5   + M_{50} Z_{50} \bar Z_{50}\;,
\end{equation}
where $H_{75}$ is a superfield transforming in the $\mathbf{75}$ representation of SU(5),
which contains a SM singlet. When $H_{75}$ gets a vacuum expectation value (VEV) SU(5)
is broken to the SM gauge group\footnote{$H_{75}$ can be replaced by the effective
combination $H_{24}^2/\Lambda$, where $H_{24}$ is the usual GUT-breaking Higgs field in the
$\mathbf{24}$ representation of SU(5) \cite{Berezhiani:1996nu}, see also section \ref{subsec:dmpm24}.}. With the triplet mass contribution from $\vev{H_{75}}$ denoted by
$V$, the mass matrices of the Higgs fields $H_5$, $\bar H_5$ and $Z_{50}$, $\bar Z_{50}$
are given by
\begin{equation}
m_D = 0\;,
\quad m_T = \begin{pmatrix}
0 & V \\ V & M_{50}
\end{pmatrix}\;,
\label{eq:mpmmass}
\end{equation}
for the doublet and triplet components $D$ and $T$ of $H_5$ and $Z_{50}$, respectively.
The dangerous terms for dimension five proton decay are obtained from the
Yukawa couplings
\begin{equation}
W_{\text{Yuk}} = \suT_i \suF_j \bar H_5 + \suT_i \suT_j H_5\;,
\end{equation}
where the families of the MSSM matter superfields are embedded in the standard way
in $\suT_i$ and $\suF_j$, transforming as $\mathbf{10}$ and $\mathbf{\bar 5}$ of SU(5),
respectively. To calculate the effective dimension five proton decay operators all Higgs
triplets from  5- and 50-dimensional representations have to be
integrated out, but only the triplets in the 5-dimensional representations dominantly
couple to matter. We denote the triplet mass eigenvalues with $\tilde M_1$ and $\tilde M_2$,
and the corresponding  mass eigenstates as $\tilde T_1$ and $\tilde T_2$, respectively. The
triplets that couple to matter are given by the combinations
\begin{equation} 
T^{(5)}=\sum_i U_{1i}^*\tilde T_i\;,\quad  \bar T^{(5)}=\sum_i V_{1i}\bar{\tilde T}_i\;,
\end{equation}
where $U$ and $V$ are unitary matrices defined by $m_T = U m_T^\text{diag} V^\dagger$.
Integrating out the triplet mass eigenstates $\tilde T_i$ leads to the effective dimension five
operator for proton decay, which is proportional to the inverse of the ``effective triplet mass'' 
\begin{equation}
(\Mf)^{-1}:=U_{1i}^* \left(m_T^\text{diag}\right)^{-1}_{ij}V_{1j}=U_{1i}^* \left(m_T^\text{diag}\right)^{-1}_{ij}V^T_{j1}=(m_T^{-1})_{11}\;.
\end{equation}
Integrating out the heavy colour triplet mass eigenstates also in the K\"ahler potential
\begin{equation}
K_T = T^{(5)}T^{(5)\dagger} + \bar T^{(5)}\bar T^{(5)\dagger} + T^{(50)}T^{(50)\dagger} + \bar T^{(50)}\bar T^{(50)\dagger}\;,
\end{equation}
effective dimension six K\"ahler operators emerge from inserting their equations of motion.
The Lagrangian obtained from the \mbox{D-terms} of $K_T$ contains baryon number
violating four fermion operators. These are proportional to
\begin{equation}
\left(\MsT\right)^{-2}:= V_{1i}\left(m_T^\text{diag}\right)^{-1}_{ij}U^\dagger_{jk}U_{km}\left(m_T^\text{diag}\right)^{-1}
_{ml}V_{l1}^\dagger = \left(m_T^{-1}m_T^{\dagger-1}\right)_{11}
\end{equation}
from $T^{(r)}T^{(r)\dagger}$ and to $\left(\MsTb\right)^{-2}=\left(m_T^{\dagger-1}m_T^{-1}\right)_{11}$ from $\bar T^{(r)}\bar T^{(r)\dagger}$.
With the mass matrix $m_T$ given in eq.~\eqref{eq:mpmmass}, the effective triplet mass is thus\footnote{
In the text when we quote numbers for $\Mf$, $\MsT$ and $\MsTb$ we will always refer to their absolute values.}
\begin{equation}
\Mf = \left(m_T^{-1}\right)_{11}^{-1}= -\frac{V^2}{M_{50}}\;,
\end{equation}
while the suppression of dimension six proton decay is given by
\begin{equation}
\left(\MsT\right)^2 = \left(\MsTb\right)^2= \left(m_T^{-1}m_T^{\dagger-1}\right)_{11}^{-1}=\frac{\abs{V}^4}{\abs{M_{50}}^2+\abs{V}^2}\;.
\end{equation}
Note that with a GUT scale value of $V \approx 10^{16}~\GeV$ and $M_{50}$ below the Planck scale, the dimension six proton decay is suppressed sufficiently with values of $\Ms$ between $10^{13}$ and $10^{16}~\GeV.$
Since the doublets obtain no mass terms, the splitting of doublet mass and effective triplet mass is achieved.
Using $\Mf \gtrsim 10^{17}~\GeV$ 
one obtains an upper bound for $M_{50} \lesssim 10^{15}~\GeV$. Having the large representations $\mathbf{50}$ and $\mathbf{\overline{50}}$ enter the Renormalization Group Equations (RGEs) at this low mass scale, however, leads to the break down of perturbativity just above the GUT scale. Thus, the MPM solves the DTS problem -- but trades it for SU(5) becoming non-perturbative much below the Planck scale $M_{\text{Pl}}$.

\subsection{The double missing partner mechanism}\label{subsec:dmpm}

This trade-off can be avoided in the double missing partner mechanism (DMPM),
where the number of Higgs fields in $\mathbf{5}$, $\mathbf{\bar 5}$, $\mathbf{50}$ and
$\mathbf{\overline{50}}$ representations gets doubled \cite{Hisano:1994fn}. The
fields $H_5$ and $\bar H_5$ couple to the matter fields $\suF_i$ and $\suT_i$,
whereas $H_5^\prime$ and $\bar H_5^\prime$ do not. The superpotential for the DMPM is given by
\begin{align}
\label{eq:DMPM75}
W_{\text{DMPM}}  & = \bar H_5H_{75} Z_{50} + \bar Z_{50} H_{75}H_5^\prime  + \bar H_5^\prime H_{75} Z_{50}^\prime + \bar Z_{50}^\prime H_{75}H_5 \nonumber \\ 
& + M_{50}Z_{50}\bar Z_{50}+ M_{50}^\prime Z^\prime_{50}\bar Z^\prime_{50}\nonumber \\
& +\mu^\prime H_5^\prime\bar H_5^\prime\;.
\end{align}
The mass matrices of the doublet and triplet components of the Higgs fields $H_5$, $H_5^\prime$,
$Z_{50}$, $Z^\prime_{50}$ and their corresponding barred fields after $H_{75}$ gets a VEV $V$
are given by
\begin{equation}
\label{eq:M_DMPM}
m_D = \begin{pmatrix}
0 & 0 \\ 
0 & \mu^\prime 
\end{pmatrix}\;,
\quad m_T = \begin{pmatrix}
0 & 0 & 0 & V\\ 
0 & \mu^\prime & V & 0 \\
V & 0 & M_{50} & 0\\
0 & V & 0 & M_{50}^\prime
\end{pmatrix}\;.
\end{equation}
While the Higgs doublets coupling to matter remain massless, the second pair of Higgs doublets
contained in $H_{5}^\prime$ and $\bar H^\prime_5$ has mass $\mu^\prime$. The improvement
of the DMPM compared to the MPM can be seen from the effective triplet mass $\Mf$
\begin{equation}
\Mf = \left(m_T^{-1}\right)_{11}^{-1} = -\frac{V^4}{\mu^\prime M_{50} M^\prime_{50}}\;.
\end{equation}
The same effective triplet mass of $\Mf\approx 10^{17}~\GeV$ can now be obtained
while keeping high masses $M_{50}\approx M_{50}^\prime \approx 10^{18}~\GeV$, provided
the heavier doublet pair has a (relatively) small mass $\mu^\prime\approx 10^{11}~\GeV$. With
the large representations of SU(5) having high masses, the perturbativity of the model
can be preserved up to (almost) the Planck scale. Dimension six proton decay is suppressed by
\begin{align}
\left(\MsT\right)^2 &= \left(m_T^{-1}m_T^{\dagger-1}\right)_{11}^{-1}=\frac{\abs{V}^8}{\abs{V}^6 + \abs{M_{50}}^2 (\abs{V}^4 + \abs{M_{50}^\prime\mu^\prime}^ 2 + \abs{V \mu^\prime}^2)}\approx (10^{14}~\GeV)^2\;,\\
(M_{\bar T}^\text{dim=6})^2 &= (m_T^{\dagger -1}m_T^{-1})_{11}^{-1}=
\frac{\abs{V}^8}{\abs{V}^6 + \abs{M_{50}^\prime}^2 (\abs{V}^4 + \abs{M_{50}\mu^\prime}^ 2 + \abs{V \mu^\prime}^2)}\approx (10^{14}~\GeV)^2\;,
\end{align}
in agreement with the bounds on proton decay.

\subsection{Planck-scale suppressed operators \label{PSO}}

The philosophy we follow in this paper is to consider all Planck-scale suppressed operators allowed by the symmetries.

The superpotentials in eq.~(\ref{eq:MPM75}) and eq.~(\ref{eq:DMPM75}) include mass terms for the 50-dimensional messengers. As such, one cannot use symmetries to forbid non-renormalizable Planck-scale suppressed operators such as $H_5 H_{75}^2 \bar{H}_5/M_{\text{Pl}}$ (for the MPM, and $H_5 H_{75}^2 \bar{H}'_5/M_{\text{Pl}}$ and $H'_5 H_{75}^2 \bar{H}_5/M_{\text{Pl}}$ for the DMPM). These Planck-scale suppressed operators do not involve the 50-dimensional messengers and therefore generate dangerously large contributions to the masses of the doublets contained in the 5-dimensional representations, effectively spoiling the mechanism. 

\begin{figure}
 \center
 \includegraphics[scale=0.75]{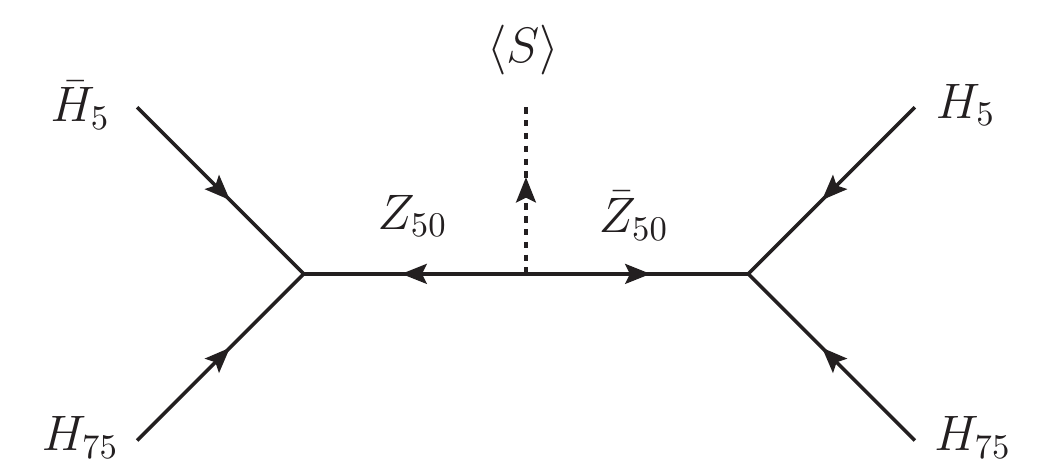}
 \caption{MPM diagram with an external $S$ field generating the mass term for the 50-dimensional messengers after getting a VEV.
 \label{fig:MPM_S}}
\end{figure}

Given our philosophy, we must forbid these operators through a shaping symmetry. The MPM and the DMPM can then be restored by adding a singlet field $S$, responsible for giving mass to the 50-dimensional superfields through couplings of the form $S Z_{50}\bar Z_{50}$, $S Z^\prime_{50}\bar Z^\prime_{50}$ and a VEV $\vev{S} \ne 0$, as seen in the diagram in figure \ref{fig:MPM_S} (note $S$ is not acting as an external field but generating the mass term). The non-trivial charge of $S$ under a shaping symmetry forbids the dangerous Planck-scale suppressed operators.

We will generalise this strategy of generating masses for the messenger
fields through an additional singlet field in sections \ref{subsec:dmpm24}
and \ref{sec:yukawas}, to avoid similarly Planck-scale suppressed operators
spoiling the DMPM or the predictions for the Yukawa coupling ratios.

\subsection{Yukawa coupling ratios}
\label{subsec:Clebsch}
The problem of non-viable GUT predictions for the fermion masses in the minimal SU(5)
model, such as $Y_e=Y_d^T$, can be solved through effective Yukawa couplings generated
from higher dimensional operators. When the higher dimensional operators contain a GUT
breaking Higgs field, new ratios between the Yukawa couplings of down-type quarks and
charged leptons can emerge once the GUT symmetry gets spontaneously broken \cite{Antusch:2009gu}.

Due to the introduction of extra SU(5) non-singlet fields, which participate in higher dimensional operators, there are in
general several ways to construct invariants, namely multiple ways to contract
the SU(5) indices. In the general case, such effective operators are not predictive and
introduce an arbitrary, linear combination of several CG factors.

This issue is generic in flavour models and can be resolved by constructing a specific UV completion
of the effective operators, see \cite{Varzielas:2010mp, Varzielas:2012ai} for mixing in lepton models, \cite{Antusch:2009gu, Antusch:2013rxa} for GUT relations, and
for applications in (GUT) flavour models, e.g.~\cite{Meroni:2012ty,Antusch:2013kna,Antusch:2013tta}. 
By introducing pairs of heavy messengers in the SU(5) representations $\mathbf{R}$ and $\mathbf{\bar R}$, the renormalizable couplings associated with the effective operators are specified and a unique contraction of SU(5) indices of the effective operator is obtained simply by integrating out the messenger fields.

\begin{figure}
    \begin{floatrow}
    \ffigbox{
        \includegraphics[page=1,scale=0.55]{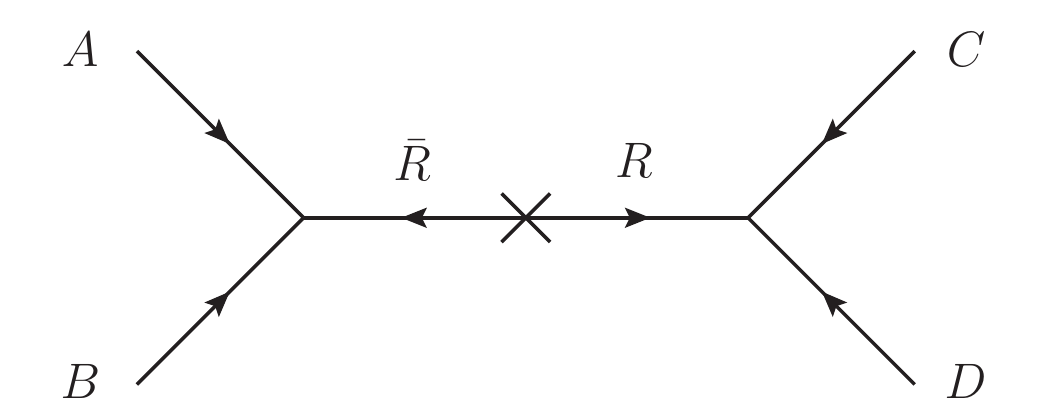}
    }{
        \caption{Supergraphs generating effectively Yukawa couplings
         upon integrating out the pair of messengers fields $R$
         and $\bar{R}$.}\label{fig:messengerdiagram1}
    }
    \capbtabbox{
        \caption{CG factors for the dimension five effective operators
        $W \supset (A B)_{R} (C D)_{\bar{R}}$. See the main text,
        Figure~\ref{fig:messengerdiagram1} and \cite{Antusch:2009gu,
        Antusch:2013rxa} for more details.} \label{tab:effectiveFToperators11}
        
    }{

        \rule{7mm}{0pt}\newline
        \begin{tabular}{cccc}
        \toprule
         $A \, B$ & $C \, D$ & $R$ & $(Y_e)_{ji} / (Y_d)_{ij}$  \\
        \midrule
         $H_{24} \, \suF$ & $\suT \, \bar H_{45}$ & $\overline{\mathbf{45}}$  &  $-\tfrac{1}{2}$  \\[0.3ex]
         $H_{24} \, \suF$ & $\suT \, \bar H_5$ & $\bar{\mathbf{5}}$          & $-\tfrac{3}{2}$  \\[0.3ex]
         $H_{24} \, \suT$ & $\suF \, \bar H_5$ & $\mathbf{10}$     &            $6$  \\
        \bottomrule
        \end{tabular}
    }
\end{floatrow}
\end{figure}

In Figure~\ref{fig:messengerdiagram1} and Table~\ref{tab:effectiveFToperators11},
we briefly review the topology of the diagram and field combinations that lead
to the CG factors $(Y_e)_{ji} / (Y_d)_{ij} = -\tfrac{1}{2}$, $-\tfrac{3}{2}$ and $6$ for
SU(5) GUTs, which have been shown to be useful for flavour model building
\cite{Antusch:2009gu, Antusch:2011qg,Marzocca:2011dh}.
To generate these relations we use a GUT breaking Higgs field $H_{24}$
in the adjoint representation of SU(5), and Higgs fields $\bar H_{5}$ and
$\bar H_{45}$ in the $\mathbf{\bar{5}}$ and $\mathbf{\overline{45}}$
representation. The factor $-\tfrac{1}{2}$ is, for instance, generated by the
coupling of $H_{24}$ and $\suF$ to a messenger field transforming as $\mathbf{45}$,
while its partner $\mathbf{\overline{45}}$ couples to $\suT$ and $\bar H_{45}$.
In our models in section \ref{sec:yukawas} the same CG factor $-\tfrac{1}{2}$ is 
obtained from a dimension six operator where $\bar{H}_{45}$ acts as heavy 
messenger field coupling to $H_{24}$ and $\bar{H}_5$.

So far we have discussed CG factors between the MSSM Yukawa couplings. For an
analysis of proton decay, the CG factors to the Yukawa couplings of the colour
Higgs triplets also play an important role. In Appendix~\ref{app:Clebsch} we present
an extensive discussion of CG factors in SU(5), including those.

\subsection{The double missing partner mechanism with an adjoint}
\label{subsec:dmpm24}

\begin{figure}
 \center
 \includegraphics[scale=0.75,page=1]{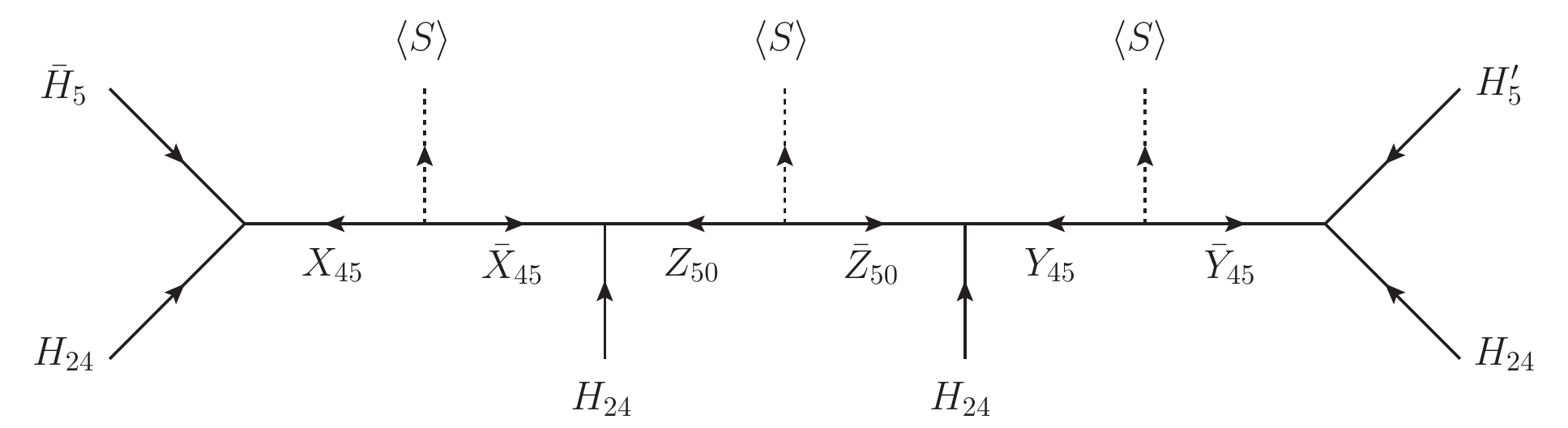}
 \includegraphics[scale=0.75,page=2]{DMPM24.pdf}
 \caption{Supergraphs generating the non-diagonal entries of the triplet mass matrix. 
 \label{Fig:Heff}}
\end{figure}

As we discussed in the previous subsection, the CG factors we want to combine
with the DMPM require the GUT breaking Higgs field to be in the adjoint representation
of SU(5), $\mathbf{24}$. This motivates us to replace the Higgs field $H_{75}$
needed for the DMPM with the effective combination $H^2_{24}/\Lambda$ \cite{Berezhiani:1996nu},
which at the renormalizable level can be obtained by integrating out heavy messenger
fields in the $\mathbf{45}$ and $\mathbf{\overline{45}}$ representations of SU(5) \cite{Zhang:2012rc}. To replace the $H_{75}$ in the MPM, we have to introduce
a set of messenger fields $X_{45}$, ${\bar X}_{45}$, $Y_{45}$ and
${\bar Y}_{45}$. For the DMPM, we also need to add a second set
$X_{45}^\prime$, ${\bar X}^\prime_{45}$, $Y_{45}^\prime$ and
${\bar Y}_{45}^\prime$. In Figure~\ref{Fig:Heff} we show the supergraphs generating
the non-diagonal entries of the triplet mass matrix in the DMPM with an adjoint $H_{24}$.

In analogy with the discussion in section \ref{PSO},
we avoid direct mass terms in order to forbid dangerous Planck-scale suppressed operators that would generate universal
mass contributions for Higgs doublets and triplets. The messenger pairs $X_{45}\bar X_{45}$,
$Y_{45}\bar Y_{45}$, $Z_{50}\bar Z_{50}$ and
their corresponding primed versions obtain masses from the VEV of a singlet
field $S$, charged under an additional shaping symmetry.
We specify these symmetries for two example models in section \ref{sec:yukawas}.

One may wonder if some of these heavy 45-dimensional messengers could be the
same, so that the number of fields in the spectrum would be reduced while preserving the structure of the mechanism.
But if either $X_{45} \equiv Y_{45}$ or
$X^\prime_{45} \equiv Y^\prime_{45}$, it can be seen from Figure~\ref{Fig:Heff} that supergraphs
without the $Z_{50}\bar Z_{50}$ mass insertion would be allowed, spoiling the splitting of
doublets and triplets and generating large non-diagonal entries in $m_D$.
In turn, if either $X_{45} \equiv X^\prime_{45}$, $Y_{45} \equiv Y^\prime_{45}$ or  $Z_{50} \equiv Z_{50}^\prime$, the DMPM is reduced to the MPM, reintroducing the issue of perturbativity.
Finally, an identification of $X_{45} \equiv Y^\prime_{45}$ would allow 
diagrams bypassing the 50-dimensional fields, generating unwanted mass term for both doublet and triplet components of $H_5$, $\bar H_5$, and thus a too large $\mu$-term.

With this messenger superfield content, we carefully checked that no dangerous Planck-scale suppressed operators spoil the mechanism. The renormalizable superpotential is:
\begin{align}
W_{\text{DMPM24}}  & = \bar H_5 H_{24}X_{45} + \bar X_{45}H_{24}Z_{50} + \bar Z_{50}H_{24}Y_{45}+\bar Y_{45}H_{24}H_5^\prime \nonumber \\
& + \bar H_5^\prime H_{24}X_{45}^\prime + \bar X_{45}^\prime H_{24}Z_{50}^\prime + \bar Z_{50}^\prime H_{24}Y_{45}^\prime+\bar Y_{45}^\prime H_{24}H_5 \nonumber \\
& + SX_{45}\bar X_{45} + SY_{45}\bar Y_{45} + SZ_{50}\bar Z_{50}
+ SX_{45}^\prime\bar X_{45}^\prime + SY_{45}^\prime\bar Y_{45}^\prime + SZ_{50}^\prime\bar Z_{50}^\prime \nonumber \\
& + \mu^\prime H_5^\prime \bar H_5^\prime\;.
\end{align}

After $H_{24}$ and $S$ obtain their VEVs and integrating out the 45-dimensional messenger fields, we find the mass matrices for the Higgs doublets
and triplets to be
\begin{equation}
m_D = \begin{pmatrix}
  0 & 0 \\
  0 & \mu^\prime
  \end{pmatrix}\;,\quad
  \quad m_T = \begin{pmatrix}
0 & 0 & 0 & -\frac{V^2}{\vev{S}}\\ 
0 & \mu^\prime & -\frac{V^2}{\vev{S}} & 0 \\
-\frac{V^2}{\vev{S}} & 0 & \vev{S} & 0\\
0 & -\frac{V^2}{\vev{S}} & 0 & \vev{S}
\end{pmatrix}\;,
\end{equation}
which one can easily compare to the ones of eq.~(\ref{eq:M_DMPM}) in section \ref{subsec:dmpm24}. Here $V$ is defined by $\vev{H_{24}} \equiv V \text{diag}(1,1,1,-\frac{3}{2},-\frac{3}{2})$. Integrating out the heavy 50-dimensional fields in a next step, the mass matrices become
\begin{equation}
m_D = \begin{pmatrix}
  0 & 0 \\
  0 & \mu^\prime
  \end{pmatrix}\;,\quad
  m_T = \begin{pmatrix}
0 & -\frac{V^4}{ \vev{S}^3} \\
-\frac{V^4}{\vev{S}^3} & \mu^\prime
\end{pmatrix}\;.
\end{equation}
Thus, the doublets in the pair $H_5 {\bar H}_{5}$ stay massless, while the doublet pair in $H^\prime_5{\bar H}^\prime_{5}$ is heavy. Using only $H_5$ and ${\bar H}_{5}$ for the Yukawa couplings of the SM fermions, the effective triplet component mass relevant for dimension 5 proton decay is given by
\begin{equation}\label{eq:MTeff24_definition}
\Mf = \left(m_T^{-1}\right)_{11}^{-1} = -\frac{V^8}{\vev{S}^6 \mu^\prime} \;.
\end{equation}
The effective masses suppressing dimension six proton decay mediated by colour triplets are given by
\begin{equation}\label{eq:MTeff624_definition}
\left(\MsT\right)^2 = \left(\MsTb\right)^2 \approx\frac{\abs{V}^8}{\abs{\vev{S}}^6}\;,
\end{equation}
where we used the fact that $\abs{\vev{S}}\gg \abs{V}$ and $\abs{\vev{S}^3\mu'}\ll \abs{V}^4$. Now the requirement of $\Ms \gtrsim 10^{12}~\GeV$ can only be obtained with $\vev{S}\approx 10^{18}~\GeV$ if the GUT scale value is larger than $V\gtrsim 10^{16}~\GeV$. In this case one needs $\mu'\approx 10^7~\GeV$ to obtain an effective triplet mass $\Mf\approx 10^{17}~\GeV$ for dimension five proton decay operators.
Therefore, if the GUT scale is high enough, the effective triplet mass can be large enough to
stabilize the proton, while the large SU(5) representations used in the DMPM can be heavy
enough to keep the theory perturbative up to the Planck scale.

When $H_{24}$ is uncharged under additional symmetries, having $\mu^\prime$ several
orders of magnitude smaller
than $V$ requires $\mu^\prime$ to arise from the spontaneous breakdown of a shaping symmetry,
to avoid the term $\vev{H_{24}}H^\prime_5\bar H^\prime_5$, which would give rise to a
much too small effective triplet mass. Note that the effective triplet masses entering dimension
five proton decay can be expressed
in terms of the mass eigenstates of doublet and triplet components as
$\Mf = - \tfrac{\tilde M_1\tilde M_2}{\mu^\prime}$. The effective triplet mass of dimension six proton decay is then excellently approximated by $\MsTb = \MsT \approx \sqrt{\Mf \mu^\prime}$.

\subsection{Introducing a second adjoint field \label{subsec:adjoints}}

We have seen that the DMPM with an adjoint GUT breaking Higgs instead of a
$\mathbf{75}$ already solves the DTS problem while providing the
necessary building block for the desirable CG factors for flavour model building, if the GUT scale is high enough. To
conclude this section, we will argue why it is compelling to further introduce a second adjoint Higgs field:

\begin{itemize}

\item 
In the minimal SUSY SU(5) model \cite{Dimopoulos:1981zb}, the single GUT breaking
$\mathbf{24}$ contains an SU(2) triplet component and an SU(3) octet component with
equal masses. Demanding gauge coupling unification, the mass of the Higgs colour
triplets is required to be about $10^{15}$ GeV \cite{Murayama:2001ur}, ruling out this model due to proton
decay. Non-renormalizable operators in the GUT breaking
superpotential can split the $\mathbf{24}$ component masses, 
allowing a higher effective triplet mass \cite{Bajc:2002pg} (see also section
\ref{sec:GUT}). An additional $\mathbf{24}$ can be used to realize this non-renormalizable superpotential 
in a renormalizable way.

\item It turns out that the introduction of an additional $\mathbf{24}$ is not just a UV-completion of the non-renormalizable superpotential of \cite{Bajc:2002pg}. When both adjoints have approximately the same mass and therefore the second $\mathbf{24}$ is not integrated out, the additional colour octet and electroweak triplet in the spectrum lead to more freedom for the GUT scale and effective triplet mass. In the following section we discuss all possible renormalizable superpotentials with two adjoints and their impact on $M_\gut$ and $\Mf$ from a gauge coupling unification analysis.

\item A renormalizable superpotential for one $\mathbf{24}$ requires it to be uncharged under shaping symmetries in order for it to obtain a VEV. 
However, such a shaping symmetry charge is vital in the type of flavour models considered here, to avoid unwanted admixtures of additional CG factors involving less insertions of $H_{24}$.
With a second $\mathbf{24}$, the adjoint fields can acquire non-vanishing VEVs even when charged under shaping symmetries.
\end{itemize}

These features of renormalizable superpotentials for two adjoints are presented in detail in the next section.

\section{Grand unification and the effective triplet mass \label{sec:GUT}}

In GUT extensions of the SM it is quite common to have additional fields below
the GUT scale that modify the RGE running, as it is the case for the class of models
in this paper. Therefore one has to study the impact of the additional fields on
the running of the gauge couplings and especially study their unification.
The modified unification condition for the gauge couplings at one-loop reads
\begin{equation}\label{eq:unification}
    \frac{1}{\alpha_u} = \frac{1}{\alpha_i} - \frac{1}{2 \pi} \left( 
        b_i^{\text{(SM)}} \log \frac{M_\susy}{M_Z} + b_i^{\text{(MSSM)}} \log \frac{M_\gut}{M_\susy} 
        + \sum\limits_f b_i^{(f)} \log \frac{M_\gut}{M_f} \right)
    \;,  
\end{equation}
where $i=1,2,3$ labels the SM gauge interaction and $f$ labels the additional
superfields (compared to the MSSM), with masses $M_f$ and $\beta$
coefficients $b_i^{(f)}$. The one-loop $\beta$-function coefficients
for the SM are $b_i^{\text{(SM)}} = (41/10,-19/6,-7)$ and for the MSSM
$b_i^{\text{(MSSM)}} = (33/5,1,-3)$. The SUSY scale $M_\susy$ is defined here as the scale 
where we make the transition from the SM $\beta$ coefficients to the MSSM ones.
The $\alpha_i$ are defined at 
low energies $\alpha_i \equiv \alpha_i(M_Z)$ while $\alpha_u$ is the unified
gauge coupling at the GUT scale $\alpha_u \equiv \alpha_i(M_\gut)$.
The GUT scale $M_\gut$ is defined here as the scale where the last SU(5)
multiplet is completed, in other words the scale where all three one-loop $\beta$
coefficients for the SM gauge couplings become equal.

From now on, we assume that the heaviest incomplete SU(5) multiplets to enter the
RGE running are the leptoquark vector bosons, such that the GUT scale corresponds
to their mass $M_\gut = M_V$. While other cases can certainly arise, we focus on
this option because it is quite common in our setup
to have heavy leptoquark vector bosons,
and furthermore we verified that in this case the triplet mass can be made
very heavy as well. 

In addition to the MSSM field content, the DMPM introduces one
additional pair of SU(2)-doublets, $D^{(5)}$ and ${\bar D}^{(5)}$ and two additional
pairs of SU(3)-triplets, $T^{(5)}_i$ and ${\bar T}^{(5)}_i$, $i = 1$, 2. They enter
the $\beta$-functions with the coefficients $b^{(5,D)}_i = (3/5,1,0)$ for the Dirac
pair of doublets and $b^{(5,T)}_i = (2/5,0,1)$ per Dirac pair of triplet and
anti-triplet.

Furthermore, we use two SU(5) breaking GUT Higgs fields $H_{24}$ and
$H'_{24}$ in the adjoint representation. They contain one SM-singlet component each, 
one SU(2)-triplet, $T^{(24)}$, with $b^{(24,T)}_i = (0,2,0)$, 
an SU(3)-octet, $O^{(24)}$ with $b^{(24,O)}_i = (0,0,3)$ and a leptoquark 
superfield pair, $L^{(24)}$ with $b^{(24,L)}_i = (5,3,2)$.
Since one leptoquark superfield pair is eaten up during the breaking of SU(5),
we are left with two triplets with masses $M_{T^{(24)}_1}$ and $M_{T^{(24)}_2}$,
two octets with masses $M_{O^{(24)}_1}$ and $M_{O^{(24)}_2}$ and one
leptoquark superfield pair with mass $M_{L^{(24)}}$.

For convenience we define the geometric means of the masses $M_{T^{(5)}}^2 = M_{T^{(5)}_1} M_{T^{(5)}_2}$
for the colour triplets and analogously $M_{T^{(24)}}^2 = M_{T^{(24)}_1} M_{T^{(24)}_2}$,
$M_{O^{(24)}}^2 = M_{O^{(24)}_1} M_{O^{(24)}_2}$ for the components of $H_{24}$ and $H'_{24}$.

Having this at hand, we can solve eq.~\eqref{eq:unification} for
$M_{D^{(5)}}$, $M_{T^{(5)}}$ and $M_\gut$, \footnote{In the following,
``$\log$'' of a mass is to be understood as the natural logarithm of the
mass divided by one common mass scale, e.g. $\log m \equiv \log(m/\GeV)$.}
\begin{align}
\log M_{D^{(5)}} &= \frac{15 \pi }{4 \alpha_1} 
        - \frac{17 \pi }{4 \alpha_2}
        - \frac{3 \pi }{2 \alpha_3} 
        + \frac{59}{3} \log M_Z \\\nonumber
        &+ \frac{2 \pi}{\alpha_u} 
        + \frac{3}{2} \log M_{L^{(24)}}
        - \frac{17}{2} \log M_{T^{(24)}}
        - \frac{9}{2} \log M_{O^{(24)}}
        - \frac{43}{6} \log M_\susy \;,\\
\log M_{T^{(5)}} &= \frac{35 \pi}{24 \alpha_1}
        - \frac{7 \pi}{8 \alpha_2}
        - \frac{19 \pi }{12 \alpha_3}
        + \frac{119}{12} \log M_Z \\\nonumber
        &+ \frac{\pi}{\alpha_u}
        + \frac{3}{4} \log M_{L^{(24)}}
        - \frac{7}{4} \log M_{T^{(24)}} 
        - \frac{19}{4} \log M_{O^{(24)}}
        - \frac{19}{6} \log M_\susy \;,\\
\log M_\gut &= \frac{5 \pi }{12 \alpha_1}
        - \frac{\pi }{4 \alpha_2}
        - \frac{\pi }{6 \alpha_3}
        + \frac{11}{6} \log M_Z \\\nonumber
        &+ \frac{1}{2} \log M_{L^{(24)}}
        - \frac{1}{2} \log M_{T^{(24)}}
        - \frac{1}{2} \log M_{O^{(24)}}
        - \frac{1}{3} \log M_\susy \;.
\end{align}

For the study of proton decay, it is more convenient to instead
solve eq.~\eqref{eq:unification} for the GUT scale gauge
coupling $\alpha_u$ and the effective triplet mass
$\Mf = M_{T^{(5)}}^2/M_{D^{(5)}}$, which gives
the suppression of the dimension 5 proton decay operators
(cf.~the discussion in section \ref{subsec:dmpm24}).
Then we get the relations
\begin{align}
\frac{\pi}{\alpha_u} &= -\frac{43 \pi}{24 \alpha_1}
        + \frac{15 \pi}{8 \alpha_2}
        + \frac{11 \pi}{12 \alpha_3}
        - \frac{197}{20} \log M_Z 
        + \frac{3}{5} \log M_{DT} \\\nonumber
        &- \frac{3}{4} \log M_{L^{(24)}}
        + \frac{15}{4} \log M_{T^{(24)}}
        + \frac{11}{4} \log M_{O^{(24)}}
        + \frac{7}{2} \log M_\susy \;,\\         
\log \Mf &= -\frac{5 \pi}{6 \alpha_1}
        + \frac{5\pi}{2 \alpha_2}
        - \frac{5\pi}{3 \alpha_3}
        + \frac{1}{6} \log M_Z \\\nonumber
        &+ 5 \log M_{T^{(24)}}
        - 5 \log M_{O^{(24)}}
        + \frac{5}{6} \log M_\susy \;,\\
\log M_\gut &= \frac{5 \pi}{12 \alpha_1}
        - \frac{\pi}{4 \alpha_2}
        - \frac{\pi }{6 \alpha_3}
        + \frac{11}{6} \log M_Z \\\nonumber
        &+ \frac{1}{2} \log M_{L^{(24)}}
        - \frac{1}{2} \log M_{T^{(24)}}
        - \frac{1}{2} \log M_{O^{(24)}}
        - \frac{1}{3} \log M_\susy \;,
\end{align}
where we have introduced the mass $M_{DT}^3 = M_{D^{(5)}}^2 M_{T^{(5)}}$. 
As one can see only $\alpha_u$ depends on $M_{DT}$, which is due to the fact that
doublets and colour triplets together form a complete representation of SU(5).
Thus, following eq.~\eqref{eq:unification}, one can see that a simultaneous rescaling
$M_{D^{(5)}} \to q^2 M_{D^{(5)}}$ and $M_{T^{(5)}} \to q M_{T^{(5)}}$ leaves the GUT
scale invariant and only shifts $\alpha_u$, while $\Mf \propto q^0$ remains
unchanged and $M_{DT} \propto q$ parametrises this rescaling.
Further interdependencies between $\alpha_u$, $\Mf$ and $M_\gut$ are then 
implicit via their shared dependence on the other masses.

Thus, unification implies that the effective triplet mass follows the relation
\begin{align}
    \Mf &= 
    \exp\left(\frac{5}{6} \pi \left(\frac{3}{\alpha_2}-\frac{2}{\alpha_3}-\frac{1}{\alpha_1}\right)\right) 
    M_Z^{1/6} M_\susy^{\frac 5 6} 
    \left(\frac{M_{T^{(24)}}}{M_{O^{(24)}}}\right)^5 \nonumber\\
    &= \label{eq:MTeff}
    2.5^{+0.6}_{-0.8} \cdot 10^{17} \; \GeV
    \left(\frac{M_\susy}{1 \, \TeV}\right)^{\frac 5 6} 
    \left(\frac{M_{T^{(24)}}}{M_{O^{(24)}}}\right)^5 \;,
\end{align}
while the GUT scale is given by
\begin{equation}\label{eq:MGUT}
    M_\gut = 
    1.37^{+0.05}_{-0.05} \cdot 10^{16} \; \GeV
    \left(\frac{M_\susy}{1 \, \TeV}\right)^{-\frac 1 3} 
    \left(\frac{M_{L^{(24)}}}{10^{16}\, \GeV}\right)^{\frac 1 2}
    \left(\frac{M_{T^{(24)}} M_{O^{(24)}}}{(10^{16}\, \GeV)^2} \right)^{-\frac 1 2} \;.
\end{equation}
For completeness, the unified gauge coupling is given by
\begin{align}\label{eq:alphau}
    \frac{1}{\alpha_u} &= 24.58 \pm 0.06 
    + \frac{7}{2\pi} \ln \frac{M_\susy}{1\, \TeV} 
    + \frac{3}{5\pi} \ln \frac{M_{DT}}{10^{14} \, \GeV} \\\nonumber 
    &\quad- \frac{3}{4\pi} \ln \frac{M_{L^{(24)}}}{10^{16}\, \GeV}
    + \frac{15}{4\pi} \ln \frac{M_{T^{(24)}}}{10^{16}\, \GeV}
    + \frac{11}{4\pi} \ln \frac{M_{O^{(24)}}}{10^{16}\, \GeV} \;.
\end{align}
For these numbers, we have used the experimental values and uncertainties for
the gauge couplings found in \cite{pdg}. Note that for all three quantities
the resulting uncertainty is dominated by the experimental error on
$\alpha_s$.
In the following we will not quote any errors on the masses anymore
since the relative uncertainty changes only negligibly for the different
superpotentials and for two-loop running.
The reference scale $10^{14}$~GeV is chosen due to the fact
that $M_{DT} = 10^{14}$~GeV and $M_{T^{(5)}} = 10^{16}$~GeV
implies $\Mf = 10^{19}$~GeV.

Since the effective triplet mass $\Mf$ receives significant
two-loop contributions (cf., for instance, \cite{Murayama:2001ur}), we have also
implemented a numerical two-loop RGE analysis using the following procedure.
We start with SM values for the gauge and Yukawa couplings \cite{Antusch:2013jca}
at $M_Z$, run up to to a scale of 1~TeV with the full two-loop SM RGEs
and match the SM to
the MSSM (including $\overline{\text{MS}}$ to $\overline{\text{DR}}$ scheme conversion). 
From there we run and match using full two-loop MSSM RGEs and one-loop gauge coupling
threshold corrections
\footnote{When one integrates out particles at a threshold
scale equal to their mass, these threshold corrections vanish, as can be seen in \cite{Hall:1980kf}.
}
while step-by-step including all additional multiplets at their mass scale via their
contributions to the one- and two-loop gauge coupling RGEs, see Appendix \ref{sec:2loopRGEs}.
The Yukawa couplings of the Higgs colour triplets are well approximated by using the
Yukawa couplings of the corresponding doublets.
We do not take into account any other Yukawa couplings. We verified this approximation
numerically and the results for the colour triplet masses are barely affected.

\subsection{Superpotentials with two adjoints of SU(5)}
\label{sec:W2Adjoints}

In this section, we systematically study all superpotentials with two adjoints
that can break SU(5) to the SM gauge group. We find only four possibilities with
non-vanishing VEVs and masses. Classified based on their symmetry, they are:
\begin{enumerate}    
    \item[(a)] $W = M_{24} \, \tr H_{24}^2 + M^\prime_{24} \tr H^{\prime 2}_{24} + \kappa^\prime \,  \tr H_{24}H_{24}^{\prime 2} + \lambda \, \tr H^3_{24}$, \\
        $\mathbb{Z}_2$ symmetry where $H_{24}$ is uncharged and $H_{24}^\prime$ charged.
    \item[(b)] $W = \tilde{M}_{24} \, \tr H_{24} H^\prime_{24} + \lambda \, \tr H_{24}^3 + \lambda^\prime \, \tr H^{\prime 3}_{24}$, \\
        $\mathbb{Z}_3$ symmetry, where $H_{24}$ has charge $2$ and $H'_{24}$ charge $1$.
    \item[(c)] $W = \tilde{M}_{24} \, \tr H_{24} H^\prime_{24} + \lambda \, \tr H_{24}^3 + \kappa^\prime \, \tr H_{24} H^{\prime 2}_{24}$, \\
        $\mathbb{Z}_4^R$ symmetry where $H_{24}$ has a charge of 2 (with $q_\theta = 1$) and $H'_{24}$ is uncharged. 
    \item[(d)] The trivial case with both fields only charged under SU(5) and all (non-linear) terms allowed.
        We will not consider this case any further.
\end{enumerate}

Since we are dealing with two adjoint Higgs fields, it is convenient to define
a quantity $\tan\beta_V$ similar to $\tan \beta$ of the MSSM, so that 
\begin{align}
    \vev{H_{24}} &= V_1 \,\, \text{e}^{\ci \phi_1} \,\, \text{diag}(1,1,1,-3/2,-3/2) \;,\label{eq:vevdef1}\\
    \vev{H'_{24}} &= V_2 \,\, \text{e}^{\ci \phi_2} \,\, \text{diag}(1,1,1,-3/2,-3/2) \;,\label{eq:vevdef2}
\end{align}
with $V_1, V_2 > 0$ and $\tan \beta_V = V_1 / V_2$. 

\subsubsection{Superpotential (a)}\label{subsec:superpotentiala}

\begin{figure}
\begin{center}
\begin{subfigure}{0.45\textwidth}
\begin{center}
\subcaption*{$\Mf$}
\includegraphics[width=1\textwidth]{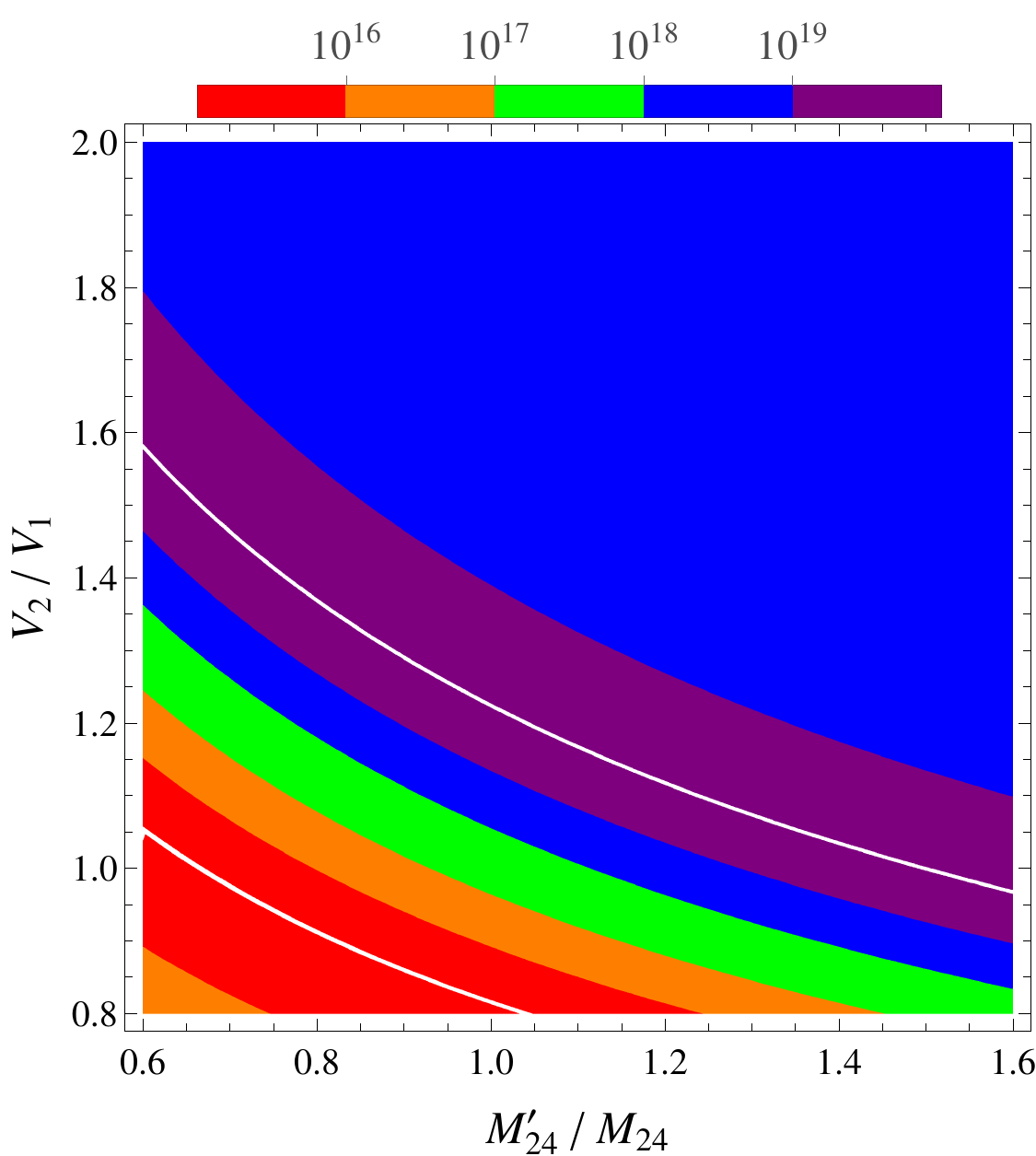}\\
\includegraphics[width=1\textwidth]{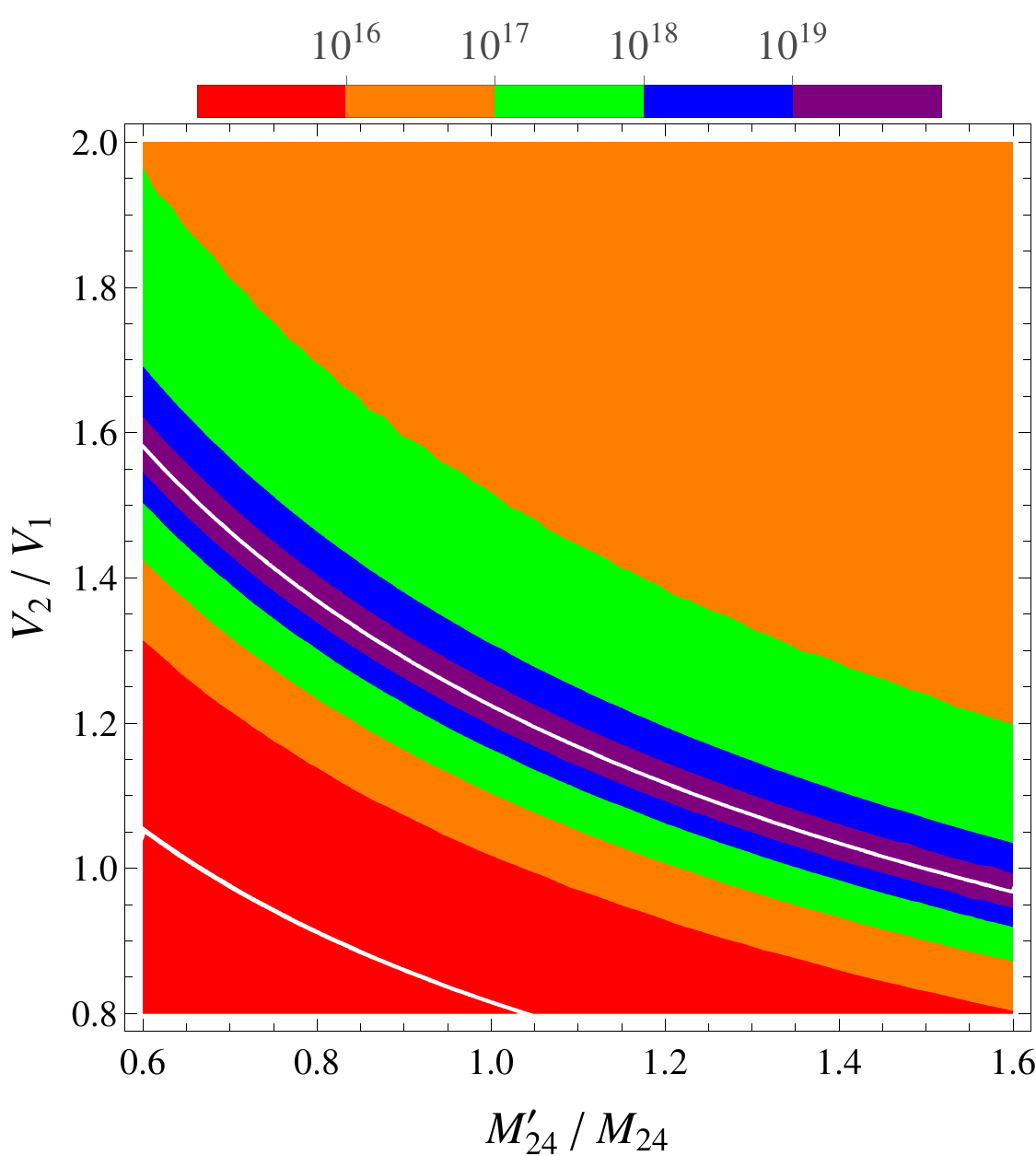}
\end{center}
\end{subfigure}
\begin{subfigure}{0.45\textwidth}
\begin{center}
\subcaption*{$M_\gut$}
\includegraphics[width=1\textwidth]{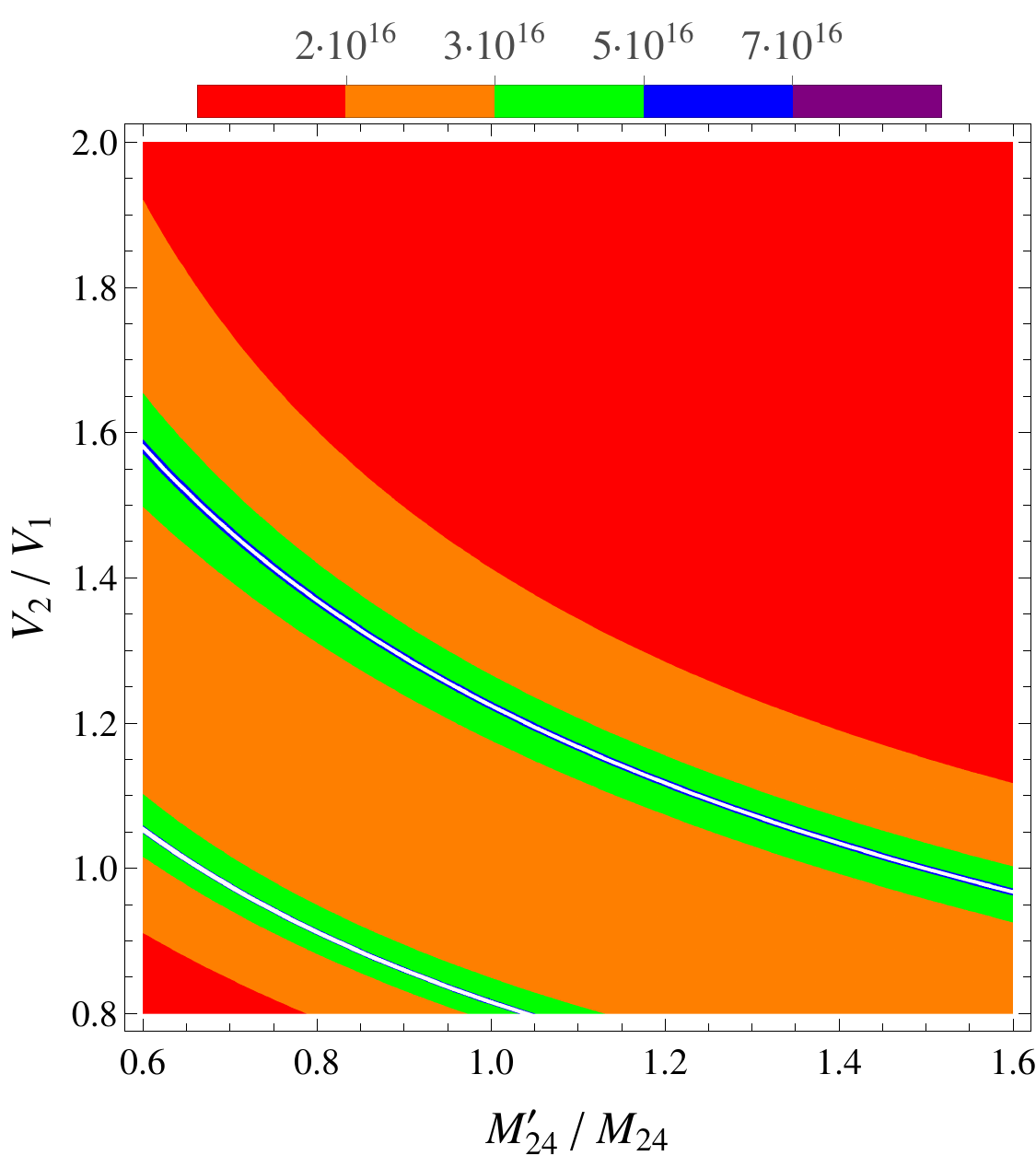}\\
\includegraphics[width=1\textwidth]{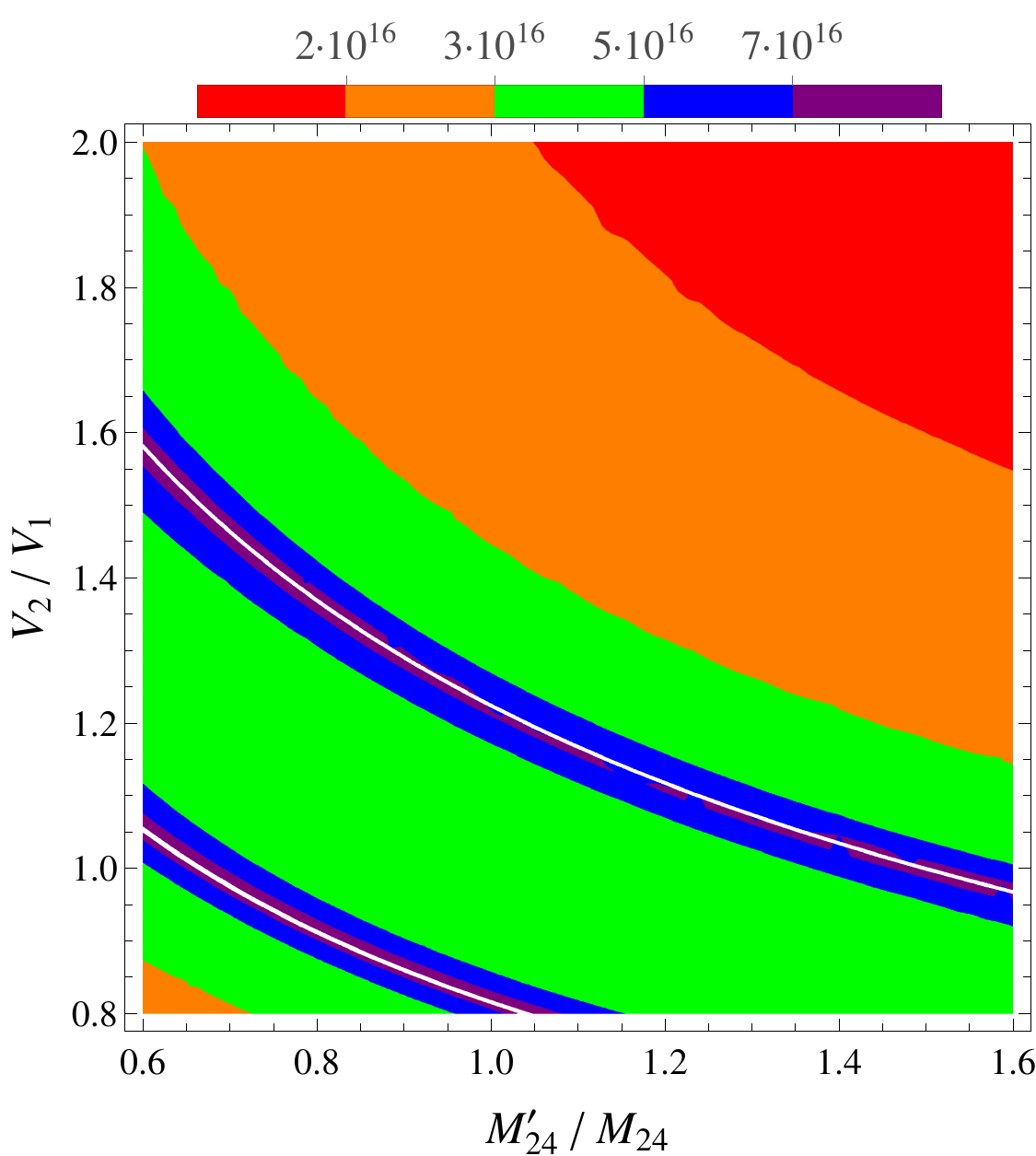}
\end{center}
\end{subfigure}
\end{center}
\caption{
The effective colour triplet mass $\Mf$ (left)
and GUT scale $M_\gut$ (right) in GeV at one-loop (upper)
and two-loop (lower) order as resulting from superpotential (a)
for $M_\susy = 1$~TeV, $M = 10^{15}$~GeV,
$M_{D^{(5)}}=1000$~TeV and $\bar{\phi} = 0$.
Note the different colour coding between left and right. 
For illustration, the white strips denote areas with light $M_{T^{(24)}}$
or $M_{O^{(24)}}$ ($<10^{13}$~GeV). Such relatively low values for these
components can arise either from cancellation between terms, or from a
generic suppression due to small parameters, cf.\ eqs.~\eqref{eq:24comps}.
}\label{fig:MTeff+MGUT}
\end{figure}

We will begin our discussion with superpotential (a) which turns out to
be the most complicated case since it has the most parameters.
As it contains two mass parameters, we introduce
a second angle $\beta_M$ and mean mass $M > 0$ such that
$M_{24} = M \text{e}^{\ci \alpha_1} \sin \beta_M$
and $M'_{24} = M \text{e}^{\ci \alpha_2} \cos \beta_M$. The vacuum expectation
values are given by
\begin{equation}
    V_1 \text{e}^{\ci \phi_1}= \frac{4 M'_{24}}{\kappa^\prime} \text{\quad and\quad}
    V_2^2 \text{e}^{2 \ci \phi_2}= 4 \, M'_{24} \frac{2 M_{24} \kappa^\prime - 3 M'_{24} \lambda}{\kappa^{\prime 3}} \;,
\end{equation}
which can as well be expressed in terms of a ratio of the coupling constants involved,
$3 \lambda / \kappa^\prime = 2 \text{e}^{\ci (\alpha_1 - \alpha_2)} \tan\beta_M - \text{e}^{-2 \ci (\phi_1 - \phi_2)} \cot^2\beta_V$. 
For the geometric means of the masses of the additional fields compared to the MSSM, we find
\begin{subequations}
\label{eq:24comps}
\begin{align}
    M_{T^{(24)}}^2 &= 5 \, M^2 \, \cos \beta_M \, 
        \sqrt{(2 \cos \beta_M - 3 \sin \beta_M \tan^2 \beta_V)^2 + \Delta} \;,\\
    M_{O^{(24)}}^2 &= 5 \, M^2 \, \cos \beta_M \, 
        \sqrt{(3 \cos \beta_M - 2 \sin \beta_M \tan^2 \beta_V)^2 + \Delta} \;, \\
    M_{L^{(24)}}^2 &= \frac{1}{4} \, M^2 \, \frac{\cos^2 \beta_M}{\sin^4 \beta_V} \;,
\end{align}
\end{subequations}
with $\Delta = 12 \, \sin(2 \beta_M) \, \cot^2(\beta_V) \, \sin^2\bar{\phi}$
and $\bar{\phi} = (\alpha_1-\alpha_2)/2 + \phi_1 - \phi_2$. Note that not only
the geometric mean masses, but also the mass eigenvalues themselves only depend
on this phase combination $\bar{\phi}$ and are invariant under
$\bar{\phi} \to \bar{\phi} + \pi$.

The effective triplet mass as of eq.~\eqref{eq:MTeff} is heaviest if the phase
$\bar{\phi}$ is $0$, $\pi$ or $2\pi$, since then the ratio
$M_{T^{(24)}} / M_{O^{(24)}}$ is not bounded from above (or below), which allows
for the maximal range for $\Mf$. Thus, in the following, we choose
$\bar{\phi}$ and thus $\Delta$ to vanish.

The resulting plots for $\Mf$ and for $M_\gut$ are shown
in Fig.~\ref{fig:MTeff+MGUT} for $M_\susy = 1$~TeV, $M= 10^{15}$~GeV,
$M_{D^{(5)}}=1000$~TeV and $\bar{\phi} = 0$,
including a comparison between one- and two-loop results.
Note however, that gauge coupling unification depends only weakly on $M_{D^{(5)}}$.
$\MsT$ and $\MsTb$ are again approximately given by $\sqrt{\Mf M_{D^{(5)}}}$.

\subsubsection{Superpotential (b) and (c)}
\label{sec:Superpotentialbc}

These two superpotentials have only one massive parameter $\tilde{M}_{24} = M \text{e}^{\ci \alpha}$
and hence the analytic results become much less cumbersome. 
The vacuum solutions are given by
\begin{equation}
    V_1 \text{e}^{\ci \phi_1}= \frac{2}{3} \frac{M \text{e}^{\ci \alpha}}{\sqrt[3]{\lambda^2 \lambda'}} \text{\quad and\quad} 
    V_2 \text{e}^{\ci \phi_2}= \frac{2}{3} \frac{M \text{e}^{\ci \alpha}}{\sqrt[3]{\lambda \lambda'^2}} \;,
\end{equation}
up to a $\mathbb{Z}_3$ symmetry transformation for superpotential (b) and 
\begin{equation}
    V_1 \text{e}^{\ci \phi_1}= \frac{1}{\sqrt{3}} \frac{M \text{e}^{\ci \alpha}}{\sqrt{\lambda \kappa'}} \text{\quad and\quad}
    V_2 \text{e}^{\ci \phi_2}= \frac{M \text{e}^{\ci \alpha}}{\kappa'} \;,
\end{equation}
up to a minus sign in $V_1$ for superpotential (c). In other words, the couplings
fulfill the relation $\lambda' / \lambda = \text{e}^{3 \ci (\phi_2 - \phi_1)}\tan^3\beta_V$
for superpotential (b) and $\kappa'/ \lambda = 3 \text{e}^{2 \ci (\phi_1 - \phi_2)}\tan^2\beta_V$
for superpotential (c).

For the geometric means of the masses of the colour triplets,
octets and the mass of the left-over leptoquark superfield in
$H_{24}$ and $H'_{24}$, we find
\begin{equation}
    M_{T^{(24)}}^2 = \frac{35}{4} \, M^2 \;,\;\;
    M_{O^{(24)}}^2 = \frac{15}{4} \, M^2 \text{ and }
    M_{L^{(24)}}^2 = \frac{1}{\sin^2 (2 \beta_V)} \, M^2 \;,
\end{equation}
for superpotential (b) and 
\begin{equation}
    M_{T^{(24)}}^2 = \frac{5}{4}                    \, M^2 \;,\;\;
    M_{O^{(24)}}^2 = \frac{5}{4}                    \, M^2 \text{ and }
    M_{L^{(24)}}^2 = \frac{1}{4 \sin^2 (2 \beta_V)} \, M^2 \;,
\end{equation}
for superpotential (c).
Note that in both cases also all mass eigenvalues turn out to be phase independent.
Therefore, applying eq.~\eqref{eq:MTeff} unification of
the gauge couplings on one-loop implies
\begin{equation} \label{eq:MTeffsup_bc}
    \Mfoloop = 2.5 \cdot 10^{17} \; \GeV \left(\frac{M_\susy}{1 \, \TeV}\right)^{\frac 5 6} 
        \times
        \begin{cases}
            \left(\frac 7 3\right)^{\frac 5 2} \approx 8.3 & \text{(b)} \\
            \quad1 & \text{(c)}
        \end{cases} \;,
\end{equation}
and
\begin{equation}\label{eq:MGUTsup_bc}
    M_\gut^{\text{1-loop}} = \frac{1.37 \cdot 10^{16} \; \GeV}{\sqrt{|\sin 2\beta_V|}}
    \left(\frac{M_\susy}{1 \, \TeV}\right)^{-\frac 1 3}
    \left(\frac{M}{10^{15} \, \GeV}\right)^{-\frac 1 2} 
    \times 
    \begin{cases}
        \sqrt{\frac{8}{\sqrt{21}}} \approx 1.32 & \text{(b)} \\
        2 & \text{(c)}
    \end{cases}
    \;.
\end{equation}
Assuming the same parameters, we find an almost
ten times heavier effective triplet mass in superpotential (b) than in (c) and hence we
focus on this case in our second example model later.

At two-loop, we find the following approximate behaviour for the masses
\begin{equation} \label{eq:MTeffsup_bc_2loop}
    \Mftloop = \left(\frac{M_\susy}{1 \, \TeV}\right)^{0.74} 
        \cdot
        \begin{cases}
            5.2 \cdot 10^{16} \; \GeV \left(\frac{M}{10^{15} \, \GeV}\right)^{-0.15} & \text{(b)} \\
            6.8 \cdot 10^{15} \; \GeV \left(\frac{M}{10^{15} \, \GeV}\right)^{-0.18} & \text{(c)}
        \end{cases} \;,
\end{equation}
and
\begin{equation}\label{eq:MGUTsup_bc_2loop}
    M_\gut^{\text{2-loop}} = |\sin 2\beta_V|^{-0.48}
    \left(\frac{M_\susy}{1 \, \TeV}\right)^{-0.4}
    \cdot 
    \begin{cases}
        2.89 \cdot 10^{16} \; \GeV \left(\frac{M}{10^{15} \, \GeV}\right)^{-0.61} & \text{(b)} \\
        4.78 \cdot 10^{16} \; \GeV \left(\frac{M}{10^{15} \, \GeV}\right)^{-0.63} & \text{(c)}
    \end{cases}
    \;.
\end{equation}
The dependence on other parameters is very small. As we can see, at two-loop, having $\Mf \gtrsim 10^{17}$ GeV requires $M_\susy \gtrsim 2.3$~TeV and $35$~TeV for superpotential (b) and (c) respectively. Again, $M_{D^{(5)}}=1000$~TeV has been fixed and the values of $\Msoloop$ and $\Mstloop$ can be approximated by the square root of the product of $M_{D^{(5)}}$ and $\Mfoloop$ or $\Mftloop$ respectively.

There are a few more comments in order. There is a claim \cite{Fallbacher:2011xg}
that the MSSM with an additional unbroken R-symmetry can not be obtained from
the spontaneous breaking of a four-dimensional (SUSY) GUT. Note that this is not in
conflict with our superpotentials, because the R-symmetry is either absent (a, b, d)
or spontaneously broken at the GUT scale (c). Superpotential (c) is particularly
interesting for model building purposes because R-symmetries are very popular in
flavour models with non-Abelian family symmetries (and spontaneous CP violation).
We will discuss this in more detail in appendix~\ref{app:SR}.

\section{Flavour Models with DMPM}
\label{sec:yukawas}

In this section we combine the DMPM (featuring two adjoints of SU(5))
with a predictive GUT flavour model for the quark-lepton Yukawa ratios at the GUT scale.
In particular, we implement
CG factors as given in \cite{Antusch:2009gu}. Two examples with different
Yukawa matrix structures are presented: in the
first model we construct diagonal down-type quark and charged lepton
Yukawa matrices $Y_d$ and $Y_e$, with all mixing originating from the up-type
quark Yukawa matrix $Y_u$. The second model realizes the attractive feature of
the Cabibbo mixing angle $\theta_C$ originating from $Y_d$. Both models 
are providing existence proofs that successful DTS and experimentally viable predictions 
for the GUT scale Yukawa coupling ratios can 
indeed be realised simultaneously in one model.

Let us be more specific on the predictions made by the two models: 
Due to the CG factors in the down-type quark and charged
lepton sector, $\tfrac{y_\tau}{y_b}$, $\tfrac{y_\mu}{y_s}$ and
$\tfrac{y_e}{y_d}$ are predicted at the GUT scale. To confront them with the 
experimental data, the RG running to low energies has to be performed, 
including in particular supersymmetric 1-loop threshold corrections \cite{SUSYthresholds}
when the  MSSM is matched to the SM. These threshold corrections can have
a sizeable impact on the low energy values of the Yukawa couplings (and thus the fermion masses), 
depending on the sparticle spectrum and $\tan \beta$.
So the predictions here are two-fold: Firstly, the predictions for the Yukawa ratios at the GUT scale imply 
constraints on the SUSY breaking parameters, which may be tested at future collider searches if 
SUSY is found\footnote{To make explicit statements about the constraints on the SUSY parameters one would have to specify the model of SUSY breaking, which is beyond the scope of this paper. A discussion and an explicit example where such constraints are worked out can be found, e.g.\ in \cite{Antusch:2011sq}.}. Secondly, the ratios $\tfrac{y_d}{y_s}$ and $\tfrac{y_e}{y_\mu}$
are not affected by RG running and by the SUSY threshold corrections (as long as the first two families 
of sfermions are almost degenerate in mass as commonly assumed). They can be directly used to
constrain GUT models. A particularly useful quantity in this context is indeed the 
double ratio
\begin{equation}
\left| \frac{y_\mu}{y_s}\frac{y_d}{y_e} \right| = 10.7^{+1.8}_{-0.8}\;,
\end{equation}
which can be checked directly at the GUT scale \cite{Antusch:2013jca}.
In our models we will use the CG factors $\tfrac{y_\tau}{y_b} = - \tfrac{3}{2}$,
$\tfrac{y_\mu}{y_s} = 6$ and $\tfrac{y_e}{y_d} = - \tfrac{1}{2}$. This leads to
$\left| \tfrac{y_\mu}{y_s} \tfrac{y_d}{y_e} \right| =12$, which is in good agreement with the
phenomenological value. On the other hand, the ubiquitous CG factors
$y_\mu = - 3 y_s$ and $y_e = \frac{1}{3} y_d$, known as Georgi-Jarlskog relations
\cite{GJ}, would give $\left| \tfrac{y_\mu}{y_s} \tfrac{y_d}{y_e} \right|=9$ and deviate
from the central value $10.7$ by more than two sigma.

In this section we explicitly construct only the Yukawa matrices of the up-
and down-type quarks and charged leptons. Adding one of the ubiquitous mechanisms
to generate neutrino masses and lepton mixing angles would be straightforward.
However, we do not consider neutrinos in this paper, since they are not directly relevant
for the discussion of proton decay, doublet-triplet splitting and the CG factors
between $Y_d$ and $Y_e$. 

\subsection[A model with diagonal $Y_d$ and $Y_e$ Yukawa matrices]
{A model with diagonal $\boldsymbol{Y_d}$ and $\boldsymbol{Y_e}$ Yukawa matrices\label{sec:modelA}}

We now turn to our first model featuring diagonal down-type quark and charged
lepton Yukawa matrices $Y_d$ and $Y_e$.\footnote{The matrices are diagonal in the
preferred basis where the different fermion generations have well defined symmetry
assignments, cf.\ Table~\ref{tab:A_MSSM}.} In this case all the mixing in the quark
sector has to come exclusively from the up-type quark Yukawa matrix $Y_u$. Explicitly,
we have the following structure for the Yukawa matrices
\begin{equation}
\label{eq:A_Y}
Y_d = \begin{pmatrix}
y_d & 0 & 0\\
0 & y_s & 0\\
0 & 0 & y_b
\end{pmatrix},\;
Y_e=\begin{pmatrix}
-\frac{1}{2}y_d & 0& 0\\
0 & 6 y_s & 0\\
0 & 0 & -\frac{3}{2}y_b
\end{pmatrix},\;
Y_u =\begin{pmatrix}
y_{11} & y_{12} & y_{13}\\
y_{12} & y_{22} & y_{23}\\
y_{13} & y_{23} & y_{33}
\end{pmatrix}\;.
\end{equation}
An approach to flavour (GUT) model building with diagonal $Y_e$ (and $Y_d$) has been
discussed recently in \cite{King:2013hoa}.

We introduce flavon fields $\theta_1$, $\theta_2$, $\theta_3$ and $\theta_4$ that obtain a VEV and generate
the hierarchical structure of the Yukawa matrices. After the flavon fields, $H_{24}$
and $H_{24}^\prime$ obtain their VEVs, the Yukawa matrices of eq.~\eqref{eq:A_Y} originate
from the following effective superpotentials
\begin{align}
W_u &= \frac{1}{\Lambda^4} H_5 {\cal T}_1 {\cal T}_1 \theta_1^2\theta_2^2 
     + \frac{1}{\Lambda^3} H_5 {\cal T}_1 {\cal T}_2 \theta_1^2 \theta_2
     + \frac{1}{\Lambda^2} H_5 {\cal T}_1 {\cal T}_3 \theta_1 \theta_2
     + \frac{1}{\Lambda^2} H_5 {\cal T}_2 {\cal T}_2 \theta_1^2 \nonumber\\
   &\phantom{=}+ \frac{1}{\Lambda}   H_5 {\cal T}_2 {\cal T}_3 \theta_1
     + H_5 {\cal T}_3 {\cal T}_3\;,\label{eq:A_Wu}\\
W_d &= \frac{1}{\vev{S^\prime}} (H_{24}^\prime {\suF}_3)_{\bar{5}} ({\bar H}_{5} {\suT}_3)_{5}
     + \frac{\theta_3}{\vev{S^\prime}^2} (H_{24}^\prime {\suT}_2)_{10} ( {\bar H}_{5} {\suF}_2)_{\overline{10}} \nonumber\\
     &\phantom{=}+ \frac{\theta_4}{\vev{S^\prime}^2 \vev{S}} (H_{24}^\prime {\suF}_1)_{\overline{45}} ( {\suT}_1 H_{24} {\bar H}_{5})_{45}\;,\label{eq:A_Wd}
\end{align}
where we do not show order one coefficients, and denote the different
messenger masses generating $W_u$ by a generic $\Lambda$. However, keep
in mind that this is just for the sake of simplicity and different entries
in the Yukawa matrix should be understood as independent parameters.
The ratios of flavon VEVs and messenger masses is small of about $0.01 - 0.1$.
For a list of all fields including their
charges under the additional discrete shaping symmetries, see Tables~\ref{tab:A_MSSM},
\ref{tab:A_VEV}, \ref{tab:A_DMPM}, and \ref{tab:A_messengers}.

\begin{table}
  \begin{tabular}{cccccccccc}
  \toprule
  & SU(5) & $\mathbb{Z}_{2}$ & $\mathbb{Z}_{4}$ & $\mathbb{Z}_{4}$ & $\mathbb{Z}_{4}$ & $\mathbb{Z}_{7}$ & $\mathbb{Z}_{7}$ & $\mathbb{Z}_{9}$ & $\mathbb{Z}_{2}$ \\
 \midrule 
$H_5$ & $\mathbf{5} $ & $.$ & $.$ & $.$ & $.$ & $.$ & $.$ & $.$ & $.$ \\
$\bar H_{5}$ & $\mathbf{\bar 5} $ & $.$ & $2$ & $.$ & $.$ & $.$ & $1$ & $2$ & $.$ \\
$\suT_1$ & $\mathbf{10} $ & $.$ & $.$ & $3$ & $.$ & $6$ & $.$ & $.$ & $1$ \\
$\suT_2$ & $\mathbf{10} $ & $.$ & $.$ & $.$ & $.$ & $6$ & $.$ & $.$ & $1$ \\
$\suT_3$ & $\mathbf{10} $ & $.$ & $.$ & $.$ & $.$ & $.$ & $.$ & $.$ & $1$ \\
$\suF_1$ & $\mathbf{\bar 5} $ & $1$ & $1$ & $1$ & $2$ & $1$ & $1$ & $2$ & $1$ \\
$\suF_2$ & $\mathbf{\bar 5} $ & $1$ & $.$ & $.$ & $2$ & $1$ & $.$ & $2$ & $1$ \\
$\suF_3$ & $\mathbf{\bar 5} $ & $1$ & $2$ & $.$ & $1$ & $.$ & $6$ & $7$ & $1$ \\
  \bottomrule
 \end{tabular}
 \caption{
   SU(5) representations and charges under discrete shaping symmetries
   of the MSSM fields and colour triplets of the model presented in subsection \ref{sec:modelA}. A dot denotes charge zero.
   \label{tab:A_MSSM}
   }
\end{table}

\begin{table}
\centering
\begin{tabular}{cccccccccc}
\toprule
& SU(5) & $\mathbb{Z}_{2}$ & $\mathbb{Z}_{4}$ & $\mathbb{Z}_{4}$ & $\mathbb{Z}_{4}$ & $\mathbb{Z}_{7}$ & $\mathbb{Z}_{7}$ & $\mathbb{Z}_{9}$ & $\mathbb{Z}_{2}$ \\
 \midrule 
$H_{24}$ & $\mathbf{24} $ & $.$ & $.$ & $.$ & $.$ & $.$ & $.$ & $.$ & $.$ \\
$H_{24}^\prime$ & $\mathbf{24} $ & $1$ & $.$ & $.$ & $.$ & $.$ & $.$ & $.$ & $.$ \\
$S$ & $\mathbf{1} $ & $.$ & $3$ & $.$ & $.$ & $.$ & $2$ & $.$ & $.$ \\
$S^\prime$ & $\mathbf{1} $ & $.$ & $.$ & $.$ & $1$ & $.$ & $.$ & $.$ & $.$ \\
$\theta_1$ & $\mathbf{1} $ & $.$ & $.$ & $.$ & $.$ & $1$ & $.$ & $.$ & $.$ \\
$\theta_2$ & $\mathbf{1} $ & $.$ & $.$ & $1$ & $.$ & $.$ & $.$ & $.$ & $.$ \\
$\theta_{3}$ & $\mathbf{1} $ & $.$ & $2$ & $.$ & $.$ & $.$ & $6$ & $5$ & $.$ \\
$\theta_{4}$ & $\mathbf{1} $ & $.$ & $.$ & $.$ & $.$ & $.$ & $.$ & $5$ & $.$ \\
\bottomrule
 \end{tabular}
 \caption{
   SU(5) representations and charges under discrete shaping symmetries
   of the superfields obtaining VEVs at around the GUT scale of the model presented in subsection \ref{sec:modelA}. A dot denotes charge zero.
   \label{tab:A_VEV}
   }
\end{table}

\begin{table}
  \begin{tabular}{cccccccccc}
  \toprule
  & SU(5) & $\mathbb{Z}_{2}$ & $\mathbb{Z}_{4}$ & $\mathbb{Z}_{4}$ & $\mathbb{Z}_{4}$ & $\mathbb{Z}_{7}$ & $\mathbb{Z}_{7}$ & $\mathbb{Z}_{9}$ & $\mathbb{Z}_{2}$ \\
 \midrule 
$H_{5}^\prime$ & $\mathbf{5} $ & $.$ & $3$ & $.$ & $.$ & $.$ & $5$ & $7$ & $.$ \\
$\bar H_{5}^\prime$ & $\mathbf{\bar 5} $ & $.$ & $1$ & $.$ & $.$ & $.$ & $6$ & $.$ & $.$ \\
$X_{45}$ & $\mathbf{45} $ & $.$ & $2$ & $.$ & $.$ & $.$ & $6$ & $7$ & $.$ \\
$\bar X_{45}$ & $\mathbf{\overline{45}} $ & $.$ & $3$ & $.$ & $.$ & $.$ & $6$ & $2$ & $.$ \\
$Y_{45}$ & $\mathbf{45} $ & $.$ & $.$ & $.$ & $.$ & $.$ & $3$ & $7$ & $.$ \\
$\bar Y_{45}$ & $\mathbf{\overline{45}} $ & $.$ & $1$ & $.$ & $.$ & $.$ & $2$ & $2$ & $.$ \\
$Z_{50}$ & $\mathbf{50} $ & $.$ & $1$ & $.$ & $.$ & $.$ & $1$ & $7$ & $.$ \\
$\bar Z_{50}$ & $\mathbf{\overline{50}} $ & $.$ & $.$ & $.$ & $.$ & $.$ & $4$ & $2$ & $.$ \\
$X_{45}^\prime$ & $\mathbf{45} $ & $.$ & $3$ & $.$ & $.$ & $.$ & $1$ & $.$ & $.$ \\
$\bar X_{45}^\prime$ & $\mathbf{\overline{45}} $ & $.$ & $2$ & $.$ & $.$ & $.$ & $4$ & $.$ & $.$ \\
$Y_{45}^\prime$ & $\mathbf{45} $ & $.$ & $1$ & $.$ & $.$ & $.$ & $5$ & $.$ & $.$ \\
$\bar Y_{45}^\prime$ & $\mathbf{\overline{45}} $ & $.$ & $.$ & $.$ & $.$ & $.$ & $.$ & $.$ & $.$ \\
$Z_{50}^\prime$ & $\mathbf{50} $ & $.$ & $2$ & $.$ & $.$ & $.$ & $3$ & $.$ & $.$ \\
$\bar Z_{50}^\prime$ & $\mathbf{\overline{50}} $ & $.$ & $3$ & $.$ & $.$ & $.$ & $2$ & $.$ & $.$ \\
  \bottomrule
  \end{tabular}
  \caption{
   SU(5) representations and charges under discrete shaping symmetries
   of the fields in the DMPM sector of the model presented in subsection \ref{sec:modelA}. A dot denotes charge zero.
   \label{tab:A_DMPM}
   }
\end{table}

\begin{table}
  \centering
  \begin{tabular}{cccccccccc}
  \toprule
& SU(5) & $\mathbb{Z}_{2}$ & $\mathbb{Z}_{4}$ & $\mathbb{Z}_{4}$ & $\mathbb{Z}_{4}$ & $\mathbb{Z}_{7}$ & $\mathbb{Z}_{7}$ & $\mathbb{Z}_{9}$ & $\mathbb{Z}_{2}$ \\
 \midrule 
$Z_{5,1}$ & $\mathbf{5} $ & $.$ & $2$ & $.$ & $3$ & $.$ & $1$ & $2$ & $1$ \\
$\bar{Z}_{5,1}$ & $\mathbf{\bar 5} $ & $.$ & $2$ & $.$ & $.$ & $.$ & $6$ & $7$ & $1$ \\
$Z_{10,1}$ & $\mathbf{10} $ & $1$ & $.$ & $.$ & $3$ & $6$ & $.$ & $.$ & $1$ \\
$\bar{Z}_{10,1}$ & $\mathbf{\overline{10}} $ & $1$ & $.$ & $.$ & $.$ & $1$ & $.$ & $.$ & $1$ \\
$Z_{10,2}$ & $\mathbf{10} $ & $1$ & $2$ & $.$ & $2$ & $6$ & $6$ & $5$ & $1$ \\
$\bar{Z}_{10,2}$ & $\mathbf{\overline{10}} $ & $1$ & $2$ & $.$ & $1$ & $1$ & $1$ & $4$ & $1$ \\
$Z_{45,1}$ & $\mathbf{45} $ & $.$ & $3$ & $3$ & $2$ & $6$ & $6$ & $7$ & $1$ \\
$\bar{Z}_{45,1}$ & $\mathbf{\overline{45}} $ & $.$ & $1$ & $1$ & $1$ & $1$ & $1$ & $2$ & $1$ \\
$Z_{45,2}$ & $\mathbf{45} $ & $.$ & $3$ & $3$ & $3$ & $6$ & $6$ & $2$ & $1$ \\
$\bar{Z}_{45,2}$ & $\mathbf{\overline{45}} $ & $.$ & $1$ & $1$ & $.$ & $1$ & $1$ & $7$ & $1$ \\
\midrule
$Z_{10,3}$ & $\mathbf{10} $ & $.$ & $.$ & $.$ & $.$ & $1$ & $.$ & $.$ & $1$ \\
$Z_{10,4}$ & $\mathbf{10} $ & $.$ & $.$ & $1$ & $.$ & $1$ & $.$ & $.$ & $1$ \\
$Z_{1}$ & $\mathbf{1} $ & $.$ & $.$ & $.$ & $.$ & $5$ & $.$ & $.$ & $.$ \\
$Z_{2}$ & $\mathbf{1} $ & $.$ & $.$ & $3$ & $.$ & $5$ & $.$ & $.$ & $.$ \\
  \bottomrule
  \end{tabular}
  \caption{
    SU(5) representations and charges under discrete shaping symmetries
    of the flavon and flavour messenger fields of the model presented in subsection \ref{sec:modelA}.
    Note that the messengers $Z_{5,1}\bar Z_{5,1}$, $Z_{10,1}\bar Z_{10,1}$,
    $Z_{10,2}\bar Z_{10,2}$, $Z_{45,1}\bar Z_{45,1}$ and $Z_{45,2}\bar Z_{45,2}$
    have no direct mass term, but get their masses through the VEVs of $S$ and $S'$.
    The other messenger fields have direct mass terms, so their corresponding barred
    field is omitted in this table. A dot denotes charge zero.
    \label{tab:A_messengers}
    }
\end{table}

The adjoint $H_{24}'$ required to
construct the desired CG factors must be charged under shaping symmetries. 
We will therefore implement superpotential (a). Note that it leaves the second
adjoint $H_{24}$ uncharged, which could in principle lead to a problem. 
A direct mass term of messenger fields $M_i Z_i\bar Z_i$ would
in this case always show a up with a term of the form $H_{24} Z_i\bar Z_i$.
Such a contribution would inevitably spoil the desired CG factors~\cite{Antusch:2013rxa}
between $Y_d$ and $Y_e^T$ as long as the mass and the
adjoint VEV are not very hierarchical. To avoid this and still generate the desired operators
the masses of the messenger fields that give rise to $W_d$ in eq.~(\ref{eq:A_Wd}) originate
from the VEVs of the fields $S$ and $S^\prime$
charged under the shaping symmetry (but with different charges than $H_{24}'$)
\footnote{The VEV of the charged singlet $S$ gives masses to
the heavy messengers of the DMPM (see section~\ref{subsec:dmpm24}).}.
 
\begin{figure}
\centering
\includegraphics[scale=0.5]{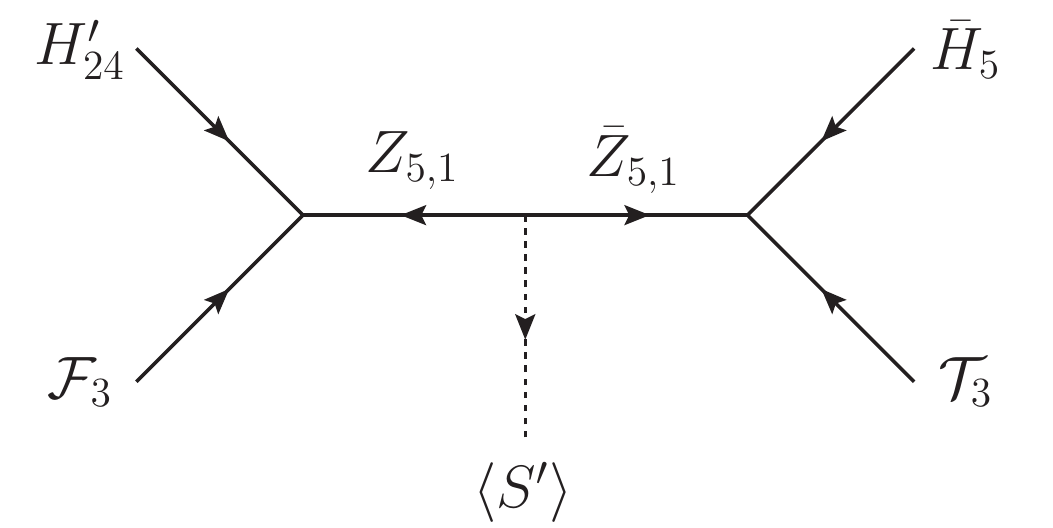}
\includegraphics[scale=0.5]{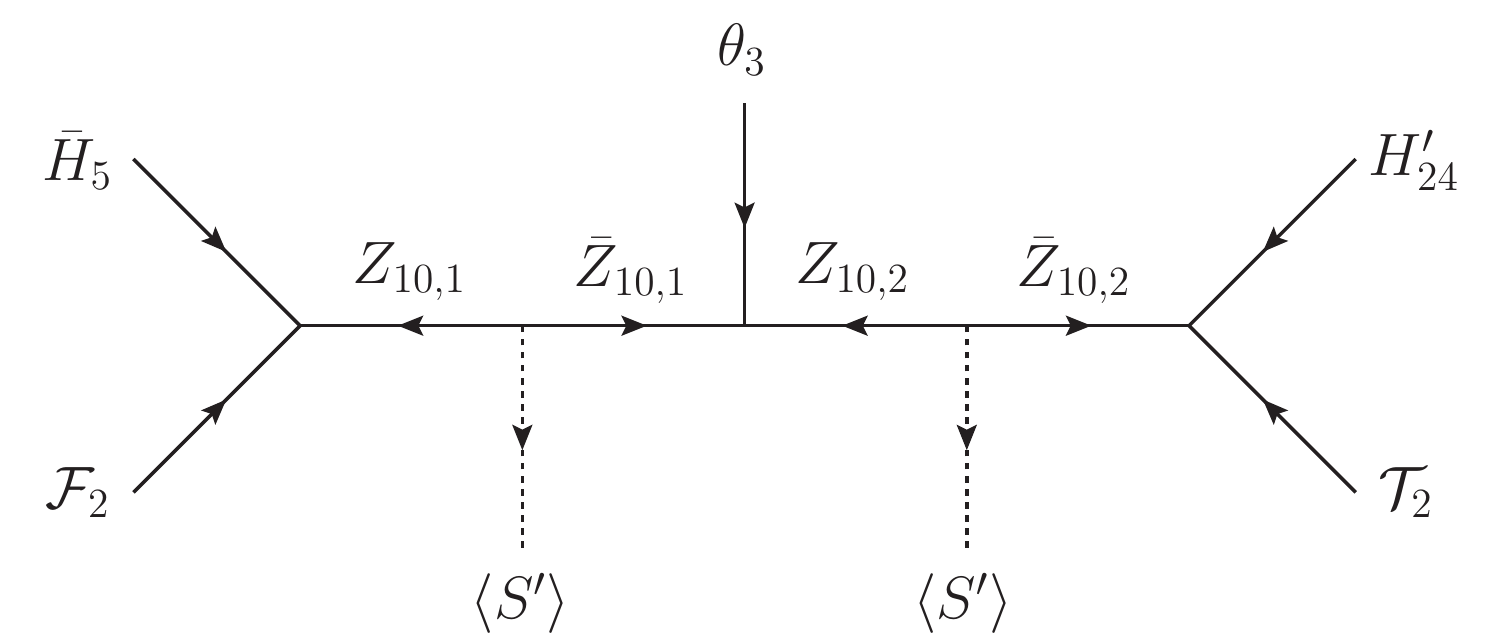}
\includegraphics[scale=0.5]{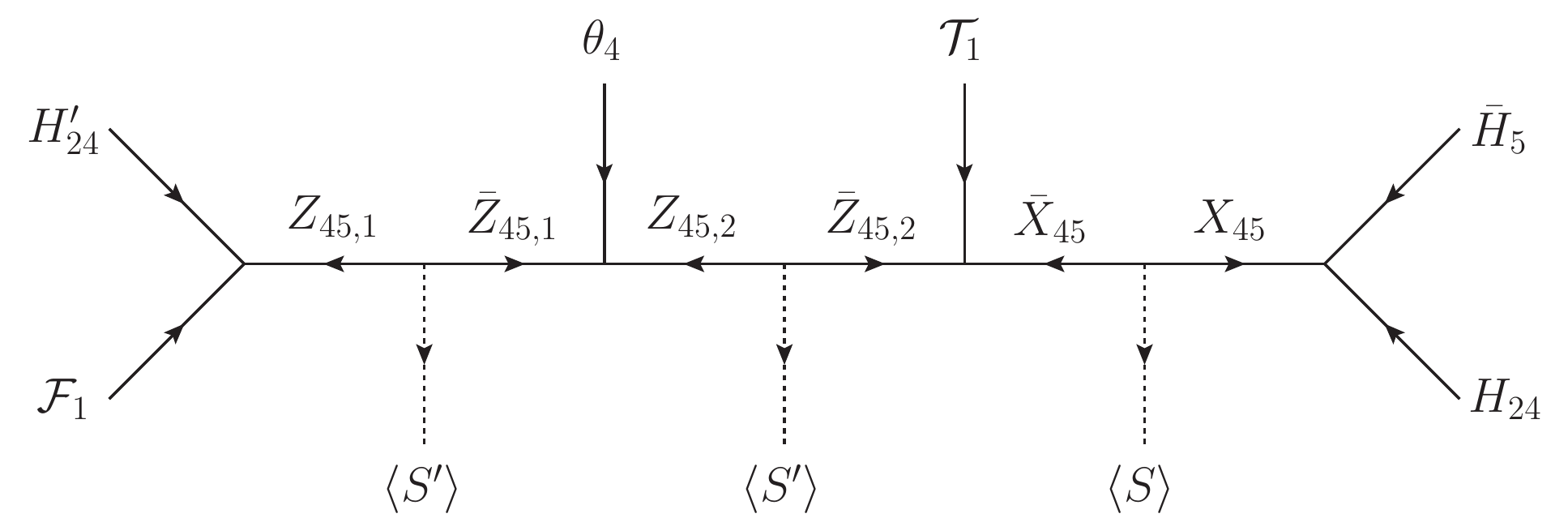}
\caption{Supergraphs leading to the effective superpotential $W_d$ of
  eq.~\eqref{eq:A_Wd} when the heavy messenger fields get integrated out in the model presented in subsection \ref{sec:modelA}.
  \label{fig:optionA_Yd}
  }
\end{figure}

\begin{figure}
\centering
\includegraphics[scale=0.5]{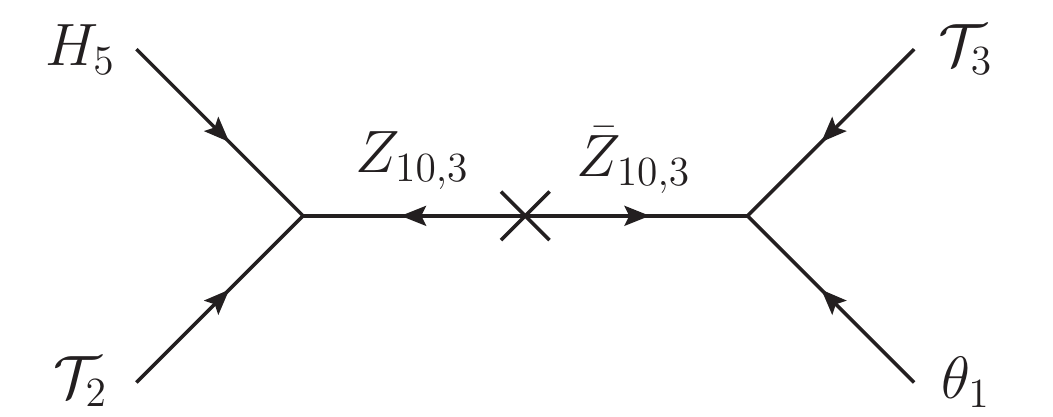}
\includegraphics[scale=0.5]{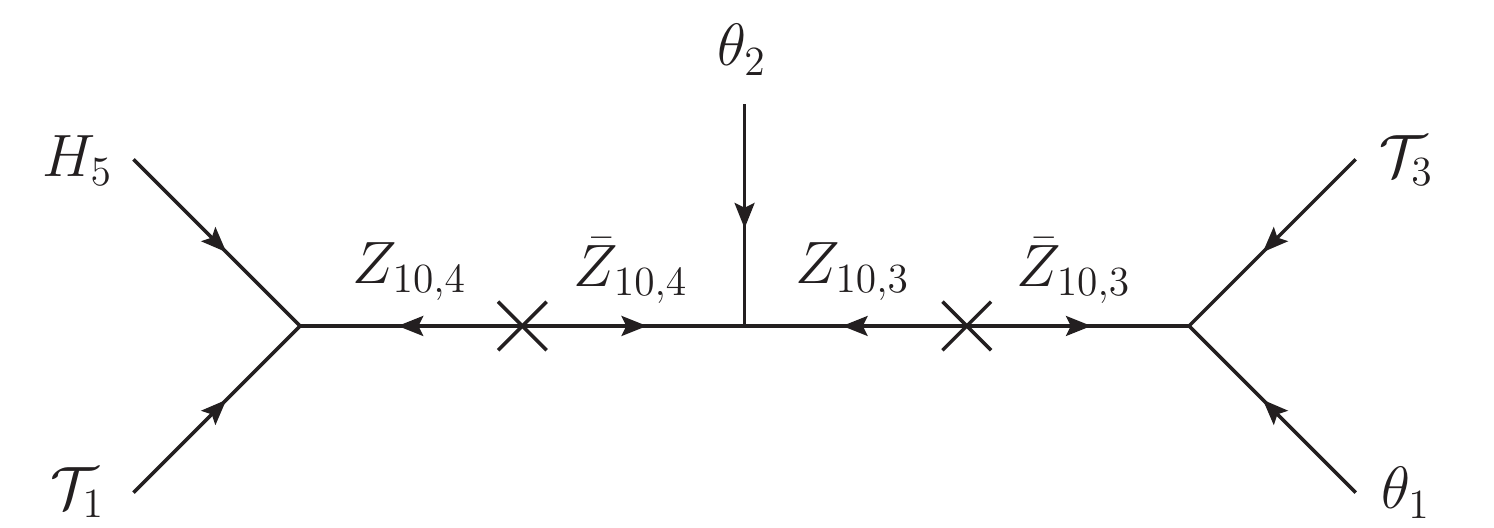}
\includegraphics[scale=0.5]{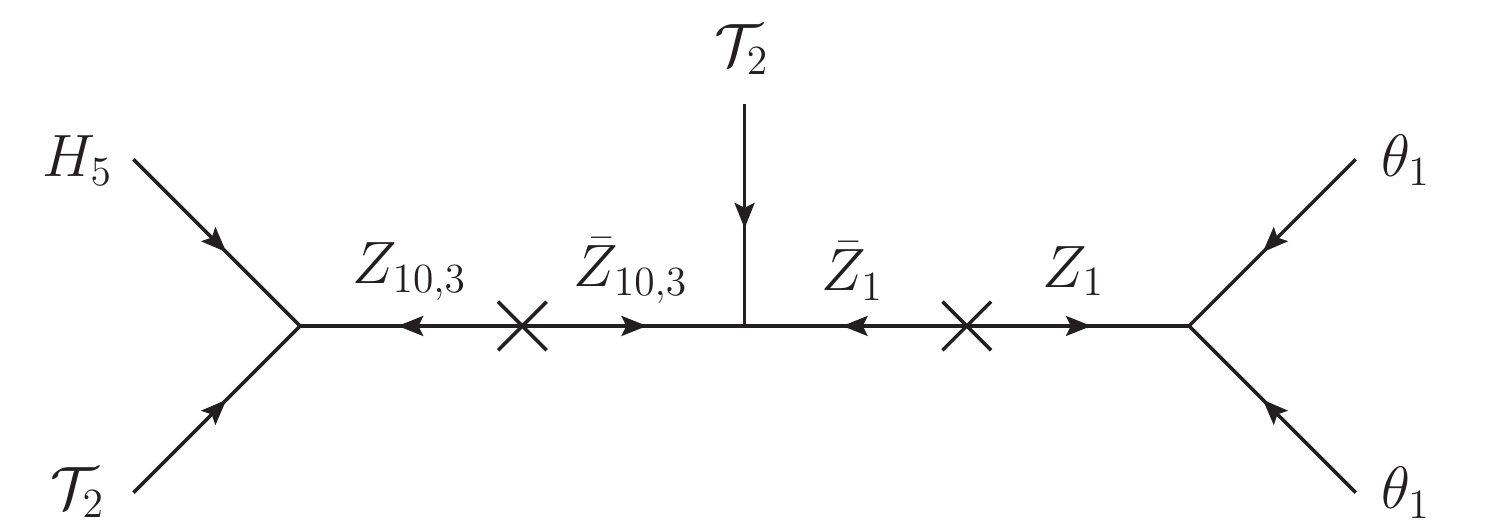}
\includegraphics[scale=0.5]{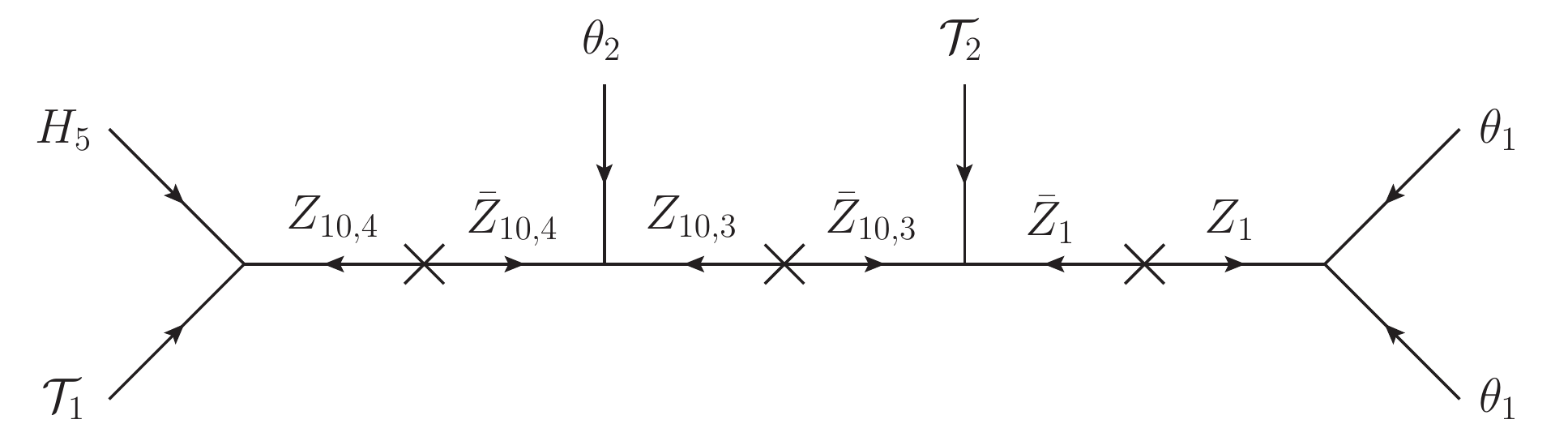}
\includegraphics[scale=0.5]{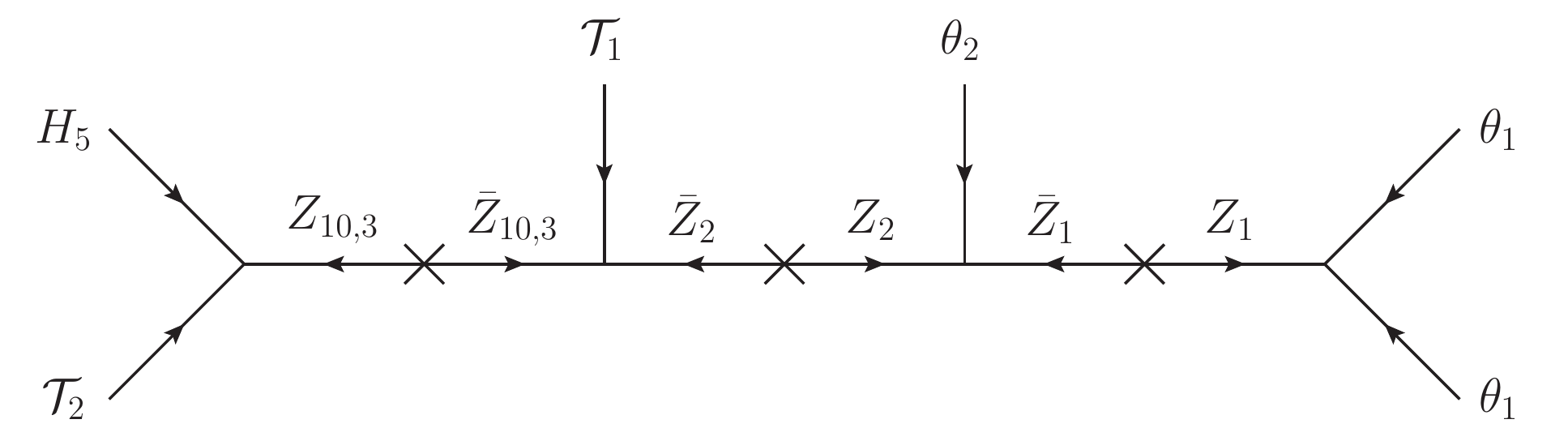}
\includegraphics[scale=0.5]{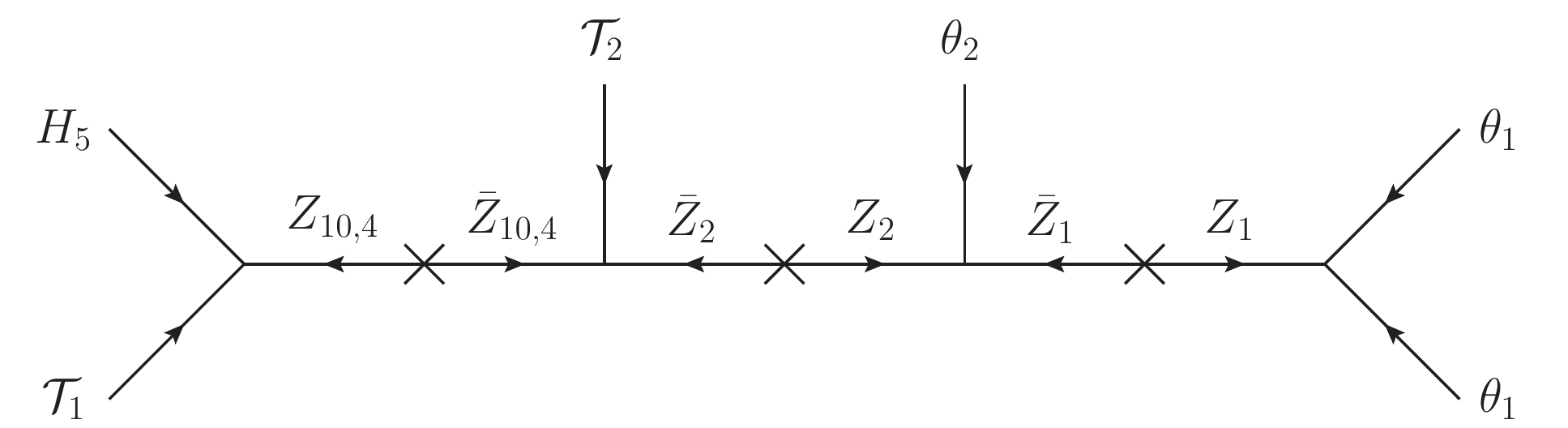}
\includegraphics[scale=0.5]{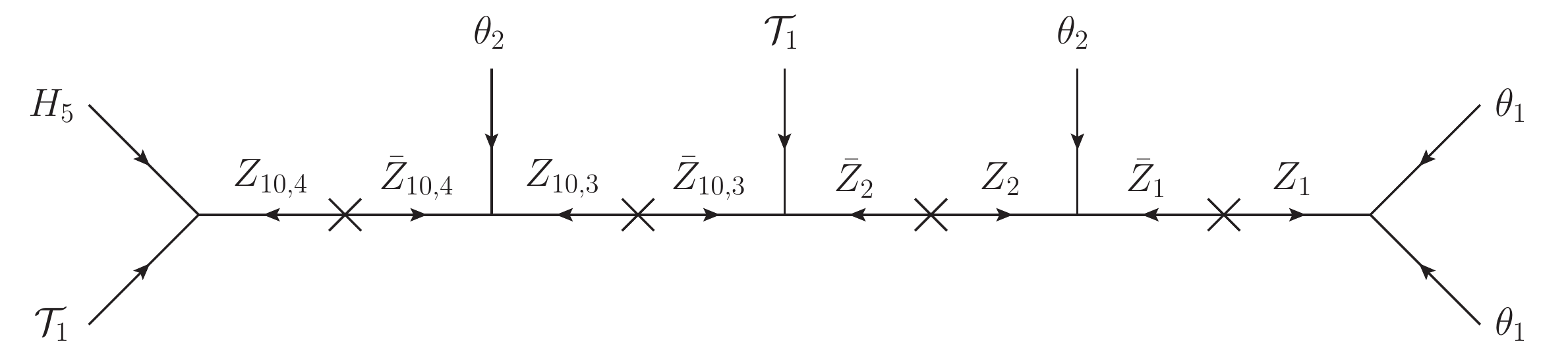}
\caption{Supergraphs leading to the effective superpotential $W_u$ of
  eq.~\eqref{eq:A_Wu} when the heavy messenger fields are integrated out in the model presented in subsection \ref{sec:modelA}. Note that there are three supergraphs contributing to the superpotential term generating $y_{12}$. A detailed discussion of the messenger sector is
presented in the appendix \ref{app:mess}.}
  \label{fig:optionA_Yu}
  \end{figure}

The full messenger sector can be read off from the supergraphs presented in Figures~\ref{fig:optionA_Yd} and \ref{fig:optionA_Yu}, see also Table~\ref{tab:A_messengers}.
After the heavy messenger fields get integrated out, the effective
superpotentials $W_d$ and $W_u$ are obtained.

The predictivity of the down-type quarks and charged lepton Yukawa
couplings has already been discussed above. Note that the hierarchical
structure is enforced due to the use of higher order effective
operators in $W_d$. In a small angle approximation the leading order
estimates for the eigenvalues of $Y_u$ and the mixing angles are given as
\begin{equation}
y_u \approx y_{11} - \frac{y_{12}^2}{y_{22}} \;,\quad
y_c \approx y_{22} \;,\quad
y_t \approx y_{33}\;,\quad
\theta_{C} \approx \frac{y_{12}}{y_{22}} \;,\quad
\theta_{23} \approx\frac{y_{23}}{y_{33}} \;,\quad
\theta_{13} \approx\frac{y_{13}}{y_{33}}\;.
\end{equation}
Phenomenology requires that all parameters $y_{ij}$ of $Y_u$ are independent, which
needs to be carefully considered in the construction of the messenger
sector, as discussed in Appendix~\ref{app:mess}.

In section \ref{subsec:dmpm24} we argued that for the case of an uncharged $H_{24}$,
as it appears in the selected superpotential (a), the mass term for the additional,
five-dimensional Higgs fields must come from the VEV of some singlet field.
In our model an effective $\mu'$ term is generated from a higher-dimensional operator and
with an even higher suppression, there is also a $\mu$-term for the Higgs fields coupling to matter,
\begin{equation}
W_5^{\text{eff}} =  \mu H_5\bar H_5 + \mu' H_5^\prime\bar H_5^\prime \;,
\end{equation}
where $\mu' \equiv \vev{\theta_3}^4/M_{\text{Pl}}^3$ and
$\mu \equiv \vev{\theta_3} \vev{\theta_4}^4/M_{\text{Pl}}^4$.\footnote{Alternatively,
we checked that a UV-complete generation of these $\mu$ and $\mu^\prime$ operators
via messenger fields would be possible. However, the necessary messenger fields
are not included in the model and it turns out that the masses are already generated
by Planck-scale suppressed effective operators.}  
The mass matrices for the doublet and triplet components are then given by 
\begin{equation}
m_D = \begin{pmatrix}
  \mu & 0 \\
  0 & \mu'
  \end{pmatrix}\;,\quad\;
\quad m_T = \begin{pmatrix}
\mu & 0 & 0 & -\frac{V_1^2}{\vev{S}}\\ 
0 & \mu' & -\frac{V_1^2}{\vev{S}} & 0 \\
-\frac{V_1^2}{\vev{S}} & 0 & \vev{S} & 0\\
0 & -\frac{V_1^2}{\vev{S}} & 0 & \vev{S}
\end{pmatrix} \;,
\end{equation}
where $V_1$ is defined in Eq.~\eqref{eq:vevdef1}.
After the heavy $50$-dimensional fields are integrated out and SU(5)
gets spontaneously broken, the mass matrices for the doublet and triplet
components of the Higgs fields $H_5$, $H^\prime_5$ and their corresponding
barred fields are given by
\begin{equation}
m_D = \begin{pmatrix}
  \mu & 0 \\
  0 & \mu'
  \end{pmatrix}\;,\quad
  m_T = \begin{pmatrix}
\mu & -\frac{V_1^4}{ \vev{S}^3} \\
-\frac{V_1^4}{\vev{S}^3} & \mu'
\end{pmatrix}\;.
\end{equation}
The effective triplet mass for dimension five proton decay is then given by
\begin{equation}
\label{eq:A_mteff}
\Mf = \left(m_T^{-1}\right)_{11}^{-1} = \mu - \frac{V_1^8}{\vev{S}^6 \mu'}
\approx -\frac{V_1^8 M_{\text{Pl}}^3}{\vev{S}^6 \vev{\theta_3}^4} \;.
\end{equation}
and the effective triplet masses suppressing dimension six proton decay are given by
\begin{equation}
\left(\MsT\right)^2 = \left(\MsTb\right)^2 \approx\frac{\abs{V_1}^8}{\abs{\vev{S}}^6}\;.
\end{equation}
Let us give an explicit example for the scales involved in the model: Because of
perturbativity the mass of the $50$-dimensional Higgs fields has to be almost at the
Planck scale. We therefore assume $\vev{S} \sim 10^{-1} M_{\text{Pl}}$. Using the known
values of the Yukawa couplings, we can estimate the values of the relevant masses of
our model. At the GUT scale with $\tan\beta = 30$ the Yukawa couplings are approximately
given by \cite{Antusch:2013jca}
\begin{equation}
\label{eq:A_yuk_num}
y_d \approx 1.6 \cdot 10^{-4}\;, \;\;
y_s \approx 3 \cdot 10^{-3}\;, \text{ and }
y_b \approx 0.18\;.
\end{equation}  
In our example model these Yukawa couplings are (up to order one couplings)
given by the operators
\begin{equation}
\label{eq:A_yuk_eqs}
y_d \sim \frac{\vev{H_{24}}\vev{H_{24}^\prime}\vev{\theta_4}}{\vev{S^\prime}^2\vev{S}}\;,\quad
y_s\sim\frac{\vev{H_{24}^\prime}\vev{\theta_3}}{\vev{S^\prime}^2}  \text{ and }
y_b\sim\frac{\vev{H_{24}^\prime}}{\vev{S^\prime}}\;.
\end{equation}
Then we find from eqs.~\eqref{eq:A_mteff}--\eqref{eq:A_yuk_eqs} for the
effective triplet masses, the $\mu$-term and the mass of
the additional heavy doublet the following values
\begin{equation}
\Mf\approx 1.4\cdot 10^{19}~\GeV\;,\quad
\Ms\approx 1.4\cdot 10^{12}~\GeV\;,\quad
\mu \approx 225~\GeV \;, \quad
\mu^\prime \approx 130~\TeV \;.
\end{equation}
Recalling from section \ref{subsec:superpotentiala} the parameters governing superpotential (a) and using here SUSY scale of $M_\susy=1~\TeV$, $\lambda\sim 0.19$, $\kappa^\prime\sim0.08$, $M_{24}^\prime = M_{24} =10^{15}~\GeV$ and $V_2/V_1=1.2$ the GUT scale is given by
\begin{equation}
M_\gut \approx 6.4\cdot 10^{16} \GeV\;.
\end{equation}
Although these numbers are only estimates, which neglect order one couplings,
they illustrate the model's features: the DTS problem is solved
with large effective triplet masses, therefore the proton decay rate is suppressed
and the fermion mass ratios are realistic. The $\mu$-term emerges from
a Planck-scale suppressed operator.
We remark that the DMPM does not suffer from any dangerous Planck-scale suppressed operators, due to the charge assignment of the singlet field $S$.

\subsection[A model with $\theta_C$ from $Y_d$]{A model with $\boldsymbol{\theta_C}$ from $\boldsymbol{Y_d}$ \label{sec:modelB}}

\begin{figure}
\begin{center}
\includegraphics[scale=0.5]{supergraph_optionA_yb.pdf}
\includegraphics[scale=0.5]{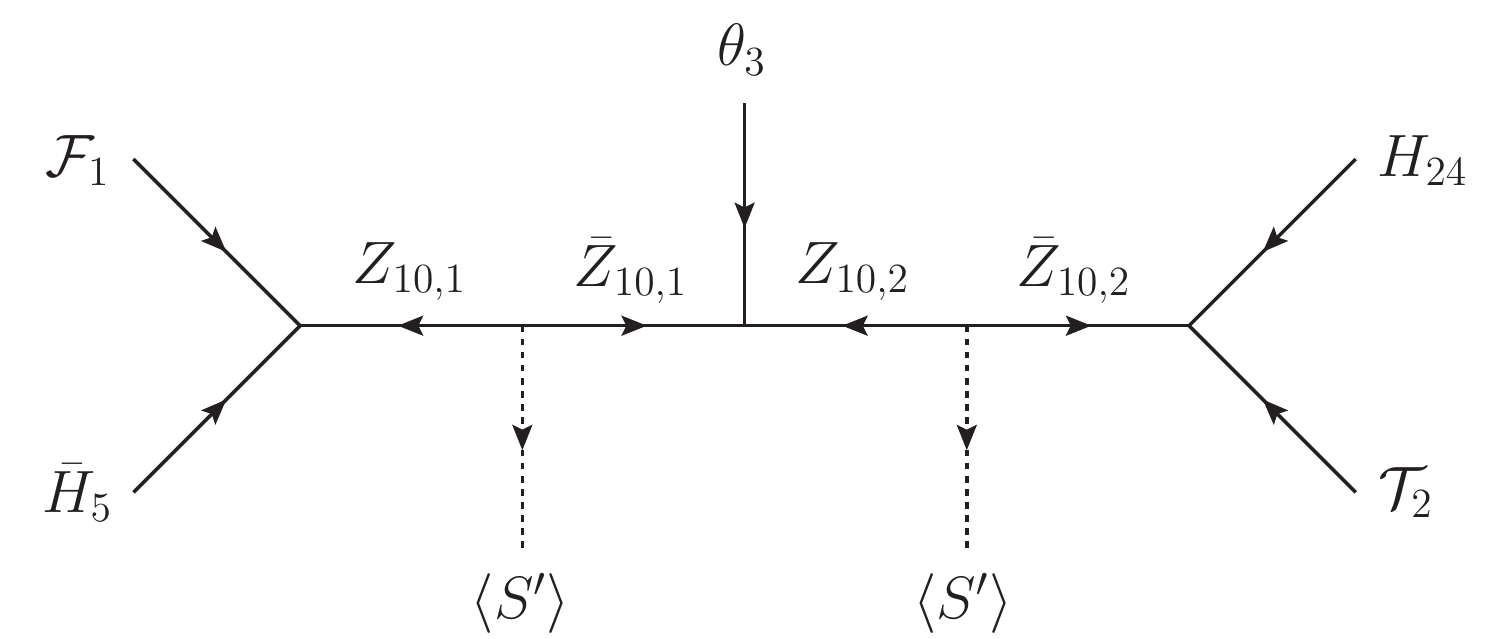}
\includegraphics[scale=0.5]{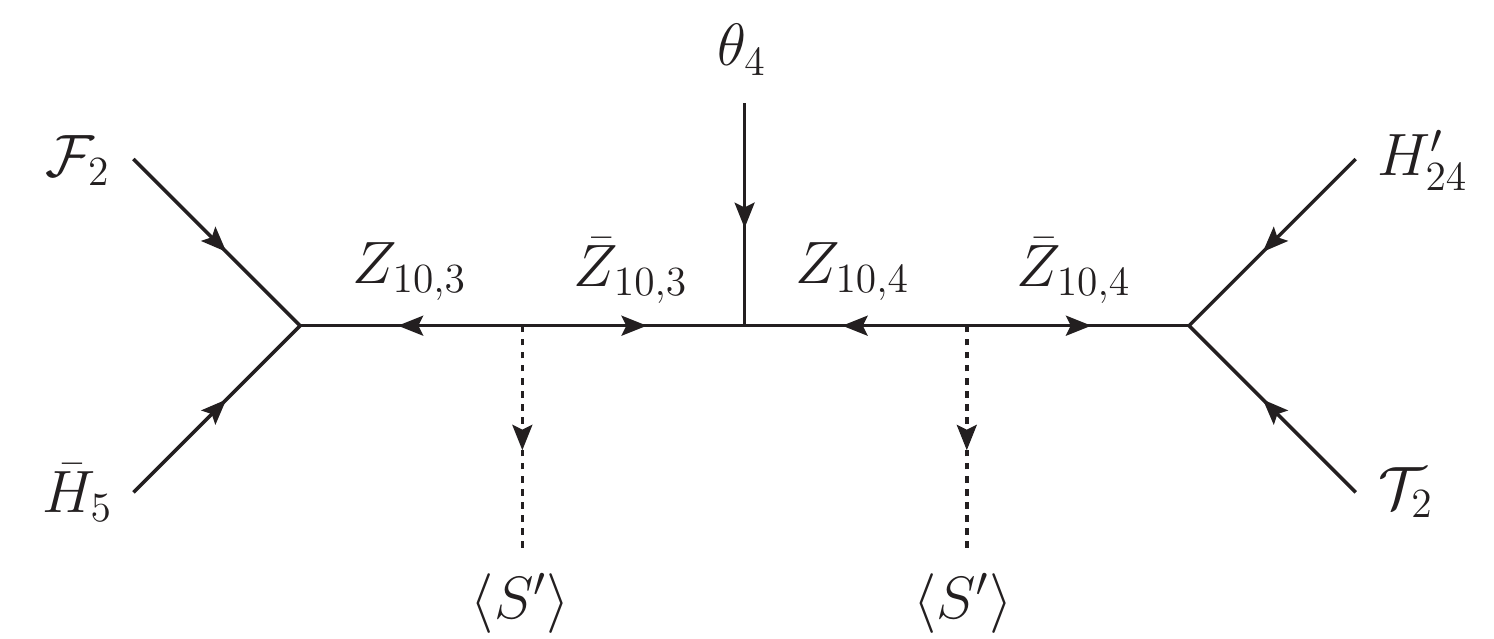}
\includegraphics[scale=0.5]{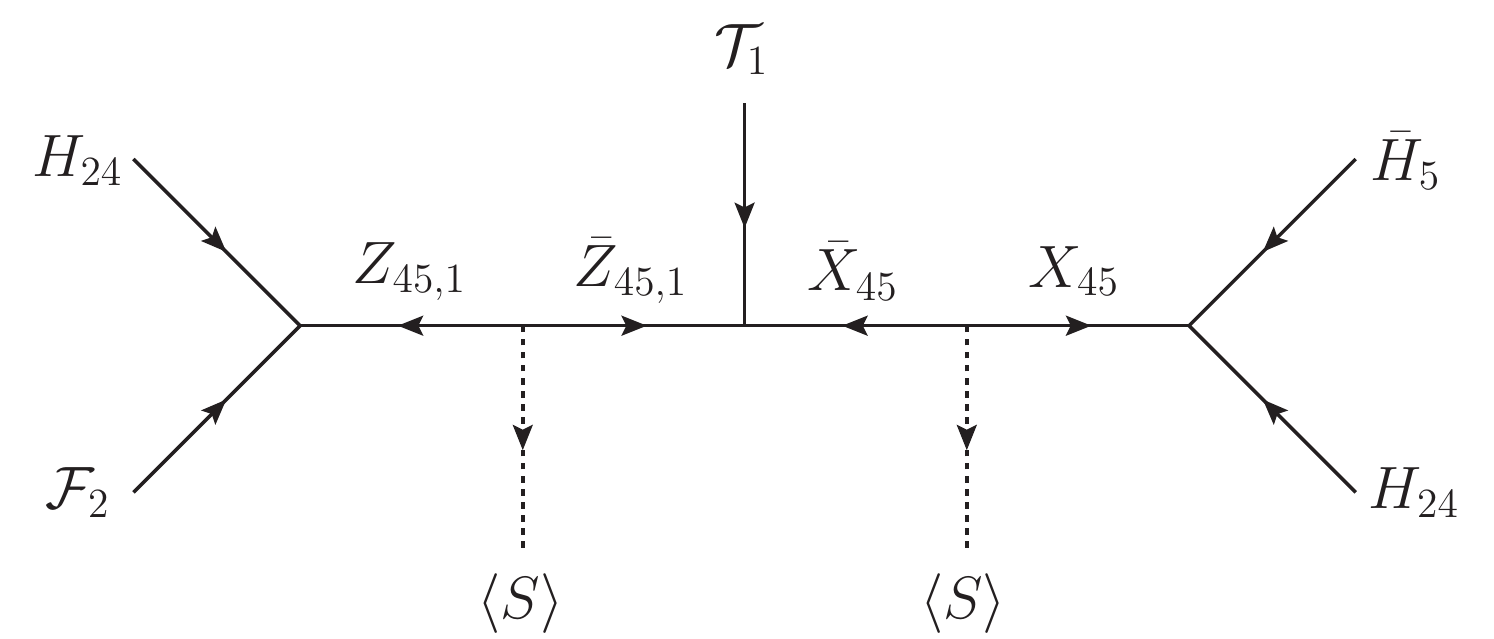}
\end{center}
\caption{Supergraphs leading to the effective superpotential $W_d$ of eq.~(\ref{eq:B_Wd}) when the heavy messenger fields get integrated out in the model presented in subsection \ref{sec:modelB}.}
\label{fig:optionB_Yd}
\end{figure}

\begin{figure}
\begin{center}
\includegraphics[scale=0.5]{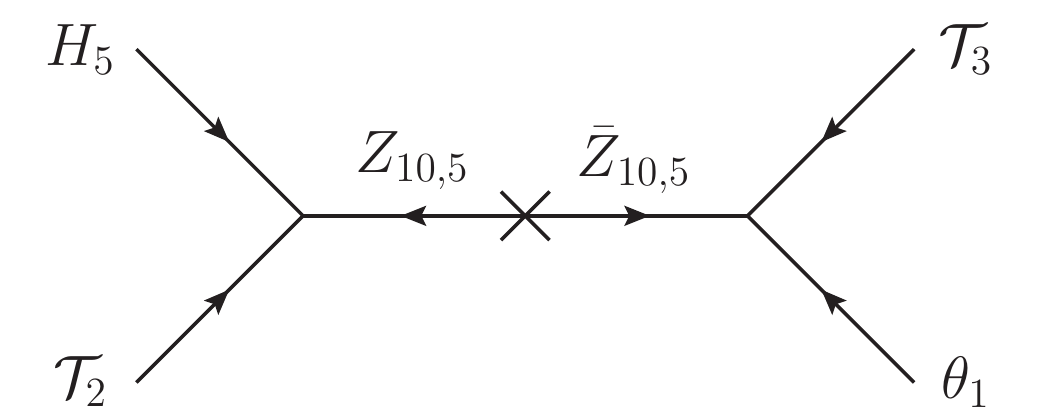}
\includegraphics[scale=0.5]{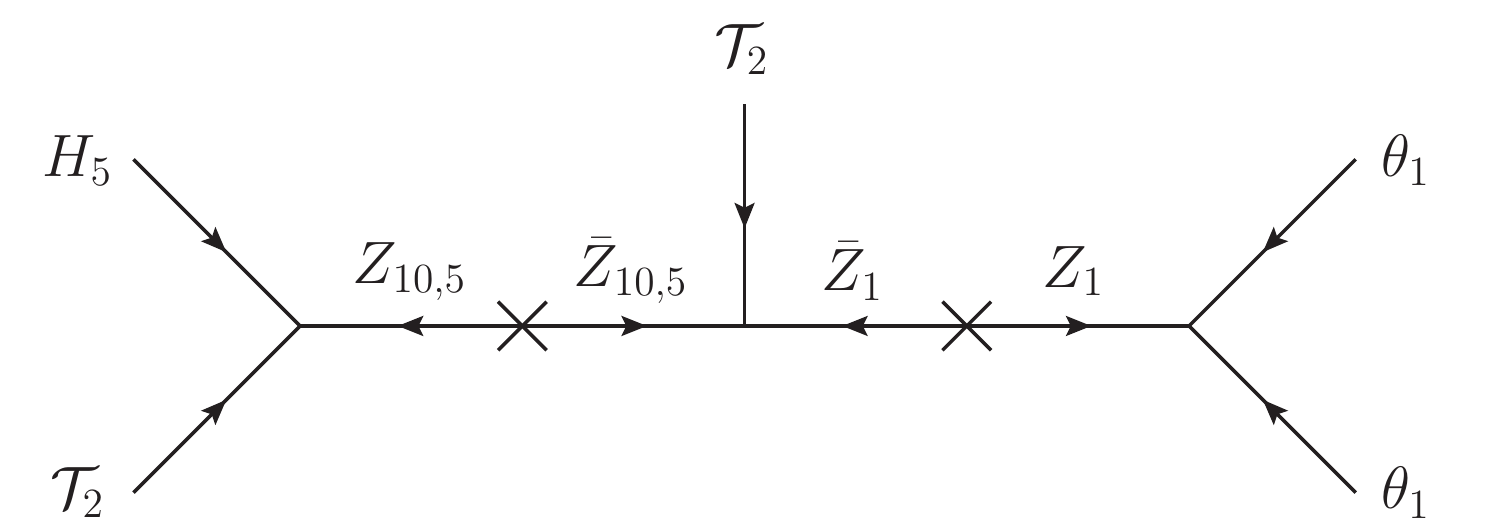}
\includegraphics[scale=0.5]{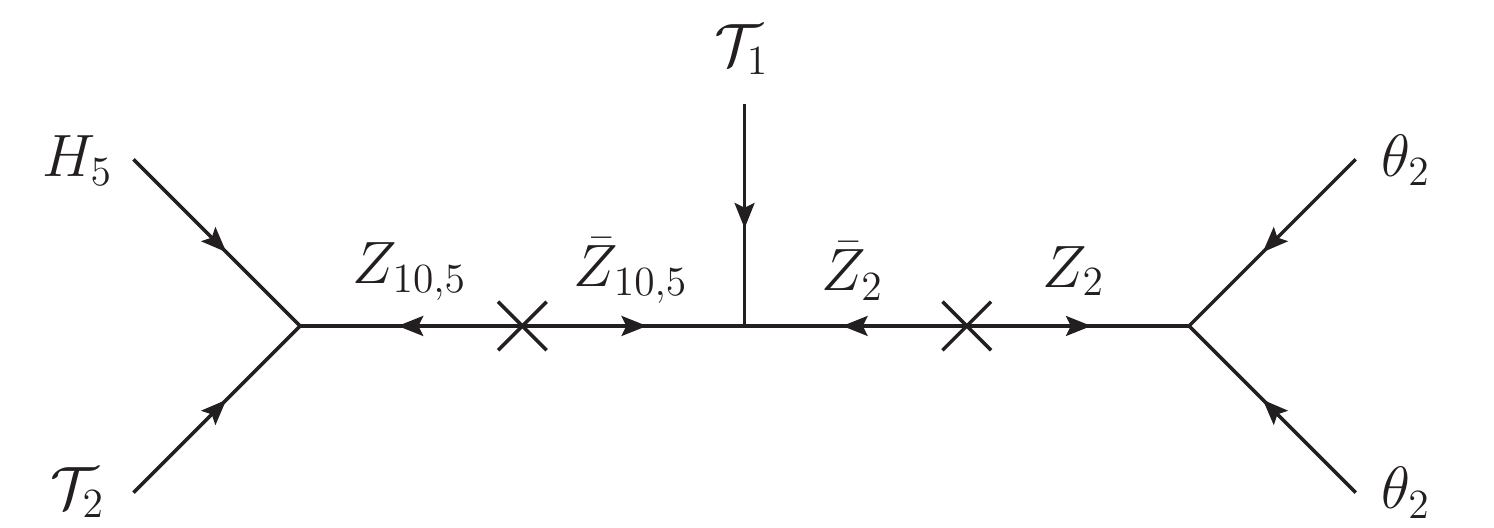}
\includegraphics[scale=0.5]{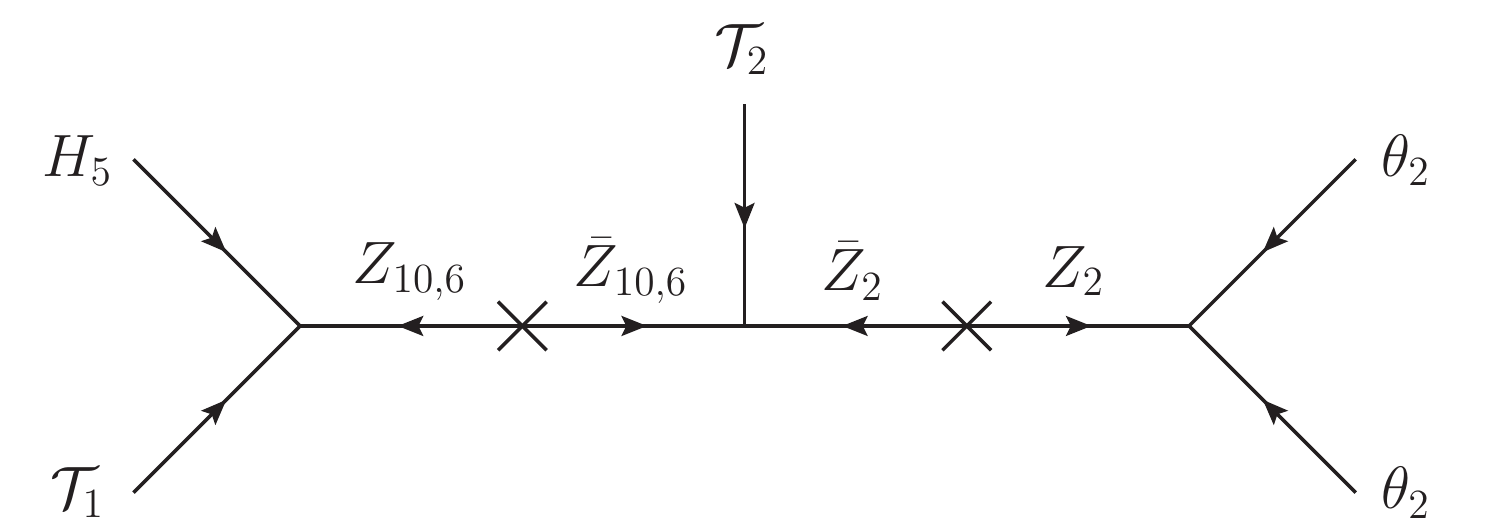}
\includegraphics[scale=0.5]{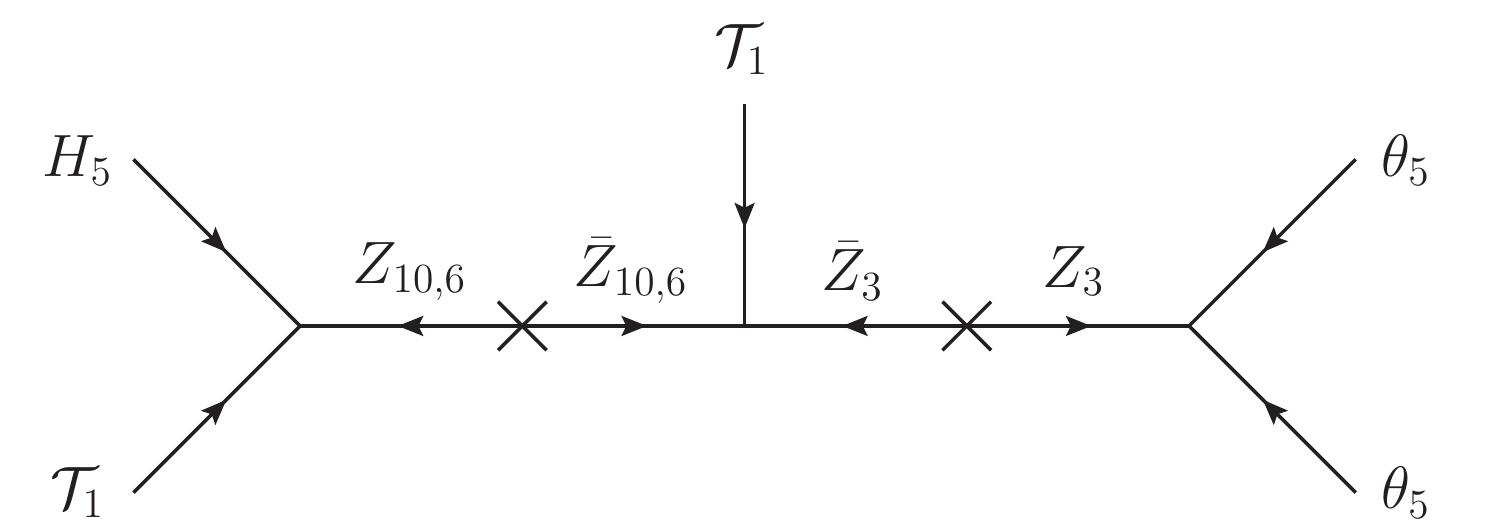}
\end{center}
\caption{Supergraphs leading to the effective superpotential $W_u$ of eq.~(\ref{eq:B_Wu}) when the heavy messenger fields get integrated out.}
\label{fig:optionB_Yu}
\end{figure}

We now turn to our second example model, which realises the attractive feature of
$\theta_C$ emerging dominantly from the down-type quark mixing $\theta_C \approx \theta_{12}^d$.
The Yukawa matrices are given by the following structure
\begin{equation}
\label{eq:B_Y}
Y_d=\begin{pmatrix}
0 & y_{d,12}& 0\\
y_{d,21} & y_s& 0\\
0 & 0& y_b
\end{pmatrix} ,
Y_e=\begin{pmatrix}
0 & 6 y_{d,21}& 0\\
-\frac{1}{2}y_{d,12} & 6 y_s& 0\\
0 & 0& -\frac{3}{2}y_b
\end{pmatrix} ,
Y_u=\begin{pmatrix}
y_{11} & y_{12} & 0\\
y_{12} & y_{22} & y_{23}\\
0 & y_{23} & y_{33}
\end{pmatrix} .
\end{equation}
A similar structure \cite{Antusch:2013kna,Antusch:2013tta} has been used to explain the relation $\theta_{13}^\text{PMNS}= \theta_C / \sqrt{2}$ via charged lepton corrections, and a right-handed quark unitarity triangle \cite{oai:arXiv.org:0910.5127}.
From the matrices we can see that we want to couple both $\suF_1$ and $\suF_2$ to $\suT_2$
but still distinguish both fields from each other to forbid $(Y_d)_{11}$. This suggests to
use at least a $\mathbb{Z}_3$ symmetry and hence we will use superpotential (b) from
section~\ref{sec:W2Adjoints} where both adjoints are charged.
The effective superpotentials that lead to the desired Yukawa matrices after integrating out
heavy messenger fields and breaking of the GUT gauge group are
\begin{align}
W_u &= \frac{1}{\Lambda^2} H_5 {\cal T}_1 {\cal T}_1 \theta_5^2 
     + \frac{1}{\Lambda^2} H_5 {\cal T}_1 {\cal T}_2 \theta_2^2
     + \frac{1}{\Lambda^2} H_5 {\cal T}_2 {\cal T}_2 \theta_1^2 
     + \frac{1}{\Lambda}   H_5 {\cal T}_2 {\cal T}_3 \theta_1
     + H_5 {\cal T}_3 {\cal T}_3\;,\label{eq:B_Wu} \\
W_d &= \frac{1}{\vev{S^\prime}} (H_{24}^\prime {\suF}_3)_{\bar{5}} ({\bar H}_{5} {\suT}_3)_{5}
     + \frac{\theta_3}{\vev{S^\prime}^2} (H_{24} {\suT}_2)_{10} ( {\bar H}_{5} {\suF}_1)_{\overline{10}}
     + \frac{\theta_4}{\vev{S^\prime}^2} (H_{24}^\prime {\suT}_2)_{10} ( {\bar H}_{5} {\suF}_2)_{\overline{10}}\nonumber\\
     &+ \frac{1}{\vev{S}^2} (H_{24} {\suF}_2)_{\overline{45}} ( {\suT}_1 H_{24} {\bar H}_{5})_{45}\;,\label{eq:B_Wd} 
\end{align}
where again VEVs of singlet fields appear in the denominators by virtue of messenger masses generated by them. Note that in comparison to $Y_u$ of the previous example model of eq.~\eqref{eq:A_Y}, the vanishing $(Y_u)_{13}$ element requires to introduce an additional flavon
field $\theta_5$. The supergraphs that generate these effective operators are shown in Figure~\ref{fig:optionB_Yd} and \ref{fig:optionB_Yu}. A complete list of all fields including their charges and representations is given in Tables~\ref{tab:B_MSSM}, \ref{tab:B_VEVS}, \ref{tab:B_DMPM} and \ref{tab:B_messengers}.

\begin{table}
  \begin{tabular}{cccccccccc}
  \toprule
 & SU(5) & $\mathbb{Z}_{2}$ & $\mathbb{Z}_{3}$ & $\mathbb{Z}_{4}$ & $\mathbb{Z}_{5}$ & $\mathbb{Z}_{6}$ & $\mathbb{Z}_{7}$ & $\mathbb{Z}_{7}$ & $\mathbb{Z}_{2}$ \\
 \midrule 
$H_5$ & $\mathbf{5} $ & $.$ & $.$ & $.$ & $.$ & $.$ & $.$ & $.$ & $.$ \\
$\bar H_{5}$ & $\mathbf{\bar 5} $ & $.$ & $2$ & $.$ & $1$ & $4$ & $6$ & $6$ & $.$ \\
$\suT_1$ & $\mathbf{10} $ & $.$ & $.$ & $.$ & $.$ & $.$ & $6$ & $5$ & $1$ \\
$\suT_2$ & $\mathbf{10} $ & $.$ & $.$ & $.$ & $.$ & $.$ & $1$ & $.$ & $1$ \\
$\suT_3$ & $\mathbf{10} $ & $.$ & $.$ & $.$ & $.$ & $.$ & $.$ & $.$ & $1$ \\
$\suF_1$ & $\mathbf{\bar 5} $ & $.$ & $.$ & $.$ & $3$ & $2$ & $5$ & $6$ & $1$ \\
$\suF_2$ & $\mathbf{\bar 5} $ & $.$ & $2$ & $.$ & $3$ & $.$ & $2$ & $3$ & $1$ \\
$\suF_3$ & $\mathbf{\bar 5} $ & $.$ & $1$ & $3$ & $4$ & $.$ & $1$ & $1$ & $1$ \\
  \bottomrule
 \end{tabular}
 \caption{
   SU(5) representations and charges under discrete shaping symmetries
   of the MSSM fields and colour triplets of the model presented in subsection \ref{sec:modelB}. A dot denotes charge zero.
   \label{tab:B_MSSM}
   }
\end{table}

\begin{table}
  \begin{tabular}{cccccccccc}
  \toprule
  & SU(5) & $\mathbb{Z}_{2}$ & $\mathbb{Z}_{3}$ & $\mathbb{Z}_{4}$ & $\mathbb{Z}_{5}$ & $\mathbb{Z}_{6}$ & $\mathbb{Z}_{7}$ & $\mathbb{Z}_{7}$ & $\mathbb{Z}_{2}$ \\
 \midrule 
$H_{24}$ & $\mathbf{24} $ & $.$ & $.$ & $.$ & $.$ & $4$ & $.$ & $.$ & $.$ \\
$H_{24}^\prime$ & $\mathbf{24} $ & $.$ & $.$ & $.$ & $.$ & $2$ & $.$ & $.$ & $.$ \\
$S$ & $\mathbf{1} $ & $1$ & $2$ & $.$ & $2$ & $.$ & $.$ & $.$ & $.$ \\
$S^\prime$ & $\mathbf{1} $ & $.$ & $.$ & $3$ & $.$ & $.$ & $.$ & $.$ & $.$ \\
$\theta_1$ & $\mathbf{1} $ & $.$ & $.$ & $.$ & $.$ & $.$ & $6$ & $.$ & $.$ \\
$\theta_2$ & $\mathbf{1} $ & $.$ & $.$ & $.$ & $.$ & $.$ & $.$ & $1$ & $.$ \\
$\theta_{3}$ & $\mathbf{1} $ & $.$ & $1$ & $2$ & $1$ & $2$ & $2$ & $2$ & $.$ \\
$\theta_{4}$ & $\mathbf{1} $ & $.$ & $2$ & $2$ & $1$ & $.$ & $5$ & $5$ & $.$ \\
$\theta_5$ & $\mathbf{1} $ & $.$ & $.$ & $.$ & $.$ & $3$ & $1$ & $2$ & $.$ \\
\bottomrule
 \end{tabular}
 \caption{
   SU(5) representations and charges under discrete shaping symmetries
   of the superfields obtaining VEVs at around the GUT scale of the model presented in subsection \ref{sec:modelB}. A dot denotes charge zero.
   \label{tab:B_VEVS}
   }
\end{table}

\begin{table}
  \begin{tabular}{cccccccccc}
  \toprule
  & SU(5) & $\mathbb{Z}_{2}$ & $\mathbb{Z}_{3}$ & $\mathbb{Z}_{4}$ & $\mathbb{Z}_{5}$ & $\mathbb{Z}_{6}$ & $\mathbb{Z}_{7}$ & $\mathbb{Z}_{7}$ & $\mathbb{Z}_{2}$ \\
 \midrule 
$H_{5}^\prime$ & $\mathbf{5} $ & $1$ & $1$ & $.$ & $.$ & $4$ & $1$ & $1$ & $.$ \\
$\bar H_{5}^\prime$ & $\mathbf{\bar 5} $ & $1$ & $.$ & $.$ & $1$ & $2$ & $.$ & $.$ & $.$ \\
$X_{45}$ & $\mathbf{45} $ & $.$ & $1$ & $.$ & $4$ & $4$ & $1$ & $1$ & $.$ \\
$\bar X_{45}$ & $\mathbf{\overline{45}} $ & $1$ & $.$ & $.$ & $4$ & $2$ & $6$ & $6$ & $.$ \\
$Y_{45}$ & $\mathbf{45} $ & $.$ & $2$ & $.$ & $3$ & $2$ & $1$ & $1$ & $.$ \\
$\bar Y_{45}$ & $\mathbf{\overline{45}} $ & $1$ & $2$ & $.$ & $.$ & $4$ & $6$ & $6$ & $.$ \\
$Z_{50}$ & $\mathbf{50} $ & $1$ & $.$ & $.$ & $1$ & $.$ & $1$ & $1$ & $.$ \\
$\bar Z_{50}$ & $\mathbf{\overline{50}} $ & $.$ & $1$ & $.$ & $2$ & $.$ & $6$ & $6$ & $.$ \\
$X_{45}^\prime$ & $\mathbf{45} $ & $1$ & $.$ & $.$ & $4$ & $.$ & $.$ & $.$ & $.$ \\
$\bar X_{45}^\prime$ & $\mathbf{\overline{45}} $ & $.$ & $1$ & $.$ & $4$ & $.$ & $.$ & $.$ & $.$ \\
$Y_{45}^\prime$ & $\mathbf{45} $ & $1$ & $1$ & $.$ & $3$ & $4$ & $.$ & $.$ & $.$ \\
$\bar Y_{45}^\prime$ & $\mathbf{\overline{45}} $ & $.$ & $.$ & $.$ & $.$ & $2$ & $.$ & $.$ & $.$ \\
$Z_{50}^\prime$ & $\mathbf{50} $ & $.$ & $2$ & $.$ & $1$ & $2$ & $.$ & $.$ & $.$ \\
$\bar Z_{50}^\prime$ & $\mathbf{\overline{50}} $ & $1$ & $2$ & $.$ & $2$ & $4$ & $.$ & $.$ & $.$ \\
  \bottomrule
  \end{tabular}
  \caption{
   SU(5) representations and charges under discrete shaping symmetries
   of the fields in the DMPM sector of the model presented in subsection \ref{sec:modelB}. A dot denotes charge zero.
   \label{tab:B_DMPM}
   }
\end{table}

\begin{table}
  \centering
  \begin{tabular}{cccccccccc}
  \toprule
 & SU(5) & $\mathbb{Z}_{2}$ & $\mathbb{Z}_{3}$ & $\mathbb{Z}_{4}$ & $\mathbb{Z}_{5}$ & $\mathbb{Z}_{6}$ & $\mathbb{Z}_{7}$ & $\mathbb{Z}_{7}$ & $\mathbb{Z}_{2}$ \\
 \midrule 
$Z_5$ & $\mathbf{5} $ & $.$ & $2$ & $1$ & $1$ & $4$ & $6$ & $6$ & $1$ \\
$\bar{Z}_5$ & $\mathbf{\bar 5} $ & $.$ & $1$ & $.$ & $4$ & $2$ & $1$ & $1$ & $1$ \\
$Z_{10,1}$ & $\mathbf{10} $ & $.$ & $1$ & $.$ & $1$ & $.$ & $3$ & $2$ & $1$ \\
$\bar{Z}_{10,1}$ & $\mathbf{\overline{10}} $ & $.$ & $2$ & $1$ & $4$ & $.$ & $4$ & $5$ & $1$ \\
$Z_{10,2}$ & $\mathbf{10} $ & $.$ & $.$ & $1$ & $.$ & $4$ & $1$ & $.$ & $1$ \\
$\bar{Z}_{10,2}$ & $\mathbf{\overline{10}} $ & $.$ & $.$ & $.$ & $.$ & $2$ & $6$ & $.$ & $1$ \\
$Z_{10,3}$ & $\mathbf{10} $ & $.$ & $2$ & $.$ & $1$ & $2$ & $6$ & $5$ & $1$ \\
$\bar{Z}_{10,3}$ & $\mathbf{\overline{10}} $ & $.$ & $1$ & $1$ & $4$ & $4$ & $1$ & $2$ & $1$ \\
$Z_{10,4}$ & $\mathbf{10} $ & $.$ & $.$ & $1$ & $.$ & $2$ & $1$ & $.$ & $1$ \\
$\bar{Z}_{10,4}$ & $\mathbf{\overline{10}} $ & $.$ & $.$ & $.$ & $.$ & $4$ & $6$ & $.$ & $1$ \\
$Z_{45,1}$ & $\mathbf{45} $ & $.$ & $1$ & $.$ & $2$ & $2$ & $5$ & $4$ & $1$ \\
$\bar{Z}_{45,1}$ & $\mathbf{\overline{45}} $ & $1$ & $.$ & $.$ & $1$ & $4$ & $2$ & $3$ & $1$ \\
\midrule
$Z_{10,5}$ & $\mathbf{10} $ & $.$ & $.$ & $.$ & $.$ & $.$ & $6$ & $.$ & $1$ \\
$Z_{10,6}$ & $\mathbf{10} $ & $.$ & $.$ & $.$ & $.$ & $.$ & $1$ & $2$ & $1$ \\
$Z_1$ & $\mathbf{1} $ & $.$ & $.$ & $.$ & $.$ & $.$ & $2$ & $.$ & $.$ \\
$Z_{2}$ & $\mathbf{1} $ & $.$ & $.$ & $.$ & $.$ & $.$ & $.$ & $5$ & $.$ \\
$Z_{3}$ & $\mathbf{1} $ & $.$ & $.$ & $.$ & $.$ & $.$ & $5$ & $3$ & $.$ \\
\bottomrule
  \end{tabular}
    \caption{
    SU(5) representations and charges under discrete shaping symmetries
    of the flavon and flavour messenger fields of the model presented in subsection \ref{sec:modelB}.
    Note that the messengers $Z_{5,1}\bar Z_{5,1}$, $Z_{10,1}\bar Z_{10,1}$,
    $Z_{10,2}\bar Z_{10,2}$, $Z_{10,3}\bar Z_{10,3}$, $Z_{10,4}\bar Z_{10,4}$
    and $Z_{45,1}\bar Z_{45,1}$ have no direct mass term, but get their masses
    through VEVs of $S$ and $S'$. The other messenger fields have direct mass
    terms, so their corresponding barred field is not shown in the table. A dot denotes charge zero.
    \label{tab:B_messengers}
    }
\end{table}

In a small angle approximation the mixing angles and Yukawa couplings are given by
\begin{align}
y_d\approx \frac{y_{d,12}y_{d,21}}{y_s}\;,\quad & \theta_C\approx\theta_{12}^d\approx\frac{y_{d,12}}{y_s}\;,\nonumber\\
y_u\approx y_{11}-\frac{y_{12}^2}{y_{22}}\;,\quad y_c\approx y_{22}\;,\quad y_t\approx y_{33}\;,\quad & \theta_{23}\approx\frac{y_{23}}{y_{33}}\;,\quad\theta_{13}\approx\frac{y_{12}}{y_{22}}\theta_{23}\;,
\end{align}
with $y_s$ and $y_b$ given as parameters of $Y_d$ in eq.~(\ref{eq:B_Y}). Thus the Yukawa matrices can fit the experimental values without tension.

We continue now by discussing additional details concerning the use of the DMPM in this model. 
Like in the previous model, Planck-scale suppressed operators generate mass terms for the
five-dimensional Higgs representations
\begin{equation}
W_5^{\text{eff}} = \mu H_5\bar H_5 +  \mu' H_5^\prime\bar H_5^\prime \;,
\end{equation}
where $\mu = \vev{\theta_3}^4 / M_{\text{Pl}}^3$ and $\mu' = \vev{\theta_4}^4 / M_{\text{Pl}}^3$. Note that although they appear at the same order, a modest hierarchy $\vev{\theta_3} < \vev{\theta_4}$, as we have in our model, will sufficiently split their masses. Furthermore, after integrating out the 45-dimensional messengers of the DMPM, the mass matrices for doublet and triplet components of the $5$- and $50$-dimensional
superfields are given by
\begin{equation}
m_D = \begin{pmatrix}
  \mu & 0 \\
  0 & \mu'
  \end{pmatrix}\;,\quad\;
\quad m_T = \begin{pmatrix}
\mu & 0 & 0 & -\frac{V_1^2}{\vev{S}}\\ 
0 & \mu' & -\frac{V_1^2}{\vev{S}} & 0 \\
-\frac{V_1^2}{\vev{S}} & 0 & \vev{S} & 0\\
0 & -\frac{V_1^2}{\vev{S}} & 0 & \vev{S}
\end{pmatrix} \;,
\end{equation}
where $V_1$ is defined in Eq.~\eqref{eq:vevdef1}.
After the heavy 50-dimensional fields get integrated out the mass matrices become
\begin{equation}
m_D = \begin{pmatrix}
  \mu & 0 \\
  0 & \mu'
  \end{pmatrix}\;,\quad
  m_T = \begin{pmatrix}
\mu & - \frac{V_1^4}{ \vev{S}^3} \\
- \frac{V_1^4}{\vev{S}^3} & \mu'
\end{pmatrix}\;.
\end{equation}
This leads to an effective triplet mass for dimension five proton decay of
\begin{equation}
\Mf = \left(m_T^{-1}\right)_{11}^{-1}
= \mu - \frac{V_1^8}{\vev{S}^6 \mu'}
\approx -\frac{V_1^8 M_{\text{Pl}}^3}{\vev{S}^6 \vev{\theta_4}^4} \;,
\end{equation}
and the effective triplet masses suppressing dimension six proton decay are given by
\begin{equation}
\left(\MsT\right)^2 = \left(\MsTb\right)^2 = \left(m_T^{-1}m_T^{\dagger-1}\right)_{11}^{-1}\approx\frac{\abs{V_1}^8}{\abs{\vev{S}}^6}\;.
\end{equation}

As for the first example model we can estimate the values of the relevant masses in an explicit example from the known Yukawa couplings. In a small angle approximation the down-type Yukawa couplings are given by 
\begin{align}
y_d & \sim \frac{\vev{H_{24}}^2}{\vev{S}^2}\frac{\vev{H_{24}}\vev{\theta_3}}{\vev{S^\prime}^2}\frac{1}{y_s}\approx 1.6\cdot 10^{-4} \text{\, with \,}y_s\sim\frac{\vev{H_{24}^\prime}\vev{\theta_4}}{\vev{S^\prime}^2}\approx 3\cdot 10^{-3}
\end{align}
and
\begin{align}
y_b&\sim\frac{\vev{H_{24}^\prime}}{\vev{S^\prime}} \approx 0.18\;,
\end{align}
where the numerical values for the Yukawa couplings taken from~\cite{Antusch:2013jca} are valid for
$\tan\beta=30$ and order one coefficients have been neglected. For the effective triplet masses of dimension five and six proton decay, respectively, the $\mu$-term and the mass of the additional, heavier Higgs doublets the following estimates emerge
\begin{equation}\label{eq:numbersB}
\Mf\approx 1.4 \cdot 10^{18}~\GeV \;,\quad \Ms\approx 10^{12}~\GeV\;,\quad \mu\approx 7~\TeV \;,\quad\mu^\prime\approx 800~\TeV\;.
\end{equation}
The parameters of superpotential (b) from section \ref{sec:Superpotentialbc} have been chosen to be real with values $\tan\beta_V= 0.5$, $\lambda=10^{-4}$ and $M=2.4\cdot 10^{12}~\GeV$. The mass of the 45-dimensional and 50-dimensional superfields has been set to $\vev{S}= 10^{18}~\GeV$. With the numbers of Eq.~\eqref{eq:numbersB} we find
\begin{equation}
M_\susy\approx 24~\TeV \;,\quad M_\gut\approx 4\cdot 10^{17}~\GeV\;.
\end{equation} 
Thus, also in this model, proton decay can be suppressed by large effective
triplet masses. DTS is achieved through the DMPM and the light doublet gets
its mass from an Planck-scale suppressed operator. The Yukawa sector of the model
features viable fermion masses and quark mixing angles.

\section{Proton Decay}
\label{sec:proton}

We have split our discussion of proton decay into two parts. In the first part we comment on
proton decay induced by dimension five operators in the superpotential.
These operators are usually considered to be more dangerous for the
validity of any SUSY GUT model. We will argue why we are more
predictive than ordinary models and still should be able to evade
current experimental bounds. In the second part we will comment
on dimension six proton decay operators, which appear also in non-SUSY GUTs. The dimension six proton decay operator emerging from the exchange of heavy gauge bosons are considered to be not as dangerous in
SUSY GUTs due to the usually higher unification scale~\cite{Murayama:2001ur}. The dimension six proton decay mediated by colour triplets, however, has to be suppressed by a high enough effective triplet mass, as for the case of dimension five proton decay.

\subsection{Proton Decay from Dimension Five Operators}

We will adopt for this section a notation similar to the SLHA
convention \cite{Skands:2003cj}, for more details see appendix \ref{app:Clebsch}.
The superpotential describing the couplings of the matter fields
to the heavy colour triplet reads
\begin{equation}
 \begin{split}
    W_T =&\; \epsilon_{\alpha\beta} \left( 
        - \frac{1}{2} (Y_{qq})_{ij} \epsilon_{abc} T^a Q_i^{\alpha b} Q_j^{\beta c} 
        + (Y_{ql})_{ij} \bar{T}^a Q_i^{\alpha a} L_j^\beta
    \right) \\
        +&\;
        (Y_{ue})_{ij} T^a \bar{U}_i^a \bar{E}_j 
        - (Y_{ud})_{ij} \epsilon_{abc} \bar{T}^a \bar{U}_i^{b} \bar{D}_j^{c} +
        M_T T^a \bar{T}^a \;.
   \end{split}     
\end{equation}
In many SU(5)
models at least for the first two generations it was assumed that the Yukawa
couplings are generated or significantly corrected by some set of higher-dimensional
operators, see for instance, \cite{Bajc:2002pg, EmmanuelCosta:2003pu, Wiesenfeldt:2004qa}. But, usually,
there is no control over which operator is the dominant one and hence it is not 
possible to calculate how strong exactly 
the heavy triplets couple to the MSSM fields.
In our setup this is not the case. For every entry of the Yukawa matrix we have specified
the operator with only very small corrections, if any. Therefore we know how strong
the MSSM fields couple to the heavy triplets if we know the MSSM Yukawa couplings
at the GUT scale. To be more precise the new Yukawa couplings are related to the MSSM
ones only via CG coefficients which are fixed by the gauge structure of the
underlying operator. We find for our first example model
\begin{align}
 Y_d &= \text{Diag}\left( y_d, y_s, y_b \right) \;,\quad
 Y_e = \text{Diag}\left( - \frac{1}{2} y_d, 6 y_s, - \frac{3}{2} y_b \right) \;, \\
 Y_{ql} &= \text{Diag}\left( y_d, y_s, -\frac{3}{2} y_b \right) \;,\quad
 Y_{ud} = \text{Diag}\left( \frac{2}{3} y_d, - 4 y_s, y_b \right) \;,\quad \\
 Y_{qq} &= Y_u \;,\quad Y_{ue} = Y_u \;,
\end{align}
where the structure of $Y_u$ can be read off from eq.~\eqref{eq:A_Y}.
For the second model we find
\begin{align}
 Y_d &= \begin{pmatrix} 0 & y_{d,12} & 0 \\ y_{d,21} & y_s & 0 \\ 0 & 0 & y_b  \end{pmatrix} \;,\quad
 Y_e^T =\begin{pmatrix} 0 & -\tfrac{1}{2} y_{d,12} & 0 \\ 6 y_{d,21} & 6 y_s & 0 \\ 0 & 0 & -\tfrac{3}{2} y_b  \end{pmatrix} \;, \\
 Y_{ql} &= \begin{pmatrix} 0 & y_{d,12} & 0 \\ y_{d,21} & y_s & 0 \\ 0 & 0 & -\tfrac{3}{2} y_b  \end{pmatrix} \;,\quad
 Y_{ud} = \begin{pmatrix} 0 & \tfrac{2}{3} y_{d,12} & 0 \\ -4 y_{d,21} & -4 y_s & 0 \\ 0 & 0 & y_b  \end{pmatrix} \;,\quad \\
 Y_{qq} &= Y_u \;,\quad Y_{ue} = Y_u \;,
\end{align}
where the structure of $Y_u$ can be read off from eq.~\eqref{eq:B_Y}.\footnote{GUT
textures for proton decay, without fully constructed models, have been considered,
for example, in \cite{Nath:1996qs, Wiesenfeldt:2004qa}.}

The dimension five operators which violate baryon number after integrating 
out the triplets read
\begin{equation}
 W_{\cancel{B}} = \frac{1}{M_T^{\text{eff}}} \left[ \frac{1}{2} Y_{qq}^{ij} Y_{ql}^{mn} Q_i Q_j Q_m L_n + Y_{ue}^{ij} Y_{ud}^{mn} \bar U_i \bar E_j \bar U_m \bar D_n \right] \;,
 \label{eq:Dim5pDecay}
\end{equation}
where we have suppressed $\epsilon$-tensors. The first operator is called
the $LLLL$ operator and the second one the $RRRR$ operator.
To make definite predictions for the proton decay rate one would have to
take the RGE evolution of these operators to low energies into account
and dress the operators with a closed loop including MSSM particles
to calculate the decay rate of the proton, see e.g.~\cite{Nath:2006ut} and
references therein. This goes clearly beyond the scope of this paper. Before we
give some more qualitative statements about what we expect for the proton decay rate in
comparison to other models we want to argue first that the operators
in eq.~\eqref{eq:Dim5pDecay} together with $M_T^{\text{eff}}$ from 
eq.~\eqref{eq:MTeff24_definition} give the dominant dimension five contribution.

Inside the additional higher dimensional representations used in the DMPM
there are additional colour triplets which could in principle give the same
operators like in eq.~\eqref{eq:Dim5pDecay} but with a weaker suppression.
We have explicitly checked that the DMPM by itself is safe: the additional
colour triplets from 5-, 45- and 50-dimensional representations can only
mediate operators that are suppressed compared to the leading contribution by one or more powers of $H_{24} / M_{\text{Pl}}$ and $H_{24}^\prime / M_{\text{Pl}}$.

We have also checked for each of our models that no other Yukawa matrix entries
(through components in the messengers) lead to a lower $\Mf$. This is,
of course, model dependent and has to be checked for any specific model.

We now turn to a qualitative discussion
of dimension five proton decay in the considered class of models. Firstly, we want to
point out that especially in the first model one could expect
some decay modes of the proton to be suppressed because $Y_{ql}$ and $Y_{ud}$
are diagonal in flavour space. Therefore, decays which need a flavour transition
in these matrices would be suppressed. This reinforces the statement that our setup is
more predictive than conventional ones, due to the better
control over the flavour structure of all the Yukawa matrices governing proton decay.

Secondly, introducing a second adjoint in a renormalizable way gives us 
enough freedom to enhance the effective triplet mass $\Mf$ 
to comfortable levels, cf.~section \ref{sec:GUT}. In the case of superpotential (b),
we get a mass of roughly $5 \cdot 10^{16} \, \GeV$ for a SUSY scale of 1~TeV
which increases with increasing $M_{\susy}$, see section~\ref{sec:Superpotentialbc}.
For superpotential (a) the situation is
more involved and depends on the specific GUT scale, and the effective
triplet mass can be easily above $10^{18} \, \GeV$.

To conclude the discussion of the dimension five operators mediating proton decay
we also want to stress that in our setup we
need only a moderate value of $\tan \beta \approx 25$. As it was pointed out
by Lucas and Raby \cite{Lucas:1996bc} especially the contribution from the
$RRRR$ operator is enhanced by $\tan^2 \beta$ for large $\tan \beta$
which poses a challenge for many GUT scenarios which rely on $\tan \beta \approx 50$.
Our setup with $y_\tau / y_b = \frac{3}{2}$ and $\tan\beta\approx 25$ has therefore a
suppression of proton decay via the $RRRR$ operator by a factor of four compared to
these models.

In summary, together with the large effective triplet mass in the double missing partner
mechanism we hence expect the proton decay rate to be sufficiently small but
possibly in the reach of the next generation of proton decay experiments.

\subsection{Proton Decay from Dimension Six Operators}

We now turn to the discussion of dimension six proton decay. As discussed in section
\ref{sec:strat},  the dimension six proton decay mediated by colour triplets originates
from the K\"ahler potential
\begin{equation}
K_T = T^a T^{\dagger a} + \bar T^a \bar T^{\dagger a}\;.
\end{equation}
When the colour triplets are integrated out, the dimension six baryon number violating K\"ahler operators emerge as 
\begin{equation}
\label{eq:Dim6pDecayTriplets}
K_{\cancel{B}} = 
- \frac{1}{\left(\MsTb\right)^2}  \frac{1}{2}Y_{qq}^{ij}Y_{ue}^{* mn}Q_iQ_j\bar U^\dagger_m\bar E^\dagger_n-
 \frac{1}{\left(\MsT\right)^2}   Y_{ql}^{ij}Y_{ud}^{* mn}Q_iL_j\bar U^\dagger_m\bar D_n^\dagger+ h.c.\;,
\end{equation}
where the Yukawa coupling matrices are defined by $W_T$ of Eq.~\eqref{eq:Dim5pDecay}.
As for the dimension five proton decay, these operators need to be subject to a RGE evolution from the GUT scale to the proton mass scale. Note however, that these operators do not need to be dressed with superparticles and therefore the proton decay rates obtained from dimension six operators are independent of the details of the SUSY spectrum.

For completeness, we will now briefly discuss as well proton decay from exchange of heavy gauge bosons.
There is no substantial difference in our models compared to other SUSY SU(5)
models since the gauge structure is exactly the same.

To make the discussion a little bit more analogous to the previous discussion
we define an effective GUT scale
\begin{equation}
    M_\gut^\text{eff} = \frac{M_\gut}{\sqrt{\alpha_u}}\;,
\end{equation}
where $\alpha_u$ is the unified gauge coupling at the GUT scale and
$M_\gut \equiv M_V$ is the  mass of the leptoquark vector bosons,
cf.~the discussion in section~\ref{sec:GUT}.
Just integrating out the heavy gauge bosons in the tree-level
diagrams governing proton decay, and from dimensional
analysis, we can estimate the lifetime of the proton to be
\begin{equation}
    \Gamma_p \approx \alpha_u^2 \frac{m_p^5}{M_V^4} =
       \frac{m_p^5}{(M_\gut^\text{eff})^4} \;,
\end{equation}
where $m_p$ is the proton mass and we have neglected RGE effects 
and order one coefficients
from nuclear matrix elements and such. The most stringent bound
on the proton lifetime is $\tau(p \to \pi^0 e^+) > 8.2 \cdot 10^{33}$~years
\cite{pdg} which yields an effective GUT scale of about $M_\gut^\text{eff}
\gtrsim 2 \cdot 10^{16} \GeV$. In section~\ref{sec:GUT} we have seen that
the GUT scale can be easily above $10^{16}$~GeV and
$\sqrt{\alpha_u} \approx 1/5$ such that the proton decay rate from dimension
six operators is sufficiently small but possibly in the reach of the next generation
of proton decay experiments.

\section{Summary and Conclusions \label{sec:conc}}

In this work we have discussed how the double missing partner mechanism solution to the
doublet-triplet splitting problem in four-dimensional supersymmetric
SU(5) Grand Unified Theories can be combined with predictive models featuring novel predictions
for the quark-lepton Yukawa coupling ratios at the GUT scale.

We have argued that towards this goal a second SU(5) breaking Higgs
field in the adjoint representation is very useful. We systematically discussed all
possible renormalizable superpotentials with two adjoint Higgs fields,
also calculating the corresponding constraints on the GUT scale and effective triplet
mass from a two-loop gauge coupling unification analysis. We found that the effective 
masses of the colour triplet, which enter dimension five and six proton decay, can easily be raised enough to avoid problems with 
proton decay (more than feasible with standard non-renormalizable Higgs potentials with only one  
adjoint GUT Higgs field). 

We have constructed two explicit flavour models with different predictions for the GUT scale Yukawa sector.
A set of shaping symmetries and a renormalizable messenger sector for the models is presented,
which guarantees that only the desired effective GUT operators are generated when the heavy degrees of freedom are integrated out.
In addition, we also include all possible effective Planck-scale suppressed operators consistent with our symmetries, 
and make sure that they do not spoil our results. The models stay perturbative until close to the Planck scale, 
such that our predictions do not suffer from large uncertainties due to these Planck-scale suppressed operators.
They serve as existence proofs that predictive models for the GUT scale quark-lepton mass relations 
can be combined successfully with the DMPM solution for solving the DTS problem.
 
We also provide several useful appendices for GUT flavour model building: For instance, one appendix contains the 
Clebsch-Gordan coefficients for the couplings of the colour triplets, which are required for calculating the rates for proton 
decay induced by their exchange.
We provide detailed tables with the Clebsch-Gordan coefficients for the possible dimension five and six GUT Yukawa operators. 
We also discuss there how one can use GUT Higgs potentials for flavour model building, where a (discrete) R-symmetry is 
broken spontaneously around the GUT scale. R-symmetries are a helpful ingredient of many flavour models, especially when they include non-Abelian family symmetries. 

In summary, we have demonstrated that four-dimensional supersymmetric SU(5) GUTs
with successful doublet-triplet splitting can be combined with predictive models
featuring promising predictions for the quark-lepton Yukawa coupling ratios at
the GUT scale. We have provided the tools for the construction of even more
ambitious GUT models of flavour with additional non-Abelian family symmetries, as 
well as towards the calculation of the predictions for the rates of the various
nucleon decay channels in such models by systematically providing the required
Clebsch-Gordan coefficients.

\section*{Acknowledgements}

This work is supported by the Swiss National Science Foundation. We thank Borut Bajc and David Emmanuel-Costa for useful discussions.

\section*{Appendix}
\begin{appendix}

\section{Yukawa Coupling Ratios including Colour Triplets}
\label{app:Clebsch}

The MSSM superpotential is given by
\begin{equation}
    W = \epsilon_{\alpha\beta} \left( 
        (Y_e)^{ij} H_d^\alpha L_i^{\beta} \bar{E}_j + 
        (Y_d)^{ij} H_d^\alpha Q_i^{\beta a} \bar{D}_j^{a} +
        (Y_u)^{ij} H_u^\beta Q_i^{\alpha a} \bar{U}_j^{a} +
        \mu H_u^\alpha H_d^\beta
    \right) \;,
\end{equation}
where $i$,$j$ are generation indices, $\epsilon_{\alpha \beta}$ the Levi-Civita
tensor ($\epsilon_{12} = 1$), $\alpha$,$\beta$ are SU(2) indices and
$a$, $b$ and $c$ are SU(3) indices\footnote{This definition coincides with the
definitions of \cite{Skands:2003cj}.}. Adding a pair of colour triplets $T$ and $\bar{T}$,
we get the additional terms
\begin{align}
    W_T =&\; \epsilon_{\alpha\beta} \left( 
        - \frac{1}{2} (Y_{qq})_{ij} \epsilon_{abc} T^a Q_i^{\alpha b} Q_j^{\beta c} 
        + (Y_{ql})_{ij} \bar{T}^a Q_i^{\alpha a} L_j^\beta
    \right) \nonumber\\
        +&\;
        (Y_{ue})_{ij} T^a \bar{U}_i^a \bar{E}_j 
        - (Y_{ud})_{ij} \epsilon_{abc} \bar{T}^a \bar{U}_i^{b} \bar{D}_j^{c} +
        M_T T^a \bar{T}^a \;,
\end{align}
where $\epsilon_{abc}$ is the three indices Levi-Civita tensor (with $\epsilon_{123} = 1$).

Extending the SM gauge group to SU(5), we embed the MSSM superfields in a $5$-plet
$H_5$, $\bar{5}$-plets ${\bar H}_{5}$ and $\suF_i$, and $10$-plets $\suT_i$ as in
\begin{align}
    H_5         &= \begin{pmatrix} T^r & T^g & T^b & H_u^+ & H_u^0 \end{pmatrix}^T \;,\\
    {\bar H}_{5} &= \begin{pmatrix} \bar{T}^r & \bar{T}^g & \bar{T}^b & H_d^- & -H_d^0 \end{pmatrix} \;,\\
    \suF_i         &= \begin{pmatrix} \bar{D}_i^r & \bar{D}_i^g & \bar{D}_i^b & E_i & -\nu_i \end{pmatrix} \;,\\
    \suT_i         &= \frac{1}{\sqrt{2}} \begin{pmatrix}
        0 & -\bar{U}_i^b & \bar{U}_i^g & -U_i^r & -D_i^r \\
        \bar{U}_i^b & 0 & -\bar{U}_i^r & -U_i^g & -D_i^g \\
        -\bar{U}_i^g & \bar{U}_i^r & 0 & -U_i^b & -D_i^b \\
        U_i^r & U_i^g & U_i^b & 0 & -\bar{E}_i \\
        D_i^r & D_i^g & D_i^b & \bar{E}_i & 0 \\
    \end{pmatrix} \;,
\end{align}
where $r,g,b$ are the SU(3) colours and $U$, $D$ and $\nu$, $E$ are the 
components of SU(2)-doublets $Q$ and $L$.\footnote{Likewise
$H_d = \begin{pmatrix}H_d^0 & H_d^- \end{pmatrix}^T$ and $H_u =
\begin{pmatrix} H_u^+ & H_u^0 \end{pmatrix}^T$.}
We can write down the renormalizable superpotential terms
\begin{equation}\label{eq:YukawaSU5}
    W = (Y_{TF})_{ij} \suT_i^{ab} (\suF_j)_a (\bar{H}_{5})_b +
        \frac{1}{2} (Y_{TT})_{ij} \epsilon_{abcde} \suT_i^{ab} \suT_j^{cd} H_5^e +
        \mu_5 H_5^a (\bar{H}_{5})_a \;,
\end{equation}
where now $a,b,c,d,e$ are SU(5)-indices and $\epsilon_{abcde}$ is the
respective Levi-Civita tensor. From the embedding of the MSSM fields,
one obtains the minimal SU(5) GUT scale relations
\begin{align}
    \mu &= M_T = \mu_5 \;, \\
    Y_d &= Y_e^T = Y_{ql} = Y_{ud} = \frac{1}{\sqrt{2}} Y_{TF} \;, \\
    Y_u &= Y_u^T = Y_{qq} = Y_{ue} = 2 \, Y_{TT} \;.
\end{align}
The relation $Y_d = Y_e^T$ is highly disfavoured as was already
discussed in section \ref{subsec:Clebsch}.
The conventional approach is to add a 45-dimensional Higgs representation
which generates a relative factor of $-3$ between the Yukawa couplings
of the charged leptons and down-type quarks \cite{GJ}.
In this work we instead focus on an approach where the ratios between
Yukawa couplings are fixed by the CG coefficients of
higher-dimensional operators where in addition an adjoint Higgs
representation of SU(5) is added \cite{Antusch:2009gu, Antusch:2013rxa}.
This approach was briefly reviewed in section \ref{subsec:Clebsch} and for
a thorough discussion we refer the interested reader to the original papers, see also
\cite{Spinrath:2010dh}.

\begin{figure}
\begin{center}
    \includegraphics[page=1,scale=0.55]{yukawaratiooperators.pdf}
    \includegraphics[page=2,scale=0.55]{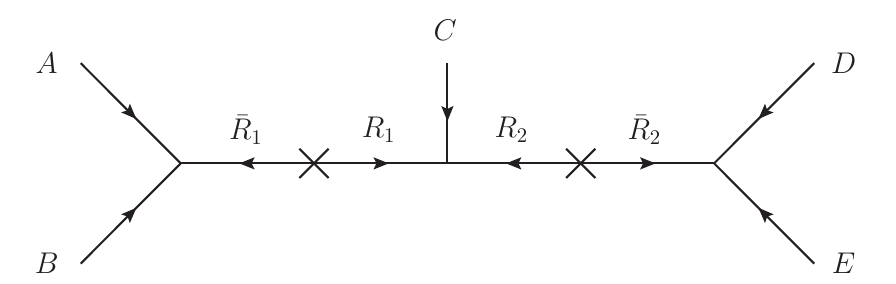}
\end{center}
\caption{Supergraphs generating Yukawa couplings upon integrating out messengers fields in representation $R$,$\bar{R}$, etc.}\label{fig:messengerdiagram}
\end{figure}

In this appendix we will extend the previous discussions to include also the
relative CG coefficents to the triplets. In \cite{Antusch:2009gu, Antusch:2013rxa}
only $Y_d$, $Y_e$ and $Y_u$ were discussed while here we will also 
discuss in detail the implications
of this approach for $Y_{ql}$, $Y_{ud}$, $Y_{qq}$, $Y_{ue}$.
The list of the resulting ratios for dimension 4 and 5 operators with Higgs fields
in a 5- and 45-dimensional representation can be found in
Tabs.~\ref{tab:effectiveFToperators1} and \ref{tab:effectiveTToperators1}, 
where the labels for the representations is defined by figure \ref{fig:messengerdiagram}.
The corresponding results for dimension six operators
are given in Tabs.~\ref{tab:effectiveFToperators2} - \ref{tab:effectiveTToperators3}.
The ones used in the example models are marked with a ``$\rightarrow$''
in the tables.

There are a few comments in order. First, note that several topologies
involving a 45-dimensional messenger field exhibit a free parameter,
simply because the tensor product $\mathbf{45} \otimes \mathbf{24}$
contains two 45-dimensional representations. Hence, there are two
operators possibly giving two different ratios so that any ratio is possible
depending on the coefficients of the two operators. For these cases we
write $x$ in the tables.

We want to mention as well that, unlike at the renormalizable level,
the up-type quark Yukawa and related matrices do not have
to be symmetric or antisymmetric.
Consider, for example, the operator
$(H_{24} \suT_1)_{\mathbf{10}} (H_5 \suT_2)_{\overline{\mathbf{10}}}$.
Due to the symmetries and messenger content the operator
$(H_{24} \suT_2)_{\mathbf{10}} (H_5 \suT_1)_{\overline{\mathbf{10}}}$
could be forbidden. In this case we find $(Y_u)_{12} / (Y_u)_{21} = -4$.
Hence we have adopted the following notation for the ratios in the tables
for the Yukawa couplings related to $Y_u$
\begin{align}
    (Y_u)_{ij} : (Y_u)_{ji} : (Y_{qq})_{ij} : (Y_{ue})_{ij} : (Y_{ue})_{ji} &= 
    a : b : c : d : e \;,
\end{align}    
which reduces for the diagonal entries of the Yukawa matrices to
\begin{align}
    (Y_u)_{ii} : (Y_{qq})_{ii} : (Y_{ue})_{ii} &= (a + b) : c : (d + e) \;.
\end{align}
The ratios related to $Y_{d}$ do not have this extra complication since none
of them could be expected to be symmetric or anti-symmetric in the first place.

\begin{table}
\begin{center}
\setlength{\extrarowheight}{3pt}
\begin{tabular}{ccccc@{ : }c@{ : }c@{ : }c}
\toprule
 &$A \, B$ & $C \, D$ & $R$ & $(Y_d)_{ij}$ & $(Y_e)_{ji}$ & $(Y_{ql})_{ij}$ & $(Y_{ud})_{ij}$ \\
\midrule
 & \multicolumn{2}{c}{$\suF_j \, \suT_i \, \bar{H}_5$}            & ---       & $1$ & $1$ & $1$ & $1$ \\
$\rightarrow$ & $H_{24} \, \suT_i$ & $\suF_j \, \bar{H}_5$ & 10                & $1$ & $6$ & $1$ & $-4$ \\
 & $H_{24} \, \suT_i$ & $\suF_j \, \bar{H}_5$ & 15                & $1$ & $0$ & $-1$ & $0$ \\
 & $H_{24} \, \bar{H}_5$ & $\suF_j \, \suT_i$ & $\bar{5}$         & $1$ & $1$ & $-\frac{2}{3}$ & $-\frac{2}{3}$ \\
 & $H_{24} \, \bar{H}_5$ & $\suF_j \, \suT_i$ & $\overline{45}$   & $1$ & $-3$ & $-2$ & $2$ \\
$\rightarrow$ & $H_{24} \, \suF_j$ & $\suT_i \, \bar{H}_5$ & $\bar{5}$         & $1$ & $-\frac{3}{2}$ & $-\frac{3}{2}$ & $1$ \\
 & $H_{24} \, \suF_j$ & $\suT_i \, \bar{H}_5$ & $\overline{45}$   & $1$ & $\frac{3}{2}$ & $-\frac{1}{2}$ & $-1$ \\
\midrule
& \multicolumn{2}{c}{$\suF_j \, \suT_i \, \bar{H}_{45}$}    & --- & $1$ & $-3$ & $\sqrt{3}$ & $-\sqrt{3}$ \\
& $H_{24} \, \suT_i$ & $\suF_j \, \bar{H}_{45}$             & 10 & $1$ & $-18$ & $\sqrt{3}$ & $4 \sqrt{3}$ \\
& $H_{24} \, \suT_i$ & $\suF_j \, \bar{H}_{45}$             & 40 & $1$ &$ 0$ & $-\frac{\sqrt{3}}{2}$ & $-\frac{\sqrt{3}}{2}$ \\
& $H_{24} \, \suT_i$ & $\suF_j \, \bar{H}_{45}$             & 175 & $1$ & $\frac{36}{23}$ & $-\frac{19 \sqrt{3}}{23}$ & $-\frac{16 \sqrt{3}}{23}$ \\
& $H_{24} \, \bar{H}_{45}$ & $\suF_j \, \suT_i$             & $\bar{5}$ & $1$ & $1$ & $-\frac{2}{\sqrt{3}}$ & $-\frac{2}{\sqrt{3}}$ \\
& $H_{24} \, \bar{H}_{45}$ & $\suF_j \, \suT_i$             & $\overline{45}$ & $1$ & $-3$ & $x$ & $-x$ \\
& $H_{24} \, \suF_j$ & $\suT_i \, \bar{H}_{45}$             & $\bar{5}$ & $1$ & $\frac{9}{2}$ & $-\frac{3 \sqrt{3}}{2}$ & $-\sqrt{3}$ \\
& $H_{24} \, \suF_j$ & $\suT_i \, \bar{H}_{45}$             & $\overline{45}$ & $1$ & $-\frac{1}{2}$ & $-\frac{\sqrt{3}}{2}$ & $-\frac{1}{\sqrt{3}}$ \\
& $H_{24} \, \suF_j$ & $\suT_i \, \bar{H}_{45}$             & $\overline{70}$ & $1$ & $\frac{9}{4}$ & $-\frac{3 \sqrt{3}}{4}$ & $-\sqrt{3}$ \\
\bottomrule
\end{tabular}
\end{center}
\caption{$Y_{TF}$-like CG ratios for the dimension 4 operator and
effective dimension 5 operators $W \supset (A B)_R (C D)_{\bar{R}}$
(involving 5- and 45-dimensional Higgs fields) corresponding to the
the left diagram in figure \ref{fig:messengerdiagram}. Note that one
combination has a free parameter $x$ due to the ambiguity of the index
contraction. See main text for more details.}\label{tab:effectiveFToperators1}
\setlength{\extrarowheight}{0pt}
\end{table}

\begin{table}
\begin{center}
\setlength{\extrarowheight}{3pt}
\begin{tabular}{ccccc@{ : }c@{ : }c@{ : }c@{ : }c}
\toprule
&$A \, B$ & $C \, D$ & $R$ & $(Y_u)_{ij}$ & $(Y_u)_{ji}$ & $(Y_{qq})_{ij}$ & $(Y_{ue})_{ij}$ & $(Y_{ue})_{ji}$\\
\midrule
$\rightarrow$ & \multicolumn{2}{c}{$\suT_i \suT_j H_5$} & --- & $1$ & $1$ & $1$ & $1$ & $1$\\
&$H_{24} H_5$ & $\suT_i \suT_j$ & $5$ & $1$ & $1$ & $-\frac{2}{3}$ & $-\frac{2}{3}$ & $-\frac{2}{3}$\\
&$H_{24} H_5$ & $\suT_i \suT_j$ & $45$ & $1$ & $-1$ & $0$ & $-2$ & $2$\\
&$H_{24} \suT_i$ & $\suT_j H_5$ & $10$ & $1$ & $-4$ & $1$ & $-4$ & $6$\\
&$H_{24} \suT_i$ & $\suT_j H_5$ & $40$ & $1$ & $\frac{1}{2}$ & $-\frac{1}{2}$ & $-1$ & $0$\\
\midrule
& \multicolumn{2}{c}{$\suT_i \suT_j H_{45}$} & --- & $1$ & $-1$ & $0$ & $\sqrt{3}$ & $-\sqrt{3}$ \\
& $H_{24} \suT_i$ & $\suT_j H_{45}$ & $10$ & $1$ & $4$ & $0$ & $-4 \sqrt{3}$ & $-6 \sqrt{3}$ \\
& $H_{24} \suT_i$ & $\suT_j H_{45}$ & $15$ & $1$ & $0$ & $-\frac{\sqrt{3}}{2}$ & $0$ & $0$ \\
& $H_{24} \suT_i$ & $\suT_j H_{45}$ & $40$ & $1$ & $-\frac{7}{2}$ & $\frac{3 \sqrt{3}}{2}$ & $-\sqrt{3}$ & $0$ \\
& $H_{24} \suT_i$ & $\suT_j H_{45}$ & $175$ & $1$ & $\frac{16}{19}$ & $-\frac{21 \sqrt{3}}{38}$ & $-\frac{16 \sqrt{3}}{19}$ & $-\frac{12 \sqrt{3}}{19}$ \\
& $H_{24} H_{45}$ & $\suT_i \suT_j$ & $5$ & $1$ & $1$ & $-\frac{2}{\sqrt{3}}$ & $-\frac{2}{\sqrt{3}}$ & $-\frac{2}{\sqrt{3}}$ \\
& $H_{24} H_{45}$ & $\suT_i \suT_j$ & $45$ & $1$ & $-1$ & $0$ & $x$ & $-x$ \\
& $H_{24} H_{45}$ & $\suT_i \suT_j$ & $50$ & $0$ & $0$ & $1$ & $-2$ & $-2$ \\
\bottomrule
\end{tabular}
\end{center}
\caption{$Y_{TT}$-like CG ratios for the dimension 4 operator and
effective dimension 5 operators $W \supset (A B)_R (C D)_{\bar{R}}$
(involving 5- and 45-dimensional Higgs fields) corresponding to the
the left diagram in figure \ref{fig:messengerdiagram}. Note that one
combination has a free parameter $x$ due to the ambiguity of the
index contraction. See main text for more details.}\label{tab:effectiveTToperators1}
\setlength{\extrarowheight}{0pt}
\end{table}

\begin{table}
\begin{center}
\setlength{\extrarowheight}{3pt}
\begin{tabular}{cccccc@{ : }c@{ : }c@{ : }c}
\toprule
& $A \, B$ & $C$ & $D \, E$ & $R_1, R_2$ & $(Y_d)_{ij}$ & $(Y_e)_{ji}$ & $(Y_{ql})_{ij}$ & $(Y_{ud})_{ij}$ \\
\midrule
& $\suT_i \, \bar{H}_5$ & $\suF_j$ & $H_{24} \, H_{24}$    & 5, 1      & $1$ & $1$ & $1$ & $1$ \\
& $\suT_i \, \bar{H}_5$ & $\suF_j$ & $H_{24} \, H_{24}$    & 5, 24     & $1$ & $-\frac{3}{2}$ & $-\frac{3}{2}$ & $1$ \\
& $\suT_i \, \bar{H}_5$ & $\suF_j$ & $H_{24} \, H_{24}$    & 45, 24    & $1$ & $\frac{3}{2}$ & $-\frac{1}{2}$ & $-1$ \\
& $\suT_i \, \bar{H}_5$ & $\suF_j$ & $H_{24} \, H_{24}$    & 45, 75    & $1$ & $-3$ & $1$ & $-1$ \\
& $H_{24} \, \bar{H}_5$ & $H_{24}$ & $\suF_j \, \suT_i$    & $\bar{5}$, 5      & $1$ & $1$ & $\frac{4}{9}$ & $\frac{4}{9}$ \\
& $H_{24} \, \bar{H}_5$ & $H_{24}$ & $\suF_j \, \suT_i$    & $\bar{5}$, 45     & $1$ & $-3$ & $\frac{4}{3}$ & $-\frac{4}{3}$ \\
& $H_{24} \, \bar{H}_5$ & $H_{24}$ & $\suF_j \, \suT_i$    & $\overline{45}$, 5     & $1$ & $1$ & $\frac{4}{3}$ & $\frac{4}{3}$ \\
& $H_{24} \, \bar{H}_5$ & $H_{24}$ & $\suF_j \, \suT_i$    & $\overline{45}$, 45    & $1$ & $-3$ & $x$ & $-x$ \\
& $H_{24} \, \bar{H}_5$ & $H_{24}$ & $\suF_j \, \suT_i$    & $\overline{70}$, 5     & $1$ & $1$ & $\frac{8}{9}$ & $\frac{8}{9}$ \\
& $H_{24} \, \bar{H}_5$ & $H_{24}$ & $\suF_j \, \suT_i$    & $\overline{70}$, 45    & $1$ & $-3$ & $\frac{8}{3}$ & $-\frac{8}{3}$ \\
& $\suF_j \, \suT_i$ & $\bar{H}_5$ & $H_{24} \, H_{24}$    & 5, 1      & $1$ & $1$ & $1$ & $1$ \\
& $\suF_j \, \suT_i$ & $\bar{H}_5$ & $H_{24} \, H_{24}$    & 5, 24     & $1$ & $1$ & $-\frac{2}{3}$ & $-\frac{2}{3}$ \\
& $\suF_j \, \suT_i$ & $\bar{H}_5$ & $H_{24} \, H_{24}$    & 45, 24    & $1$ & $-3$ & $-2$ & $2$ \\
& $\suF_j \, \suT_i$ & $\bar{H}_5$ & $H_{24} \, H_{24}$    & 45, 75    & $1$ & $-3$ & $1$ & $-1$ \\
& $H_{24} \, \bar{H}_5$ & $\suT_i$ & $H_{24} \, \suF_j$    & $\bar{5}$, $\bar{5}$      & $1$ & $-\frac{3}{2}$ & $1$ & $-\frac{2}{3}$ \\
& $H_{24} \, \bar{H}_5$ & $\suT_i$ & $H_{24} \, \suF_j$    & $\bar{5}$, $\overline{45}$     & $1$ & $\frac{3}{2}$ & $\frac{1}{3}$ & $\frac{2}{3}$ \\
& $H_{24} \, \bar{H}_5$ & $\suT_i$ & $H_{24} \, \suF_j$    & $\overline{45}$, $\bar{5}$     & $1$ & $\frac{9}{2}$ & $3$ & $2$ \\
$\rightarrow$ & $H_{24} \, \bar{H}_5$ & $\suT_i$ & $H_{24} \, \suF_j$    & $\overline{45}$, $\overline{45}$    & $1$ & $-\frac{1}{2}$ & $1$ & $\frac{2}{3}$ \\
& $H_{24} \, \bar{H}_5$ & $\suT_i$ & $H_{24} \, \suF_j$    & $\overline{45}$, $\overline{70}$    & $1$ & $\frac{9}{4}$ & $\frac{3}{2}$ & $2$ \\
& $H_{24} \, \bar{H}_5$ & $\suT_i$ & $H_{24} \, \suF_j$    & $\overline{70}$, $\overline{45}$    & $1$ & $\frac{3}{2}$ & $\frac{2}{3}$ & $\frac{4}{3}$ \\
& $H_{24} \, \bar{H}_5$ & $\suT_i$ & $H_{24} \, \suF_j$    & $\overline{70}$, $\overline{70}$    & $1$ & $\frac{3}{4}$ & $1$ & $\frac{2}{3}$ \\ 
& $\suF_j \, \bar{H}_5$ & $H_{24}$  & $H_{24} \, \suT_i$   & $\overline{10}$, 10    & $1$ & $36$ & $1$ & $16$ \\
& $\suF_j \, \bar{H}_5$ & $H_{24}$  & $H_{24} \, \suT_i$   & $\overline{10}$, 15    & $1$ & $0$ & $1$ & $0$ \\
& $\suF_j \, \bar{H}_5$ & $H_{24}$  & $H_{24} \, \suT_i$   & $\overline{10}$, 40    & $1$ & $0$ & $1$ & $1$ \\
& $\suF_j \, \bar{H}_5$ & $H_{24}$  & $H_{24} \, \suT_i$   & $\overline{10}$, 175   & $1$ & $\frac{72}{61}$ & $1$ & $\frac{64}{61}$ \\
& $\suF_j \, \bar{H}_5$ & $H_{24}$  & $H_{24} \, \suT_i$   & $\overline{15}$, 10    & $1$ & $0$ & $-1$ & $0$ \\
& $\suF_j \, \bar{H}_5$ & $H_{24}$  & $H_{24} \, \suT_i$   & $\overline{15}$, 15    & $1$ & $0$ & $-1$ & $0$ \\
& $\suF_j \, \bar{H}_5$ & $H_{24}$  & $H_{24} \, \suT_i$   & $\overline{15}$, 175   & $1$ & $0$ & $-1$ & $0$ \\
\bottomrule
\end{tabular}
\end{center}
\caption{$Y_{TF}$-like CG ratios for the effective dimension 6 operators $W \supset (A B)_{R_1} C (D E)_{R_2}$ corresponding to the the right diagram in figure \ref{fig:messengerdiagram}. Note one combination has a free parameter $x$. 
See main text for more details.
}\label{tab:effectiveFToperators2}
\setlength{\extrarowheight}{0pt}
\end{table}

\begin{table}
\begin{center}
\setlength{\extrarowheight}{3pt}
\begin{tabular}{cccccc@{ : }c@{ : }c@{ : }c}
\toprule
& $A \, B$ & $C$ & $D \, E$ & $R_1, R_2$ & $(Y_d)_{ij}$ & $(Y_e)_{ji}$ & $(Y_{ql})_{ij}$ & $(Y_{ud})_{ij}$ \\
\midrule
& $H_{24} \, \suF_j$ & $H_{24}$  & $\suT_i \, \bar{H}_5$   & $\bar{5}$, 5      & $1$ & $\frac{9}{4}$ & $\frac{9}{4}$ & $1$ \\
& $H_{24} \, \suF_j$ & $H_{24}$  & $\suT_i \, \bar{H}_5$   & $\bar{5}$, 45     & $1$ & $-\frac{9}{4}$ & $\frac{3}{4}$ & $-1$ \\
& $H_{24} \, \suF_j$ & $H_{24}$  & $\suT_i \, \bar{H}_5$   & $\overline{45}$, 5     & $1$ & $\frac{3}{4}$ & $\frac{3}{4}$ & $1$ \\
& $H_{24} \, \suF_j$ & $H_{24}$  & $\suT_i \, \bar{H}_5$   & $\overline{45}$, 45    & $1$ & $x$ & $-\frac{x}{3}$ & $-1$ \\
& $H_{24} \, \suF_j$ & $H_{24}$  & $\suT_i \, \bar{H}_5$   & $\overline{70}$, 5     & $1$ & $\frac{9}{8}$ & $\frac{9}{8}$ & $1$ \\
& $H_{24} \, \suF_j$ & $H_{24}$  & $\suT_i \, \bar{H}_5$   & $\overline{70}$, 45    & $1$ & $-\frac{9}{8}$ & $\frac{3}{8}$ & $-1$ \\
& $H_{24} \, \suF_j$ & $\bar{H}_5$ & $H_{24} \, \suT_i$    & $\bar{5}$, 10     & $1$ & $-9$ & $-\frac{3}{2}$ & $-4$ \\
& $H_{24} \, \suF_j$ & $\bar{H}_5$ & $H_{24} \, \suT_i$    & $\bar{5}$, 15     & $1$ & $0$ & $\frac{3}{2}$ & $0$ \\
& $H_{24} \, \suF_j$ & $\bar{H}_5$ & $H_{24} \, \suT_i$    & $\overline{45}$, 10    & $1$ & $9$ & $-\frac{1}{2}$ & $4$ \\
& $H_{24} \, \suF_j$ & $\bar{H}_5$ & $H_{24} \, \suT_i$    & $\overline{45}$, 40    & $1$ & $0$ & $1$ & $1$ \\
& $H_{24} \, \suF_j$ & $\bar{H}_5$ & $H_{24} \, \suT_i$    & $\overline{45}$, 175   & $1$ & $\frac{18}{19}$ & $\frac{23}{38}$ & $\frac{16}{19}$ \\
& $H_{24} \, \suF_j$ & $\bar{H}_5$ & $H_{24} \, \suT_i$    & $\overline{70}$, 15    & $1$ & $0$ & $\frac{3}{4}$ & $0$ \\
& $H_{24} \, \suF_j$ & $\bar{H}_5$ & $H_{24} \, \suT_i$    & $\overline{70}$, 175   & $1$ & $\frac{9}{7}$ & $\frac{33}{28}$ & $\frac{8}{7}$ \\
& $H_{24} \, \bar{H}_5$ & $\suF_j$ & $H_{24} \, \suT_i$    & $\bar{5}$, 10     & $1$ & $6$ & $-\frac{2}{3}$ & $\frac{8}{3}$ \\
& $H_{24} \, \bar{H}_5$ & $\suF_j$ & $H_{24} \, \suT_i$    & $\bar{5}$, 15     & $1$ & $0$ & $\frac{2}{3}$ & $0$ \\
& $H_{24} \, \bar{H}_5$ & $\suF_j$ & $H_{24} \, \suT_i$    & $\overline{45}$, 10    & $1$ & $-18$ & $-2$ & $-8$ \\
& $H_{24} \, \bar{H}_5$ & $\suF_j$ & $H_{24} \, \suT_i$    & $\overline{45}$, 40    & $1$ & $0$ & $1$ & $1$ \\
& $H_{24} \, \bar{H}_5$ & $\suF_j$ & $H_{24} \, \suT_i$    & $\overline{45}$, 175   & $1$ & $\frac{36}{23}$ & $\frac{38}{23}$ & $\frac{32}{23}$ \\
& $H_{24} \, \bar{H}_5$ & $\suF_j$ & $H_{24} \, \suT_i$    & $\overline{70}$, 15    & $1$ & $0$ & $\frac{4}{3}$ & $0$ \\
& $H_{24} \, \bar{H}_5$ & $\suF_j$ & $H_{24} \, \suT_i$    & $\overline{70}$, 175   & $1$ & $\frac{12}{11}$ & $\frac{28}{33}$ & $\frac{32}{33}$ \\
& $\suF_j \, \bar{H}_5$ & $\suT_i$ & $H_{24} \, H_{24}$    & $\overline{10}$, 1     & $1$ & $1$ & $1$ & $1$ \\
& $\suF_j \, \bar{H}_5$ & $\suT_i$ & $H_{24} \, H_{24}$    & $\overline{10}$, 24    & $1$ & $6$ & $1$ & $-4$ \\
& $\suF_j \, \bar{H}_5$ & $\suT_i$ & $H_{24} \, H_{24}$    & $\overline{10}$, 75    & $1$ & $-3$ & $1$ & $-1$ \\
& $\suF_j \, \bar{H}_5$ & $\suT_i$ & $H_{24} \, H_{24}$    & $\overline{15}$, 24    & $1$ & $0$ & $-1$ & $0$ \\
\bottomrule
\end{tabular}
\end{center}
\caption{Continuation of table \ref{tab:effectiveFToperators2}: $Y_{TF}$-like CG ratios for the effective dimension 6 operators $W \supset (A B)_{R_1} C (D E)_{R_2}$ corresponding to the the right diagram in figure \ref{fig:messengerdiagram}. Note another combination with a free parameter $x$. See main text for more details.}\label{tab:effectiveFToperators3}
\setlength{\extrarowheight}{0pt}
\end{table}

\begin{table}
\begin{center}
\setlength{\extrarowheight}{3pt}
\begin{tabular}{cccccc@{ : }c@{ : }c@{ : }c@{ : }c}
\toprule
& $A \, B$ & $C$ & $D \, E$ & $R_1,R_2$ & $(Y_u)_{ij}$ & $(Y_u)_{ji}$ & $(Y_{qq})_{ij}$ & $(Y_{ue})_{ij}$ & $(Y_{ue})_{ji}$\\
\midrule
& $H_{24} \suT_i$ & $H_ 5$ & $H_{24} \suT_j$ & $10, 10$ & $1$ & $1$ & $-\frac{1}{4}$ & $6$ & $6$ \\
& $H_{24} \suT_i$ & $H_ 5$ & $H_{24} \suT_j$ & $10, 40$ & $1$ & $-8$ & $-1$ & $0$ & $-12$ \\
& $H_{24} \suT_i$ & $H_ 5$ & $H_{24} \suT_j$ & $15, 40$ & $1$ & $0$ & $1$ & $0$ & $0$ \\
& $H_{24} \suT_i$ & $H_ 5$ & $H_{24} \suT_j$ & $40, 10$ & $1$ & $-\frac{1}{8}$ & $\frac{1}{8}$ & $\frac{3}{2}$ & $0$ \\
& $H_{24} \suT_i$ & $H_ 5$ & $H_{24} \suT_j$ & $40, 15$ & $0$ & $1$ & $1$ & $0$ & $0$ \\
& $H_{24} \suT_i$ & $H_ 5$ & $H_{24} \suT_j$ & $40, 175$ & $1$ & $\frac{23}{32}$ & $\frac{19}{32}$ & $\frac{3}{4}$ & $0$ \\
& $H_{24} \suT_i$ & $H_ 5$ & $H_{24} \suT_j$ & $175, 40$ & $1$ & $\frac{32}{23}$ & $\frac{19}{23}$ & $0$ & $\frac{24}{23}$ \\
& $H_{24} \suT_i$ & $H_ 5$ & $H_{24} \suT_j$ & $175, 175$ & $1$ & $1$ & $\frac{41}{40}$ & $\frac{6}{5}$ & $\frac{6}{5}$ \\
& $\suT_i \suT_j$ & $H_ 5$ & $H_{24} H_{24}$ & $\bar{5}, 1$ & $1$ & $1$ & $1$ & $1$ & $1$ \\
& $\suT_i \suT_j$ & $H_ 5$ & $H_{24} H_{24}$ & $\bar{5}, 24$ & $1$ & $1$ & $-\frac{2}{3}$ & $-\frac{2}{3}$ & $-\frac{2}{3}$ \\
& $\suT_i \suT_j$ & $H_ 5$ & $H_{24} H_{24}$ & $\overline{45}, 24$ & $1$ & $-1$ & $0$ & $-2$ & $2$ \\
& $\suT_i \suT_j$ & $H_ 5$ & $H_{24} H_{24}$ & $\overline{45}, 75$ & $1$ & $-1$ & $0$ & $1$ & $-1$ \\
& $\suT_i \suT_j$ & $H_ 5$ & $H_{24} H_{24}$ & $\overline{50}, 75$ & $0$ & $0$ & $1$ & $-2$ & $-2$ \\
& $\suT_i H_ 5$ & $\suT_j$ & $H_{24} H_{24}$ & $\overline{10}, 1$ & $1$ & $1$ & $1$ & $1$ & $1$ \\
& $\suT_i H_ 5$ & $\suT_j$ & $H_{24} H_{24}$ & $\overline{10}, 24$ & $1$ & $-\frac{1}{4}$ & $-\frac{1}{4}$ & $-\frac{3}{2}$ & $1$ \\
& $\suT_i H_ 5$ & $\suT_j$ & $H_{24} H_{24}$ & $\overline{10}, 75$ & $1$ & $-1$ & $-1$ & $3$ & $1$ \\
& $\suT_i H_ 5$ & $\suT_j$ & $H_{24} H_{24}$ & $\overline{40}, 24$ & $1$ & $2$ & $-1$ & $0$ & $-2$ \\
& $\suT_i H_ 5$ & $\suT_j$ & $H_{24} H_{24}$ & $\overline{40}, 75$ & $1$ & $-1$ & $\frac{1}{2}$ & $0$ & $-2$ \\
\bottomrule
\end{tabular}
\end{center}
\caption{$Y_{TT}$-like CG ratios for the effective dimension 6 operators $W \supset (A B)_{R_1} C (D E)_{R_2}$ corresponding to the the right diagram in figure \ref{fig:messengerdiagram}. }\label{tab:effectiveTToperators2}
\setlength{\extrarowheight}{0pt}
\end{table}

\begin{table}
\begin{center}
\setlength{\extrarowheight}{3pt}
\begin{tabular}{cccccc@{ : }c@{ : }c@{ : }c@{ : }c}
\toprule
& $A \, B$ & $C$ & $D \, E$ & $R_1,R_2$ & $(Y_u)_{ij}$ & $(Y_u)_{ji}$ & $(Y_{qq})_{ij}$ & $(Y_{ue})_{ij}$ & $(Y_{ue})_{ji}$\\
\midrule
& $H_{24} \suT_i$ & $\suT_j$ & $H_{24} H_ 5$ & $10, 5$ & $1$ & $-4$ & $-\frac{2}{3}$ & $\frac{8}{3}$ & $-4$ \\
& $H_{24} \suT_i$ & $\suT_j$ & $H_{24} H_ 5$ & $10, 45$ & $1$ & $4$ & $0$ & $8$ & $12$ \\
& $H_{24} \suT_i$ & $\suT_j$ & $H_{24} H_ 5$ & $15, 45$ & $1$ & $0$ & $1$ & $0$ & $0$ \\
& $H_{24} \suT_i$ & $\suT_j$ & $H_{24} H_ 5$ & $40, 5$ & $1$ & $\frac{1}{2}$ & $\frac{1}{3}$ & $\frac{2}{3}$ & $0$ \\
& $H_{24} \suT_i$ & $\suT_j$ & $H_{24} H_ 5$ & $40, 45$ & $1$ & $-\frac{7}{2}$ & $-3$ & $2$ & $0$ \\
& $H_{24} \suT_i$ & $\suT_j$ & $H_{24} H_ 5$ & $40, 70$ & $1$ & $\frac{1}{2}$ & $\frac{2}{3}$ & $\frac{4}{3}$ & $0$ \\
& $H_{24} \suT_i$ & $\suT_j$ & $H_{24} H_ 5$ & $175, 45$ & $1$ & $\frac{16}{19}$ & $\frac{21}{19}$ & $\frac{32}{19}$ & $\frac{24}{19}$ \\
& $H_{24} \suT_i$ & $\suT_j$ & $H_{24} H_ 5$ & $175, 70$ & $1$ & $\frac{8}{7}$ & $\frac{20}{21}$ & $\frac{16}{21}$ & $\frac{8}{7}$ \\
& $\suT_i \suT_j$ & $H_{24}$ & $H_{24} H_ 5$ & $\bar{5}, 5$ & $1$ & $1$ & $\frac{4}{9}$ & $\frac{4}{9}$ & $\frac{4}{9}$ \\
& $\suT_i \suT_j$ & $H_{24}$ & $H_{24} H_ 5$ & $\bar{5}, 45$ & $1$ & $1$ & $\frac{4}{3}$ & $\frac{4}{3}$ & $\frac{4}{3}$ \\
& $\suT_i \suT_j$ & $H_{24}$ & $H_{24} H_ 5$ & $\bar{5}, 70$ & $1$ & $1$ & $\frac{8}{9}$ & $\frac{8}{9}$ & $\frac{8}{9}$ \\
& $\suT_i \suT_j$ & $H_{24}$ & $H_{24} H_ 5$ & $\overline{45}, 5$ & $1$ & $-1$ & $0$ & $\frac{4}{3}$ & $-\frac{4}{3}$ \\
& $\suT_i \suT_j$ & $H_{24}$ & $H_{24} H_ 5$ & $\overline{45}, 45$ & $1$ & $-1$ & $0$ & $x$ & $-x$ \\
& $\suT_i \suT_j$ & $H_{24}$ & $H_{24} H_ 5$ & $\overline{45}, 70$ & $1$ & $-1$ & $0$ & $\frac{8}{3}$ & $-\frac{8}{3}$ \\
& $\suT_i \suT_j$ & $H_{24}$ & $H_{24} H_ 5$ & $\overline{50}, 45$ & $0$ & $0$ & $1$ & $-2$ & $-2$ \\
& $H_{24} \suT_i$ & $H_{24}$ & $\suT_j H_ 5$ & $10, \overline{10}$ & $1$ & $16$ & $1$ & $16$ & $36$ \\
& $H_{24} \suT_i$ & $H_{24}$ & $\suT_j H_ 5$ & $10, \overline{40}$ & $1$ & $-2$ & $-\frac{1}{2}$ & $4$ & $0$ \\
& $H_{24} \suT_i$ & $H_{24}$ & $\suT_j H_ 5$ & $15, \overline{10}$ & $1$ & $0$ & $1$ & $0$ & $0$ \\
& $H_{24} \suT_i$ & $H_{24}$ & $\suT_j H_ 5$ & $40, \overline{10}$ & $1$ & $1$ & $1$ & $1$ & $0$ \\
& $H_{24} \suT_i$ & $H_{24}$ & $\suT_j H_ 5$ & $40, \overline{40}$ & $1$ & $x$ & $-\frac{1}{2}$ & $-2 x$ & $0$ \\
& $H_{24} \suT_i$ & $H_{24}$ & $\suT_j H_ 5$ & $175, \overline{10}$ & $1$ & $\frac{64}{61}$ & $1$ & $\frac{64}{61}$ & $\frac{72}{61}$ \\
& $H_{24} \suT_i$ & $H_{24}$ & $\suT_j H_ 5$ & $175, \overline{40}$ & $1$ & $4$ & $-\frac{1}{2}$ & $-8$ & $0$ \\
\bottomrule
\end{tabular}
\end{center}
\caption{Continuation of table \ref{tab:effectiveTToperators2}: $Y_{TT}$-like CG ratios for the effective dimension 6 operators $W \supset (A B)_{R_1} C (D E)_{R_2}$ corresponding to the the right diagram in figure \ref{fig:messengerdiagram}. Note two combinations have a free parameter $x$. See main text for more details.}\label{tab:effectiveTToperators3}
\setlength{\extrarowheight}{0pt}
\end{table}

If the considered model contains Higgs fields in 5- and 45-dimensional representations
\footnote{One could also imagine using 45-dimensional Higgs fields exclusively. However, this severely exacerbates the doublet-triplet splitting problem as now one has to split the doublets from even more component fields that can generate proton decay operators.},
there are two Higgs doublet pairs in the spectrum and
care should be taken that unification is still possible. One solution is mixing both and making one linear combination heavy while one stays at the electroweak scale. The simplest term generating such a mixing is
\begin{equation}\label{eq:5to45mixing}
    W \supset H_{24} H_5 \bar{H}_{45}
      \propto H_u H_d^{\overline{45}} - \frac{2}{\sqrt{3}} T \bar{T}^{\overline{45}} \;,
\end{equation}
where we suppressed any additional MSSM multiplets in $\bar{H}_{45}$. 
Since a $\overline{\mathbf{45}}$ contains more potentially dangerous MSSM multiplets, 
it is natural to have the heavy linear combination be predominantly in the $\overline{\mathbf{45}}$. 
Then it is possible to treat $\bar{H}_{45}$ like a messenger field and the 
renormalizable operator $\suF \suT \bar{H}_{45}$ turns into the non-renormalizable 
operator $(\suF \suT)_{45} (H_{24} \bar{H}_5)_{{\overline{45}}}$, cf.\ table \ref{tab:effectiveFToperators1}. 
Analogous limits can be deduced trivially.
If the approximation $m_{45} \gg \vev{H_{24}}$ does not hold, one has to 
take into account the full mass matrix for the Higgs doublets including the term in eq.~\eqref{eq:5to45mixing}.

\section{Two-loop RGEs in Extensions to the MSSM \label{sec:2loopRGEs}}
\def\arraystretch{1.2}

The renormalization group equations for gauge couplings at two-loop are given by \cite{Jones:1974pg+1983vk}
\begin{equation}
    \mu \frac{d}{d\mu} g_a= \frac{g_a^3}{16 \pi^2} b_a + \frac{g_a^3}{(16\pi^2)^2} \left(
        \sum\limits_{b=1}^3 B_{ab} g_b^2 - \sum\limits_{f} C_a^f \tr(Y_f^\dagger Y_f) 
    \right)\;,
\end{equation}
in the $\overline{\text{DR}}$ renormalization scheme, where $\mu$ is the renormalization
scale and $f$ runs over all Yukawa coupling matrices. In the MSSM, the beta function coefficients
are given by (in GUT normalisation for $g_1$) 
\begin{equation}
    b_a = \begin{pmatrix} \frac{33}{5} \\ 1 \\ -3 \end{pmatrix}, \;
    B_{ab} = \begin{pmatrix} 
        \frac{199}{25} & \frac{27}{5} & \frac{88}{5} \\
        \frac{9}{5} & 25 & 24 \\
        \frac{11}{5} & 9 & 14 \\
    \end{pmatrix} \;,
\end{equation}
and
\begin{equation}
    C_a^{u,d,e} = \begin{pmatrix}
        \frac{26}{5} & \frac{14}{5} & \frac{18}{5} \\
        6 & 6 & 2 \\
        4 & 4 & 0 \\
    \end{pmatrix} \;.
\end{equation}
where the first column stands for $u$, the second column for $d$
and the third for $e$. Additional colour triplet and weak doublet pairs,
as those contained in $\mathbf{5}$, $\bar{\mathbf{5}}$ representations,
contribute (per pair) at one-loop with 
\begin{equation}
    b^{(5,T)}_a = \begin{pmatrix} \frac{2}{5} \\ 0 \\ 1 \end{pmatrix}, \;
    b^{(5,D)}_a = \begin{pmatrix} \frac{3}{5} \\ 1 \\ 0 \end{pmatrix}, \;
\end{equation}
and at two-loop with
\begin{equation}
    B^{(5,T)}_{ab} = \begin{pmatrix} 
        \frac{8}{75} & 0 & \frac{32}{15} \\
        0 & 0 & 0 \\
        \frac{4}{15} & 0 & \frac{34}{3} 
    \end{pmatrix}, \;
    B^{(5,D)}_{ab} = \begin{pmatrix} 
        \frac{9}{25} & \frac{9}{5} & 0 \\
        \frac{3}{5} & 7 & 0 \\
        0 & 0 & 0 \\
    \end{pmatrix} \;,
\end{equation}
SU(2) triplets, SU(3) octets and leptoquark superfields\footnote{Note that leptoquark superfields can only appear in Dirac pairs due to their charges.} from an adjoint contribute
(per chiral superfield) at one-loop with
\begin{equation}
    b^{(24,T)}_a = \begin{pmatrix} 0 \\ 2 \\ 0 \end{pmatrix}, \;
    b^{(24,O)}_a = \begin{pmatrix} 0 \\ 0 \\ 3 \end{pmatrix}, \; 
    b^{(24,L)}_a = \begin{pmatrix} \frac{5}{2} \\ \frac{3}{2} \\ 1 \end{pmatrix}, \; 
\end{equation}
and at two-loop with
\begin{equation}
    B^{(24,T)}_{ab} = \begin{pmatrix} 
        0 & 0 & 0 \\
        0 & 24 & 0 \\
        0 & 0 & 0 
    \end{pmatrix} ,\;
    B^{(24,O)}_{ab} = \begin{pmatrix} 
        0 & 0 & 0 \\
        0 & 0 & 0 \\
        0 & 0 & 54 
    \end{pmatrix} ,\;
    B^{(24,L)}_{ab} = \begin{pmatrix}
        \frac{25}{6} & \frac{15}{2} & \frac{40}{3} \\
        \frac{5}{2} & \frac{21}{2} & 8 \\
        \frac{5}{3} & 3 & \frac{34}{3} 
    \end{pmatrix} \;.
\end{equation}
Additional Yukawa couplings between SM fermion superfields and a pair of 
colour triplet/anti-triplet contribute with
\begin{equation}
C_a^{qq,ue,ql,ud} = \begin{pmatrix}
        \frac{6}{5} & \frac{28}{5} & \frac{14}{5} & \frac{24}{5} \\
        6 & 0 & 6 & 0 \\
        6 & 2 & 4 & 6 
    \end{pmatrix} \;.
\end{equation} 
In our numerical analysis we have assumed that $Y_{qq} = Y_{ue} = Y_u$
and $Y_{ql} = Y_{ud} = Y_d$ (as motivated by minimal SU(5)). We have checked
that this approximation changes our results only negligibly.

\def\arraystretch{1}

\section{Discussion of messenger fields}
\label{app:mess}
We discuss now the messenger fields appearing in Figure~\ref{fig:optionA_Yu} of the model presented in section~\ref{sec:modelA}. There are three supergraphs generating the superpotential term $y_{12}$. The renormalizable superpotential corresponding to Figure~\ref{fig:optionA_Yu} is
\begin{align}
W \ \supset\ & \gamma_1 \, H_5 \suT_1 Z_{10,4} + \gamma_2 \, H_5 \suT_2 Z_{10,3} + \gamma_3 \, \suT_3 \theta_1 \bar Z_{10,3} + \gamma_\theta \, \theta_1^2 Z_1\nonumber \\
+\ & \lambda_1 \, \theta_2 Z_2 \bar Z_1 + \lambda_{10} \, \theta_2 Z_{10,3} \bar Z_{10,4}  \nonumber \\
+\ & \eta_1 \, \suT_1 \bar Z_{10,3} \bar Z_2 + \eta_2 \, \suT_2 \bar Z_{10,4} \bar Z_2 + \eta'_2 \, \suT_2 \bar Z_{10,3} \bar Z_1  \nonumber \\
 +\ & M_1 \, Z_1\bar Z_1 + M_2 \, Z_2\bar Z_2 + M_{10,3} \, Z_{10,3}\bar Z_{10,3} + M_{10,4} \, Z_{10,4}\bar Z_{10,4}\;,
\end{align}
where we explicitly denote the coupling constants at the ends of a diagram with $\gamma_x$, 
coupling constants in the middle of the diagrams with $\lambda_x$ if they involve $\theta_j$ and $\eta_i$ if they involve $\suT_i$ and messenger masses with $M_x$.
After 
$Z_x$ and $\bar Z_x$ are integrated out and the flavons $\theta_i$ obtain VEVs, the elements of the up-type Yukawa matrix $Y_u$ of eq.~\eqref{eq:A_Y} are given by
\begin{align}
y_{11} &= \gamma_1 \, \lambda_{10} \, \eta_1 \, \lambda_1 \, \gamma_\theta \, \frac{\vev{\theta_1}^2 \vev{\theta_2}^2}{M_1 M_2 M_{10,3} M_{10,4}} \;,\nonumber \\
y_{12} &= \gamma_2 \, \eta_1 \, \lambda_1 \, \gamma_\theta \, \frac{\vev{\theta_1}^2 \vev{\theta_2}}{M_1 M_2 M_{10,3}} + \gamma_1 \, \eta_2 \, \lambda_1 \, \gamma_\theta \, \frac{\vev{\theta_1}^2 \vev{\theta_2}}{M_1 M_2 M_{10,4}} + \gamma_1 \, \lambda_{10} \, \eta'_2 \, \gamma_\theta \, \frac{\vev{\theta_1}^2 \vev{\theta_2}}{M_1 M_{10,3} M_{10,4}} \;,\nonumber \\
y_{22} &= \gamma_2 \, \eta'_2 \, \gamma_\theta \frac{\vev{\theta_1}^2}{M_1 M_{10,3}} \;,\nonumber \\
y_{13} &= \gamma_1 \, \lambda_{10} \, \gamma_3 \, \frac{\vev{\theta_1}\vev{\theta_2}}{M_{10,3} M_{10,4}} \;, \nonumber \\
y_{23} &= \gamma_2 \, \gamma_3 \, \frac{\vev{\theta_1}}{M_{10,3}}\;,
\end{align}
and $y_{33}$ is a renormalizable Yukawa coupling coefficient. 

Removing the messenger pair $Z_2$, $\bar Z_2$ from the spectrum eliminates two
supergraphs and thus the first two terms contributing to $y_{12}$.\footnote{A different pair of fields $Z_2'$, $\bar Z_2'$ would need to be introduced anyway (using different charge assignment than $Z_2$,$\bar Z_2$) in order to generate $y_{11}$. The charges can be assigned such that no extra contributions to $y_{12}$ appear.} However, without $Z_2 \bar Z_2$, evaluating $y_{12}$ yields
\begin{align}
y_{12} & = \gamma_1 \, \lambda_{10} \, \eta'_2 \, \gamma_\theta \, \frac{\vev{\theta_1}^2\vev{\theta_2}}{M_1 M_{10,3} M_{10,4}} = y_{13} \, \frac{\vev{\theta_1}}{M_1} \, \frac{\eta'_2\, \gamma_\theta}{\gamma_3} \nonumber \\
& = y_{13} \, y_{22} \, \frac{M_{10,3}}{\vev{\theta_1}}\frac{1}{\gamma_2\gamma_3} = \frac{y_{13}y_{22}}{y_{23}}\;.
\end{align}
This relation is not phenomenologically viable as it would imply $\theta_C = \theta_{13} / \theta_{23}$. To fit $Y_u$ to the observed data an additional degree of freedom is needed. In our model this is realised through $Z_2\bar Z_2$ enabling additional diagrams contributing to $y_{12}$.

\section{Simultaneous R-symmetry and GUT breaking}
\label{app:SR}

It was shown in the literature that the MSSM with an additional R-symmetry cannot
be obtained from the spontaneous breaking of a four-dimensional (SUSY) GUT
\cite{Fallbacher:2011xg}. On the other hand, in flavour models, R-symmetries
are often used in superpotentials that generate the required VEVs for the family symmetry
breaking Higgs
fields (i.e. the flavons) as well as for spontaneous breaking of CP. In the following we
present some simple examples that show that it is possible - in particular with discrete
R-symmetries - to simultaneously break the GUT gauge group and the R-symmetry.
Without an R-symmetry below the Planck scale, there is no conflict to the statement of
\cite{Fallbacher:2011xg}. We will also illustrate that GUT flavour models can rely on
such a discrete R-symmetry for the flavon VEV alignment, such that our setup can be
used to construct flavour models with non-Abelian family symmetries and spontaneous
CP violation.

\paragraph{Simple Example}
Consider as example a discrete $\mathbb{Z}_{4}^R$ R-symmetry under which
the superfield $S$ is charged and the superfield $H$ is uncharged (for the
beginning we assume them to be SU(5) singlets). The fields
are also charged under an additional conventional $\mathbb{Z}_4$ symmetry
with charges $1$ and $3$ respectively. We consider only the
required lowest order superpotential terms:
\begin{equation}
W_R =\mu_1 S H + \lambda_1 S^3 H^3/M_{\text{Pl}}^3 + \lambda_2 S H^5/M_{\text{Pl}}^3 + \lambda_3 S^5 H/M_{\text{Pl}}^3\,.
\end{equation}
The F-terms for this simple example lead to the conditions
\begin{align}
\mu_1 H + 3 \lambda_1 S^2 H^3/M_{\text{Pl}}^3 + \lambda_2 H^5/M_{\text{Pl}}^3 + 5 \lambda_3 S^4 H/M_{\text{Pl}}^3 =& 0 \\
\mu_1 S + 3 \lambda_1 S^3 H^2/M_{\text{Pl}}^3 + 5 \lambda_2 S H^4/M_{\text{Pl}}^3 + \lambda_3 S^5/M_{\text{Pl}}^3  =& 0 \,.
\end{align}
These equations have several solutions but here we are only interested in the
non-trivial solution $\vev{S}^4 = (\lambda_2/\lambda_3) \langle H \rangle^4$  and
$\langle H \rangle^4 = -\mu_1 M_{\text{Pl}}^3/(3\lambda_1 \sqrt{\lambda_2/\lambda_3} + 6 \lambda_2)$.
If we assign $S$ and $H$ under SU(5) as adjoints (similarly to
superpotential (c) where we had the two fields $H_{24}$ and $H'_{24}$)
$W_R$ is SU(5) invariant (by taking the appropriate contractions) and
importantly, the R-symmetry and SU(5) are simultaneously broken by the
non-trivial VEV configuration. We can achieve a phenomenological viable model by breaking a discrete R-symmetry at the GUT scale.

In general, leaving additional symmetry (or symmetries) unspecified, $S$ may
be a singlet of the GUT symmetry group and we denote the (polynomial) functions of the superfield
$H$ that make the respective terms invariant as $A(H)$, $C(H)$ (which include
the associated couplings and $M_{\text{Pl}}$ suppressions). We write
\begin{equation}
W_R = S A(H) + S^3 C(H) \,,
\end{equation}
leading to the F-term equations
\begin{align}
A(H) + 3 \, C(H) S^2 &= 0 \;,\\
\frac{\text{d}A(H)}{\text{d}H} S + \frac{\text{d}C(H)}{\text{d}H} S^3 &= 0 \,.
\end{align}
The generalised solution breaking the R-symmetry and the GUT gauge group is then
\begin{align}
\vev{S}^2 &= - \VEV{\frac{ A(H)}{ 3 \, C(H)}} \;, \\
\VEV{\frac{\text{d}A(H)}{\text{d}H}} &= \VEV{\frac{A(H)}{ 3 \, C(H)} \frac{ \text{d}C(H)}{\text{d}H}} \,.
\end{align}
Note the second equation, $\tfrac{\text{d}A}{A}=\tfrac{\text{d}C}{3 C}$, constrains the allowed
functions $A(H)$ and $C(H)$ which indirectly imposes conditions on the unspecified
additional symmetries.

\paragraph{Generalisations}
While the R-symmetry must be discrete for the crucial interplay between two
terms, it needs not be a $\mathbb{Z}_4^R$ and generalising to other
$\mathbb{Z}_N^R$  symmetries is straightforward
\begin{equation}
W_R^N = S A(H) + S^{N-1} C(H) \,.
\end{equation}
This can be further generalised to multiple fields. Consider, for instance,
$S$, $R$ and $T$ charged under a $\mathbb{Z}_4^R$, some unspecified
symmetries with fields $H$, $I$, $J$, and generalized functions
$A_{S,R,T}(H,I,J)$ as well as $C(H,I,J)$, such that keeping only the
necessary lowest order terms in $S$, $R$, $T$ we write
\begin{equation}
W_R = S \, A_S(H,I,J) + R \, A_R(H,I,J) + T \, A_T(H,I,J)+ S \, R \, T \,C(H, I, J) \,.
\end{equation}
After some manipulation we find again a non-trivial solution breaking
$\mathbb{Z}_4^R$ and the GUT symmetry
\begin{equation}
\frac{A_S}{\langle RT \rangle} = \frac{A_R}{\langle T S \rangle} = \frac{A_T}{\langle SR \rangle} = - C \,,
\end{equation}
provided the derivatives with respect to each of the GUT superfields $H$, $I$, $J$ fulfill
\begin{equation}
\frac{\partial A_S}{A_S} + \frac{\partial A_R}{A_R} + \frac{\partial A_T}{A_T} + \frac{\partial C}{C}=0 \,.
\end{equation}
The simple examples above illustrate that a simultaneous breaking of
R- and GUT symmetry is possible. VEV alignments required by flavour
models can therefore still be obtained within GUTs by having additional
superfields charged under the R-symmetry.

\paragraph{Flavour Alignment}
As a very simple example
we discuss the ``alignment'' of a GUT singlet flavon $\phi$ charged under a non-Abelian
family symmetry via a driving field $P$. As additional symmetries we impose
$\mathbb{Z}_4^R$ and a conventional $\mathbb{Z}_n$. The allowed
renormalizable superpotential is
\begin{equation}
\label{eq:DVAM}
W = P \left( \frac{ \phi^n }{\Lambda^{n-2}} + M^2  \right) + \kappa P^3\;,
\end{equation}
where $\Lambda$ is a generic messenger scale and $M$ a mass parameter.
Minimising the F-term conditions we find two possible solutions:
\begin{align}
\text{solution A: }& \langle P \rangle = 0  \text{ and } \langle \phi \rangle^n = - M^2 \Lambda^{n-2}  \;, \\
\text{solution B: }& \langle P \rangle^2 = - \frac{M^2}{3 \, \kappa}  \text{ and } \langle \phi \rangle = 0 \;.
\end{align}
In a flavour model we want the flavon to get a non-vanishing VEV so that
we would adopt solution A there. Furthermore solution A shows how the
discrete vacuum alignment method~\cite{Antusch:2011sx} can be generalised
to models with discrete R-symmetries. This method was invented in the context
of spontaneous CP violation. If CP is promoted to be fundamental all the phases
of the parameters in eq.~\eqref{eq:DVAM} are fixed and hence the phase of the
flavon VEV is fixed as well (up to a discrete choice).

\end{appendix}


\end{document}